\documentclass[10pt]{article}

\usepackage{graphicx}

\usepackage[all,cmtip]{xy}

\usepackage{setspace}
\usepackage{bbm,youngtab}
\usepackage{amscd,mathrsfs}
\usepackage{amssymb,amsmath,dsfont,amsfonts,amsthm}
\usepackage{euscript,enumerate}
\usepackage{stmaryrd}
\usepackage[cbgreek]{textgreek} 
\usepackage{booktabs}
\usepackage{multirow}
\usepackage{hhline} 
\usepackage{color}

\DeclareMathAlphabet{\mathpzc}{OT1}{pzc}{m}{it}

\allowdisplaybreaks 

\usepackage{authblk}
\usepackage{natbib}
\usepackage{color, soul}
\usepackage{enumerate}
\usepackage{empheq}
\usepackage{rotating}
\usepackage{ftnxtra}
\usepackage{fnpos}
\usepackage[titletoc,toc,title]{appendix}
\usepackage{euscript}
\usepackage{graphicx}
\usepackage{epsfig}
\usepackage{epstopdf}
\DeclareGraphicsExtensions{.pdf,.png,.jpg,.eps}
\usepackage{pstool}
\usepackage{upgreek}
\usepackage{mathrsfs}
\usepackage{tikz-cd}

\usepackage{psfrag}

\numberwithin{equation}{section}

\theoremstyle{plain}	
\newtheorem{thm}{Theorem}[section]

\newtheorem*{prop*}{Proposition}
\theoremstyle{definition}	

\newtheorem{remark}[thm]{Remark}
\newtheorem{example}[thm]{Example}

\usepackage{caption}
\usepackage{subcaption}

\usepackage{hyperref}
\hypersetup{colorlinks=true, linkcolor=blue}
\hypersetup{colorlinks=true,citecolor=blue}

\DeclareMathAlphabet{\mathpzc}{OT1}{pzc}{m}{it}

\usepackage{amsmath, amsthm, amssymb}

\usepackage{cleveref}

\DeclarePairedDelimiter\abs{\lvert}{\rvert}

\makeatletter
\newsavebox{\@brx}
\newcommand{\llangle}[1][]{\savebox{\@brx}{\(\m@th{#1\langle}\)}%
  \mathopen{\copy\@brx\mkern2mu\kern-0.9\wd\@brx\usebox{\@brx}}}
\newcommand{\rrangle}[1][]{\savebox{\@brx}{\(\m@th{#1\rangle}\)}%
  \mathclose{\copy\@brx\mkern2mu\kern-0.9\wd\@brx\usebox{\@brx}}}%
\let\oldabs\abs
\def\abs{\@ifstar{\oldabs}{\oldabs*}}
\makeatother

\usepackage{accents}
\newcommand{\Fe}{\accentset{e}{\mathbf{F}}}

\newcommand{\Fa}{\accentset{a}{\mathbf{F}}}

\newcommand{\tac}{\accentset{+}{\tau}}
\newcommand{\tab}{\accentset{-}{\tau}}
\newcommand{\phiac}{\accentset{+}{\varphi}}
\newcommand{\phiab}{\accentset{-}{\varphi}}
\newcommand{\Fac}{\accentset{+}{\mathbf{F}}}
\newcommand{\Fab}{\accentset{-}{\mathbf{F}}}
\newcommand{\Vac}{\accentset{+}{\mathbf{V}}}
\newcommand{\Vab}{\accentset{-}{\mathbf{V}}}
\newcommand{\rac}{\accentset{+}{r}}
\newcommand{\rab}{\accentset{-}{r}}
\newcommand{\lambdac}{\accentset{+}{\lambda}}
\newcommand{\lambdab}{\accentset{-}{\lambda}}
\newcommand{\velac}{\accentset{+}{\upsilon}}
\newcommand{\velab}{\accentset{-}{\upsilon}}

\newcommand{\Phiac}{\accentset{+}{\Phi}}
\newcommand{\Phiab}{\accentset{-}{\Phi}}
\newcommand{\Uac}{\accentset{+}{\mathbf{U}}}
\newcommand{\Uab}{\accentset{-}{\mathbf{U}}}
\newcommand{\uac}{\accentset{+}{\mathbf{u}}}
\newcommand{\uab}{\accentset{-}{\mathbf{u}}}
\newcommand{\Wac}{\accentset{+}{\mathbf{W}}}
\newcommand{\Wab}{\accentset{-}{\mathbf{W}}}
\newcommand{\wac}{\accentset{+}{\mathbf{w}}}
\newcommand{\wab}{\accentset{-}{\mathbf{w}}}
\newcommand{\Nac}{\accentset{+}{\mathbf{N}}}
\newcommand{\Nab}{\accentset{-}{\mathbf{N}}}
\newcommand{\nac}{\accentset{+}{\mathbf{n}}}
\newcommand{\nab}{\accentset{-}{\mathbf{n}}}

\newcommand{\Omegaac}{\accentset{+}{\Omega}}
\newcommand{\Omegaab}{\accentset{-}{\Omega}}
\newcommand{\Aac}{\accentset{+}{\mathcal{A}}} 
\newcommand{\Aab}{\accentset{-}{\mathcal{A}}} 
\newcommand{\omegaac}{\accentset{+}{\omega}}
\newcommand{\omegaab}{\accentset{-}{\omega}}
\newcommand{\Xac}{\accentset{+}{X_0}}
\newcommand{\Xab}{\accentset{-}{X_0}}

\newcommand{\timeablation}{t_{\text{ablation}}}

\usepackage{mathtools}

    {\end{bmatrix}}%

\usepackage[all,cmtip]{xy}
\usepackage{bm}

\usepackage{enumitem}

\begin{document}
\bibliographystyle{abbrvnat}

\title{\textbf{Accretion-Ablation Mechanics}}

\author[1]{Satya Prakash Pradhan}
\author[1,2]{Arash Yavari\thanks{Corresponding author, e-mail: arash.yavari@ce.gatech.edu}}
\affil[1]{\small \textit{School of Civil and Environmental Engineering, Georgia Institute of Technology, Atlanta, GA 30332, USA}}
\affil[2]{\small \textit{The George W. Woodruff School of Mechanical Engineering, Georgia Institute of Technology, Atlanta, GA 30332, USA}}

\maketitle

\begin{abstract}
In this paper we formulate a geometric nonlinear theory of the mechanics of accreting-ablating bodies. This is a generalization of the theory of accretion mechanics of \citet{Sozio2019}. More specifically, we are interested in large deformation analysis of bodies that undergo a continuous and simultaneous accretion and ablation on their boundaries while under external loads. In this formulation the natural configuration of an accreting-ablating body is a time-dependent Riemannian $3$-manifold with a metric that is an unknown a priori and is determined after solving the accretion-ablation initial-boundary-value problem. In addition to the time of attachment map, we introduce a time of detachment map that along with the time of attachment map, and the accretion and ablation velocities describes the time-dependent reference configuration of the body. The kinematics, material manifold, material metric, constitutive equations, and the balance laws are discussed in detail. As a concrete example and application of the geometric theory, we analyze a thick hollow circular cylinder made of an arbitrary incompressible isotropic material that is under a finite time-dependent extension while undergoing  continuous ablation on its inner cylinder boundary and accretion on its outer cylinder boundary. The state of deformation and stress during the accretion-ablation process, and the residual stretch and stress after the completion of the accretion-ablation process are computed.
\end{abstract}

\begin{description}
\item[Keywords:] Accretion, ablation, surface growth, nonlinear elasticity, residual stress, geometric mechanics.
\end{description}

\tableofcontents

\section{Introduction}  \label{Sec:Introduction}

There are numerous examples of structures in Nature and engineering that are built through an accretion process and/or during their lifetime experience ablation.
As examples in Nature one can mention growth of biological tissues and crystals, formation of planetary objects, volcanic and sedimentary rock formations, snow and ice cover build-up, glacier accumulation and ablation, etc.
As engineering applications one can mention additive manufacturing and $3$D printing, metal solidification, construction of concrete structures in successive layers, construction of masonry structures, the deposition of thin films, ice accretion on an aircraft wing that may lead to degradation of aerodynamic performance, and laser ablation of polymers, etc.

The first accretion mechanics problem was solved by \citet{Southwell1941}, namely the analysis of thick-walled cylinders manufactured by wire winding of an initial elastic tube.
The problem of a growing planet subject to self-gravity was studied in a seminal paper by \citet{brown1963gravitational}. Their analysis was done in the setting of linear elasticity and it was observed that accretion may induce residual stresses. In the seminal work of \citet{Skalak1982,skalak1997kinematics}, the kinematics of surface growth was formulated and  a time of attachment map was introduced.
Another notable work is the paper of \citet{metlov1985accretion} who formulated a large-deformation theory of aging viscoelastic solids undergoing accretion. He also introduced a time of attachment map as part of the kinematic description of accretion.
There have been many more works on the mechanics of surface growth in the literature  \citep{arutyunyan1990mathematical,manzhirov1995general,Drozdov1998a,Drozdov1998b,ong2004equations,kadish2005stresses,hodge2010continuum,lychev2011universal,lychev2013mathematical,manzhirov2014mechanics,lychev2013mathematical,lychev2013reference,Tomassetti2016,Sozio2017,Zurlo2017,Zurlo2018,Abi2018,Truskinovsky2019,Sozio2019,Sozio2020,Abi2020,Bergel2021,Lychev2021,Yavari2023Accretion}, see \citep{naumov1994mechanics} and \citep{Sozio2017} for more detailed reviews. 

In classical nonlinear elasticity the reference configuration is a fixed manifold (a fixed set of material points equipped with a metric inherited from the Euclidean ambient space). Motion is a one-parameter family of maps from the fixed reference configuration to the Euclidean ambient space. In anelasticity (in the sense of \citet{Eckart1948}) the reference configuration is a fixed manifold equipped with a metric that explicitly depends on the source of anelasticty, e.g., temperature changes, swelling, bulk growth, remodeling, defects, etc. In this more general setting, motion is still a map from the fixed reference configuration to the Euclidean ambient space. However, the natural distances in the reference configuration are measured using the non-flat metric of the material manifold.
Theory of anelasticity has been used in modeling a very large class of problems, e.g., thermoelasticity \citep{stojanovic1964finite,ozakin2010geometric,Sadik2017Thermoelasticity}, mechanics of distributed defects \citep{YavariGoriely2012a, YavariGoriely2012b, YavariGoriely2013, Yavari2014Discombinations, Sadik2016, Golgoon2017Defects}, swelling and cavitation \citep{pence2005swelling,pence2006swelling,pence2007bulk,goriely2010elastic,moulton2011anticavitation}, and bulk growth \citep{Epstein2000,amar2005growth,Yavari2010Growth,Goriely2017}.
Anealsticity has traditionally been formulated using the multiplicative decomposition of deformation gradient into an elastic and an anelastic part: $\mathbf{F}=\Fe\Fa$.\footnote{For detailed historical accounts of this decomposition see \citep{Sadik2017,YavariSozio2023}.}
This leads to the notion of the so-called ``intermediate configuration", which is usually defined only locally.\footnote{See \citep{Goodbrake2021,YavariSozio2023} for discussions on global intermediate configurations (manifolds).} A more natural approach would be to follow \citet{Eckart1948} and \citet{Kondo1949,Kondo1950} and use a Riemannian material manifold, which is a fixed manifold with a possibly time-dependent metric if the source of anealsticity is time dependent \citep{Sozio2020Anelasticity}.\footnote{Anelastic bodies are non-Euclidean in the sense that their reference configurations are not Euclidean, in general. The term \emph{Non-Euclidean solids} has been used interchangeably for anelastic bodies in the recent literature \citep{Zurlo2017,Zurlo2018,Truskinovsky2019} (this term was coined by Henri Poincar\'e \citep{Poincare1905}). Also, anelastic plates have been called \emph {non-Euclidean plates} in the literature \citep{Efrati2009}.} 
For example, in the case of bodies undergoing bulk growth the reference configuration is a Riemannian manifold $(\mathcal{B},\mathbf{G}_t)$, where $\mathcal{B}$ is a fixed $3$-manifold with a time-dependent Riemannian metric $\mathbf{G}_t$ \citep{Yavari2010Growth}.

For accreting-ablating bodies the set of material points is time dependent; the reference configuration is a time-dependent set $\mathcal{B}_t$. In the absence of any other anelastic source, the initial body $\mathcal{B}_0$ has a flat metric $\mathbf{G}_0$ that is inherited from the Euclidean ambient space (the induced Euclidean metric).
A point on the boundary of the body is either non-active or active. At an active point of the boundary and at a given time $t$ there is either accretion (addition of new material) or ablation (removal of material); accretion and ablation cannot occur simultaneously at the same point and at the same time.
Stress-free or pre-stressed material can be added on the accretion boundary.
The material metric is controlled by both the state of stress of the material before joining the body and the state of deformation of the body at the time of attachment. It is known that accretion may induce residual stresses due to the non-flatness of the material metric \citep{Sozio2017,Sozio2019} and this is the nonlinear analogue of \citet{brown1963gravitational}'s observation.

Most of the existing works in the literature of surface growth mechanics are restricted to accretion problems without ablation. Exceptions are \citep{Papadopoulos2010,Bergel2021}, and \citep{naghibzadeh2022accretion}. \citet{naghibzadeh2022accretion} considered both accretion and ablation in a Eulerian large-deformation setting and analyzed accretion and ablation problems without explicitly using an evolving reference configuration. They used the governing equations in a Eulerian setting with  $\Fe$ (the elastic part of the deformation gradient) as their kinematic descriptor. They recovered some of the results from \citep{Sozio2017} in their Eulerian formulation for an infinite thick hollow cylinder accreting on its outer surface with a hydrostatic pressure applied on its inner surface. 
They also analyzed a thick hollow spherical shell undergoing accretion through its fixed inner boundary while ablation takes place on its traction-free outer boundary.

Recently, the nonlinear geometric accretion theory developed in \citep{Sozio2017,Sozio2019} was used to solved two classes of problems: i) time-dependent finite extension of incompressible isotropic accreting circular cylindrical bars \citep{Yavari2023Accretion}, and ii) time-dependent finite torsion of incompressible isotropic accreting circular cylindrical bars \citep{Yavari2022Torsion}. In the absence of accretion, these deformations are subsets of Family $3$ universal deformations \citep{Ericksen1954}. It was shown that even in the presence of cylindrically-symmetric accretion these deformations are universal \citep{Yavari2022Torsion,Yavari2023Accretion}. 

Let us consider a body in a time-dependent motion. This body while undergoing large deformations is simultaneously growing on part of its boundary by absorbing mass while it is losing mass on another part of its boundary. We are interested in understanding the mechanics of such accreting-ablating systems. There are four questions that any mechanical/mathematical model of accretion-ablation should be able to answer: 
\begin{itemize}[topsep=0pt,noitemsep, leftmargin=18pt]
\item [i)] What is the state of deformation and stresses during a process of accretion-ablation? 
\item [ii)] At the end of an accretion-ablation process and after removing the external loads, what is the state of deformation and internal stresses (residual stresses) in the body? 
\item [iii)] Now the final structure is put under service loads. How can one analyze such a residually-stressed structure? 
\item [iv)] How should an accretion-ablation process be designed in order to build structures with the desired distribution of residual stresses, and the optimum stiffness/compliance under service loads? 
\end{itemize}
In the second part of this paper we show how our theory can be used in answering the first three questions. The fourth problem will be studied in a future communication.

This paper is organized as follows. In \S2, a geometric theory of accretion-ablation mechanics is formulated. The kinematics, material manifold, material metric, constitutive equations, and the balance laws are discussed in detail. An example is analyzed in detail in \S\ref{Sec:Example} as an application of the geometric theory. More specifically, a thick hollow cylinder under finite time-dependent extension while undergoing simultaneous accretion and ablation is analyzed. Both displacement and force-control loadings are considered. The state of deformation and stress during accretion-ablation and the effect of loading during accreting-ablation on the residual stress distribution are analyzed. Several numerical examples are presented in the case of incompressible neo-Hookean solids.
Conclusions are given in \S\ref{Sec:Conclusions}.

\section{A continuum theory of accreting-ablating bodies}

\begin{figure}[t!]
\centering
\vskip 0.0in
\includegraphics[width=.65\textwidth]{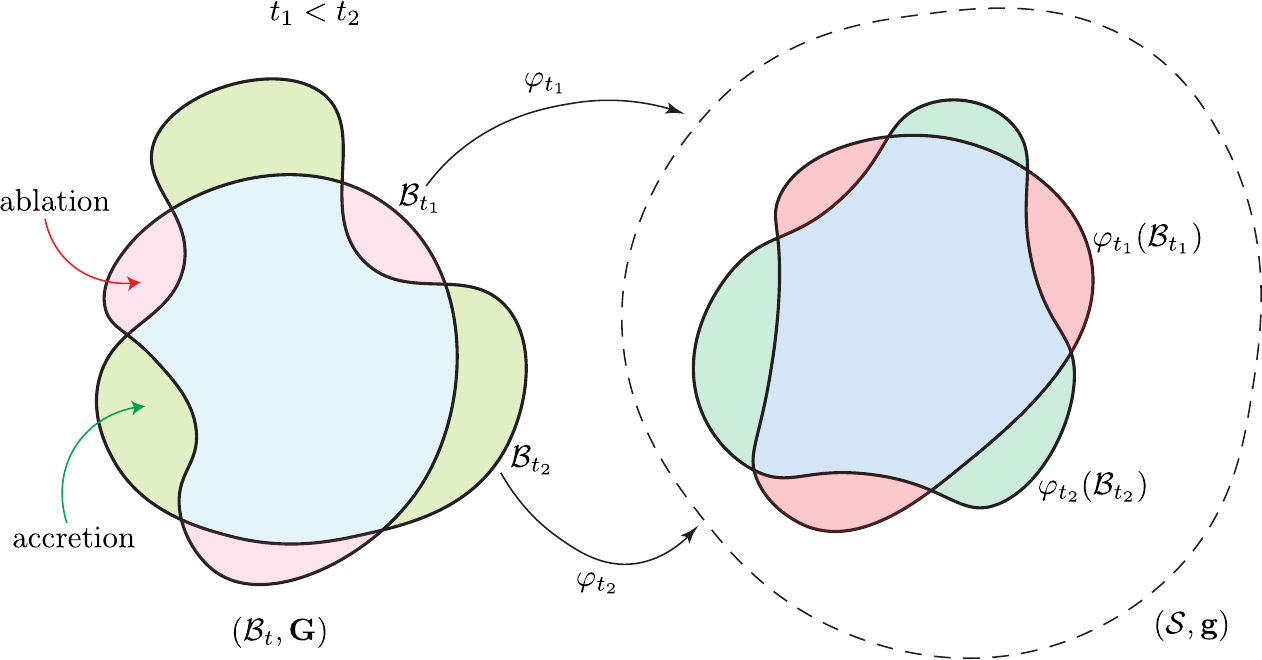}
\vskip 0.1in
\caption{Motion of an accreting-ablating elastic body. The material manifold is a time-dependent Riemannian manifold $(\mathcal{B}_t,\mathbf{G})$ whose metric is to be determined. The deformed body is embedded in the Euclidean ambient space $(\mathcal{S},\mathbf{g})$, where $\mathbf{g}$ is a time-independent background metric.}
\label{fig:motion}
\end{figure}

In this section we formulate a large-deformation continuum theory of accreting-ablating bodies. This theory is a generalization of the accretion theory of \citet{Sozio2019}. More specifically, we are interested in the state of deformation and internal stresses of a body that while under external loads undergoes simultaneous ablation and accretion on its boundary, see Fig.~\ref{fig:motion}.

\subsection{Riemannian geometry and kinematics of finite deformations}

In this section we tersely review some basic concepts of Riemannian geometry that are used in the description of the kinematics of accreting-ablating bodies. 
A smooth $3$-manifold $\mathcal{B}$ is a topological space that locally looks like the three-dimensional Euclidean space $\mathbb{R}^3$. A coordinate chart $\{X^A\}:\mathcal{B}\to\mathbb{R}^3$ is a local diffeomorphism.
The body in its reference configuration is identified with the $3$-manifold $\mathcal{B}$. The tangent space of $\mathcal{B}$ at a point $X \in \mathcal{B}$ is a three-dimensional linear space and is denoted by $T_X\mathcal{B}$. The three-dimensional Euclidean ambient space is denoted by $\mathcal{S}$. A local chart on $\mathcal{S}$ is denoted by $\{x^a\}$. Motion $\varphi_t:\mathcal{B}\to\mathcal{S}$ is a one-parameter family of smooth, invertible, and orientation preserving maps between the reference configuration and the ambient space (more precisely, motion is a curve $t\mapsto\varphi_t$ in the space of all configurations of $\mathcal{B}$). 
The derivative (tangent) map of $\varphi_t$ is denoted by $\mathbf{F}_t=T\varphi_t$, and is a two-point tensor, i.e., is a linear map between tangent spaces of two different manifolds, $\mathbf{F}(X,t):T_X\mathcal{B}\to T_x\mathcal{C}_t$, where $x=\varphi(X)$ and $\mathcal{C}_t=\varphi_t(\mathcal{B})$. 
In the continuum mechanics literature $\mathbf{F}$ is called deformation gradient. This term may be misleading as gradient is a metric-dependent operator while the tangent map is not.
With respect to the local coordinate charts $\{X^A\}$ and $\{x^a\}$ for $\mathcal{B}$ and $\mathcal{S}$, respectively, deformation gradient has the following representation 
\begin{equation} \label{Deformation-Gradient}
    \mathbf{F}=F^a{}_A\,\frac{\partial}{\partial x^a}\otimes dX^A,\qquad 
    F^a{}_A=\frac{\partial \varphi^a}{\partial X^A}\,.
\end{equation}
The set of vectors $\big\{\frac{\partial}{\partial x^a}\big\}$ and co-vectors ($1$-forms) $\{dX^A\}$ are bases for the tangent space $T_x\mathcal{C}$ and co-tangent space $T^*_X\mathcal{B}$, respectively.
A smooth vector field on $\mathcal{B}$ is a smooth assignment of vectors (elements of the tangent space) to points of $\mathcal{B}$. Thus, for a smooth vector field $\mathbf{W}$ on $\mathcal{B}$, $\mathbf{W}_X\in T_X\mathcal{B}$, $X\mapsto \mathbf{W}_X\in T_X\mathcal{B}$ varies smoothly. 
Push-forward of a vector field on $\mathcal{B}$ by the deformation mapping is a vector field in the ambient space and is defined as $\varphi_*\mathbf{W} = T\varphi \cdot \mathbf{W} \circ \varphi^{-1}$.
Pull-back of a vector field $\mathbf{w}$ on $\mathcal{C}=\varphi(\mathcal{B})$ is a vector field on $\mathcal{B}$ and is defined as $\varphi^*\mathbf{w}=T(\varphi^{-1}) \cdot \mathbf{w} \circ \varphi$.
The push-forward and pull-back of vectors have the following coordinate representations: $(\varphi_*\mathbf{W})^a=F^a{}_A \,W^A$, and $(\varphi^*\mathbf{w})^A=F_a{}^{-A} \,w^a$.

A bilinear map $\mathbf{T}: T_{X}\mathcal{B}\times T_{X}\mathcal{B} \to  \mathbb{R}$ is a $({}^{0}_{2})$-tensor at $X \in \mathcal{B}$. 
In a local coordinate chart $\{X^A\}$ for $\mathcal{B}$ and arbitrary vectors $\mathbf{U}$ and $\mathbf{W}$, one writes $\mathbf{T}(\mathbf{U},\mathbf{W})=   T_{AB}\,U^{A}\,W^{B}$.
Let us consider an inner product $\mathbf{G}_X$ on the tangent space $T_X\mathcal{B}$ that varies smoothly, i.e., if $\mathbf{U}$ and $\mathbf{W}$ are vector fields on $\mathcal{B}$, then $X \mapsto \mathbf{G}_X(\mathbf{U}_X,\mathbf{W}_X)$ is a smooth function. A positive-definite bilinear form $\mathbf{G}$ is called a metric. The inner product induced by the metric $\mathbf{G}_X$ is denoted by $\llangle .,.\rrangle_{\mathbf{G}_X}$. The manifold $\mathcal{B}$ equipped with a smooth metric $\mathbf{G}$ is called a Riemannian manifold and is denoted as $(\mathcal{B},\mathbf{G})$.
Distances in the ambient space are calculated using the background Euclidean metric $\mathbf{g}$. The Riemannian manifold $(\mathcal{S},\mathbf{g})$ is called the ambient space manifold.
In elasticity, the reference configuration inherits a flat metric $\mathbf{G}_0$ from the ambient space and the material manifold $(\mathcal{B},\mathbf{G}_0)$ is flat.
In anelasticity the natural distances explicitly depend on the source of anelasticity and so does the material metric $\mathbf{G}$, which is non-flat, in general.
For the two Riemannian manifolds $(\mathcal{B},\mathbf{G})$ and $(\mathcal{C},\mathbf{g})$, and the deformation mapping $\varphi:\mathcal{B}\to\mathcal{C}$, push-forward of the metric $\mathbf{G}$ is denoted by $\varphi_*\mathbf{G}$, which is a metric on $\mathcal{C}=\varphi(\mathcal{B})$, and is defined as 
\begin{equation}
	\llangle \mathbf{u}_{x},\mathbf{w}_{x}\rrangle_{(\varphi*\mathbf{G})_{x}}
	=\llangle (\varphi^*\mathbf{u})_X,(\varphi^*\mathbf{w})_X\rrangle_{\mathbf{G}_X}\,,
\end{equation}
where $x=\varphi(X)$. It has components $(\varphi_*\mathbf{G})_{ab} = F_a{}^{-A}\, F_b{}^{-B}\, G_{AB}$.
Similarly, the pull-back of the metric $\mathbf{g}$ is denoted by $\varphi^*\mathbf{g}$ and is a metric in $\varphi^{-1}(\mathcal{C})=\mathcal{B}$ defined as 
\begin{equation} \label{pulled-back-metric}
	\llangle\mathbf{U}_X,\mathbf{W}_X\rrangle_{(\varphi^*\mathbf{g})_X}
	=\llangle (\varphi_*\mathbf{U})_{x},(\varphi_*\mathbf{W})_{x} \rrangle_{\mathbf{g}_{x}} \,.
\end{equation}
It has components, $(\varphi^*\mathbf{g})_{AB} = F^a{}_A \,F^b{}_B \,g_{ab}$.
The two Riemannian manifolds $(\mathcal{B},\mathbf{G})$ and $(\mathcal{C},\mathbf{g})$ are called isometric if $\mathbf{G}=\varphi^*\mathbf{g}$, or equivalently, $\mathbf{g}=\varphi_*\mathbf{G}$. In this case, $\varphi$ is called an isometry. 

The adjoint of deformation gradient $\mathbf{F}^{\star}(X,t): T^*_x\mathcal{C}_t\to T^*_X\mathcal{B}$ is defined such that 
\begin{equation} 
	\langle \boldsymbol{\alpha},\mathbf{F}\mathbf{W}\rangle
	=\langle \mathbf{F}^{\star}\boldsymbol{\alpha},\mathbf{W}\rangle\,,\quad 
	\forall \mathbf{W}\in T_X\mathcal{B}\,,~\boldsymbol{\alpha}\in T^*_X\mathcal{C}
	\,,
\end{equation}
where $\langle .,. \rangle$ is the natural paring of $1$-forms and vectors, e.g., $\langle \boldsymbol{\alpha},\mathbf{w}\rangle=\alpha_a\,w^a$, and $T^*_X\mathcal{B}$ and $T^*_x\mathcal{C}_t$ are the co-tangent spaces of $\mathcal{B}$ at $X$ and $\mathcal{C}_t$ at $x$, respectively. $\mathbf{F}^{\star}$ has the following coordinate representation 
\begin{equation} 
	\mathbf{F}^{\star}(X,t)=\frac{\partial\varphi^a(X,t)}{\partial X^A} \,dX^A 
	\otimes \frac{\partial}{\partial x^a}\,.
\end{equation}
The transpose of the deformation gradient $\mathbf{F}^{\mathsf{T}}(X,t): T_x\mathcal{C}_t\to T_X\mathcal{B}$ is defined such that
\begin{equation} 
	\llangle \mathbf{F}\mathbf{U},\mathbf{w} \rrangle_{\mathbf{g}}
	=\llangle \mathbf{U},\mathbf{F}^{\mathsf{T}}\mathbf{w} \rrangle_{\mathbf{G}}\,,\quad 
	\forall \mathbf{U}\in T_X\mathcal{B}\,,~\mathbf{w}\in T_X\mathcal{C}
	\,.
\end{equation}
It has the components $\big(F^{\mathsf{T}}\big)^A{}_a=G^{AB}F^b{}_B\,g_{ba}$, or $\mathbf{F}^{\mathsf{T}}=\mathbf{G}^{\sharp}\mathbf{F}^{\star}\mathbf{g}$. 

The Jacobian of deformation $J(X,t)$ relates the undeformed and deformed volume elements as $dv\circ\varphi(X,t)=J(X,t)\,dV(X,t)$.\footnote{Any Riemannian manifold has a natural volume form. For the Riemannian manifolds $(\mathcal{B},\mathbf{G})$ and $(\mathcal{C},\mathbf{g})$ the volume $3$-forms are denoted by $\boldsymbol{\mu}_{\mathbf{G}}$ and $\boldsymbol{\mu}_{\mathbf{g}}$, respectively. With respect to the coordinate charts $\{X^A\}$ and $\{x^a\}$ for $\mathcal{B}$ and $\mathcal{C}$, respectively, they have the following representations 
\begin{equation} 
	\boldsymbol{\mu}_{\mathbf{G}}=\sqrt{\det\mathbf{G}}\,dX^1\wedge dX^2\wedge dX^3\,,\qquad
	\boldsymbol{\mu}_{\mathbf{g}}=\sqrt{\det\mathbf{g}}\,dx^1\wedge dx^2\wedge dx^3\,,
\end{equation}
where $\wedge$ is the wedge product of differential forms. In terms of the volume forms the Jacobian is defined as $\varphi^*\boldsymbol{\mu}_{\mathbf{g}}=J\,\boldsymbol{\mu}_{\mathbf{G}}$.} 
It is defined as
\begin{equation}
	J(X,t)=\sqrt{\frac{\det \mathbf{g}(\varphi(X,t))}{\det\mathbf{G}(X)}}\det \mathbf{F}(X,t)\,.
\end{equation}
For incompressible materials, $J(X,t)=1$.

The material velocity is defined as $\mathbf{V}(X,t)=\frac{\partial \varphi(X,t)}{\partial t}\in T_{\varphi_t(X)}\mathcal{C}_t$. The spatial velocity is defined as $\mathbf{v}_t(x)=\mathbf{V}_t\circ\varphi_t^{-1}(x)\in T_x\mathcal{C}_t$, where $x=\varphi_t(X)$. 
The material acceleration is defined as $\mathbf{A}(X,t)=D^{\mathbf{g}}_{t}\mathbf{V}(X,t)=\nabla^{\mathbf{g}}_{\mathbf{V}(X,t)}\mathbf{V}(X,t)\in T_{\varphi_t(X)}\mathcal{C}_t$, where $D^{\mathbf{g}}_{t}$ is the covariant derivative along the curve $\varphi_t(X)$ in $\mathcal{C}_t$. In components, $A^a=\frac{\partial V^a}{\partial t}+\gamma^a{}_{bc}V^bV^c$. 
The spatial acceleration is defined as $\mathbf{a}_t(x)=\mathbf{A}_t\circ\varphi_t^{-1}(x)\in T_x\mathcal{C}_t$. It has components, $a^a=\frac{\partial v^a}{\partial t}+\frac{\partial v^a}{\partial x^b}v^b+\gamma^a{}_{bc}v^bv^c$. Equivalently, the spatial acceleration is the material time derivative of $\mathbf{v}$, i.e., $\mathbf{a}=\dot{\mathbf{v}}=\frac{\partial \mathbf{v}}{\partial t}+\nabla_{\mathbf{v}}^{\mathbf{g}}\mathbf{v}$.

\begin{figure}[t!]
\centering
\vskip 0.0in
\includegraphics[width=.25\textwidth]{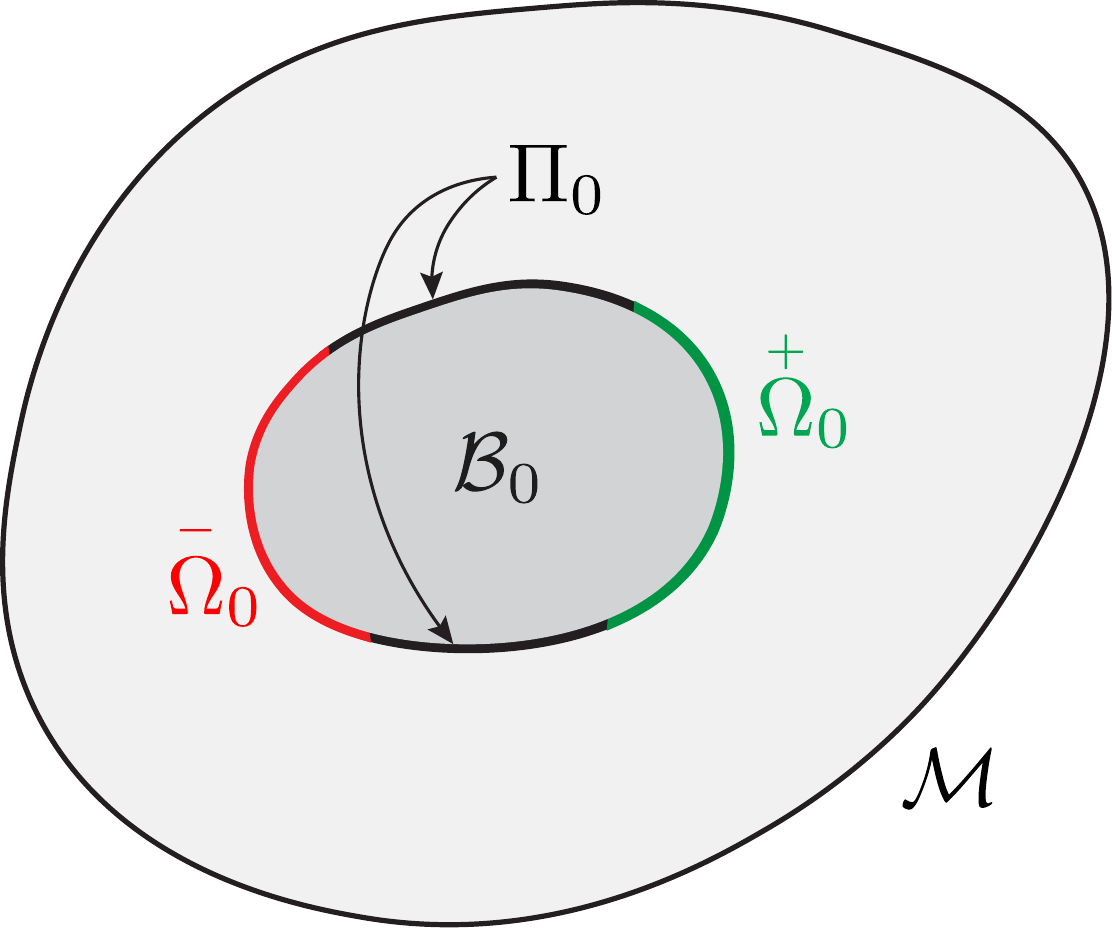}
\vskip 0.1in
\caption{An initial body $\mathcal{B}_0$ in the material ambient space $\mathcal{M}$ and partitioning of its boundary into accretion, ablation, and inactive surfaces.}
\label{fig:InitialBody}
\end{figure}

\subsection{Kinematics of accreting-ablating bodies}

Consider an initial stress-free body $\mathcal{B}_0$ that inherits the (flat) metric $\mathbf{G}_0$ from the Euclidean ambient space. 
The boundary of the initial body is partitioned as $\partial\mathcal{B}_0=\Omegaac_0\sqcup\Omegaab_0\sqcup\Pi_0$, where $\Omegaac_0$, $\Omegaab_0$, and $\Pi_0$ are the accretion, ablation, and inactive parts of the boundary, respectively, and $\sqcup$ denotes the disjoint union of sets (see Fig.~\ref{fig:InitialBody}).
It should be noted that $\mathcal{B}_0$ is the initial body prior to the onset of accretion-ablation, $\Omegaac_0$ represents the portion of $\partial\mathcal{B}_0$ where accretion is about to occur, and $\Omegaab_0$ comprises of the points that are about to leave once the process starts at $t=0$.
In other words, as soon as accretion-ablation begins, the body ceases to be the set $\mathcal{B}_0$; the points on $\Omegaab_0$ leave and new points are attached onto $\Omegaac_0$.

\begin{figure}[t!]
\centering
\vskip 0.0in
\includegraphics[width=.65\textwidth]{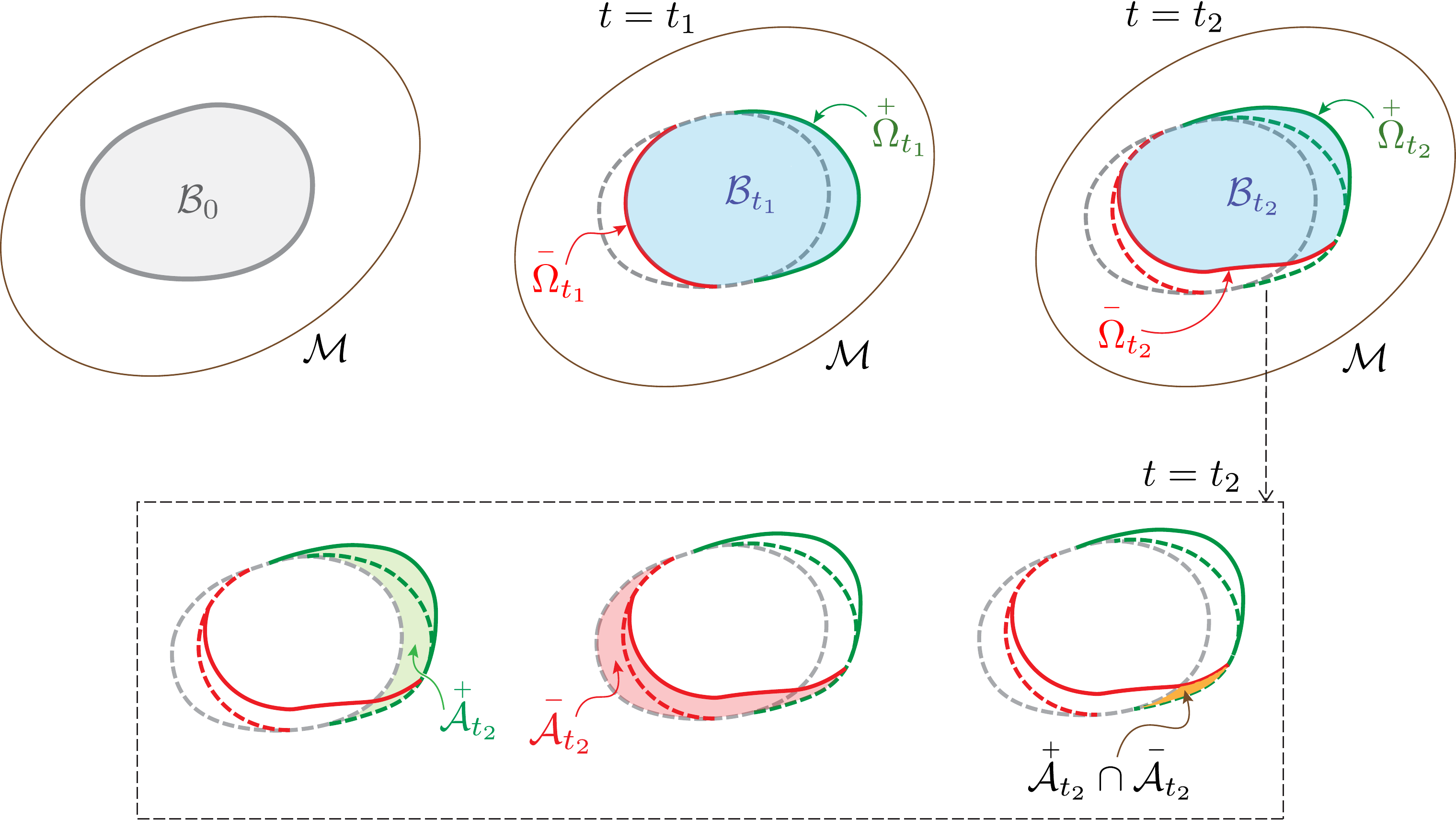}
\vskip 0.1in
\caption{The material manifold is shown at times $t_1$ and $t_2$ where $t_1<t_2$. The accretion boundaries are shown in green and the ablation boundaries are shown in red. It should be noted that the set of all those points that have been ablated until time $t_2$ may include points other than those in the initial body.}
\label{fig:Rev2}
\end{figure}

Let $\mathcal{M} \supset \mathcal{B}_0$ be a connected and orientable $3$-manifold with boundary embeddable in $\mathbb{R}^3$. 
This is called the \emph{material ambient space}.
The accretion-ablation process occurs in the time interval $t \in [0,T]$, where $T$ is the final time. Let us define a time of attachment map $\tac: (\mathcal{M}\setminus \mathcal{B}_0 )\cup \Omegaac_0  \to [0,T]$ and a time of detachment map $\tab: \Aab_T \subset \mathcal{M} \to [0,T]$, where $\Aab_T$ is the subset of $ \mathcal{M}$ that is ablated in the time interval $[0,T]$. Note that $\tac^{-1}(0)=\Omegaac_0$, and $\tab^{-1}(0)=\Omegaab_0$.
Let us define the reference configuration at time $t$ as
\begin{equation}
	\mathcal{B}_t:=\left( \mathcal{B}_0 \cup \tac^{-1}(0,t] \right)\setminus \tab^{-1}[0,t)\,,
	\quad \forall t\in[0,T]\,,
\end{equation}
which is constructed by removing all the points that have been ablated out before time $t$ from the union of $\mathcal{B}_0$ and the points accreted onto $\mathcal{B}_0$ up to time $t$.\footnote{Note that $\tac^{-1}[0,T]= \tac^{-1}(0) \cup \tac^{-1}(0,T]= \Omegaac_0  \cup  (\mathcal{M}\setminus \mathcal{B}_0 )=  \mathcal{M}\setminus (\text{int}\,\mathcal{B}_0 \cup \Omegaab_0 \cup \Pi_0)$, and
$\tab^{-1}(0,T]=   \tab^{-1}[0,T] \setminus \tab^{-1}(0) = \Aab_T \setminus\Omegaab_0$.}
It is assumed that the differentials d$\tac$ and d$\tab$ never vanish. Let us introduce the level sets 
\begin{equation}
	\Omegaac_t= \tac^{-1}(t)\,, \qquad \Omegaab_t= \tab^{-1}(t)\,,\quad t \in [0,T]	\,,
\end{equation}
and define
\begin{equation}
	\Aac_t=\displaystyle \bigcup\limits_{\tac\in [0,t]} \Omegaac_{\tac}\,,\qquad 
	\Aab_t=\displaystyle \bigcup\limits_{\tab\in [0,t]} \Omegaab_{\tab}\,, 
	\quad  t \in [0,T]\,.
\end{equation}
It is assumed that $\Omegaac_t$ and $\Omegaab_t$ are $2$-manifolds. It is also assumed that all $\Omegaac_t$'s are diffeomorphic to $\Omegaac_0$ and all $\Omegaab_t$'s are diffeomorphic to $\Omegaab_0$.\footnote{Clearly, this is a restrictive assumption. One way to remove this restriction is to divide the analysis into time intervals in which there is no change in the topology of either $\Omegaac_t$ or $\Omegaab_t$.}
Since accretion and ablation cannot take place at a given point and at the same time, it is required that
\begin{equation}
	\Omegaac_t \cap \Omegaab_t = \emptyset\,, \quad \forall\, t\in [0,T]\,.
\end{equation}
It should be emphasized that the intersection $\Aac_t \cap \Aab_t $ may be nonempty,\footnote{For a point $X\in\Aac_t \cap \Aab_t$, which is accreted and subsequently ablated out, $\tac(X) <\tab(X)$. This also implies that as soon as a layer has been ablated it cannot be reattached to the body.} see Fig.~\ref{fig:Rev2}. Note that $\mathcal{M}=(\Aac_T\setminus \Omegaac_0) \cup \mathcal{B}_0$, and
\begin{equation}
	\mathcal{B}_t= \left(\mathcal{B}_0 \cup (\Aac_t\setminus \Omegaac_0) \right)\setminus 
	(\Aab_t\setminus \Omegaab_t ) \,, 	\quad t\in [0,T]\,.
\end{equation}

The deformation mapping $\varphi_t:\mathcal{B}_t \to \mathcal{C}_t$ is assumed to be a $C^1$ homeomorphism for each $t$, where $\mathcal{C}_t=\varphi_t(\mathcal{B}_t)$ is the current configuration of the accreting-ablating body. The deformed level sets are defined as $\omegaac_t=\varphi_t(\Omegaac_t)$, and $\omegaab_t=\varphi_t(\Omegaab_t)$.
We denote the non-active portion of the boundary by $\Pi_t=\partial \mathcal{B}_t \setminus (\Omegaac_t \cup \Omegaab_t)$ in the reference configuration and $\pi_t=\varphi_t(\Pi_t)$ in the deformed configuration, so that
\begin{equation}
	\partial \varphi_t (\mathcal{B}_t)=\varphi_t(\Omegaac_t \cup \Omegaab_t \cup \Pi_t)
	= \omegaac_t \cup \omegaab_t \cup \pi_t \,.
\end{equation} 
For $X\in \mathcal{M}$ and $\tac (X)\leq t < \tab(X)$, denote $\varphi(X,t)=\varphi_t(X)$. 
Let us define the following two time-independent maps
\begin{equation}
	\phiac(X):=\varphi(X,\tac(X))\,,  \qquad  \phiab(X):=\varphi(X,\tab(X)) \,,
\end{equation}
where $\phiac$ is defined on $\Aac_T$ and $\phiab$ is defined on $\Aab_T$. Note that for $t \in [0,T] $
\begin{equation}
	\phiac(\Aac_t)=\displaystyle \bigcup\limits_{{\tac}\in [0,t]} \omegaac_{\tac}\,, \qquad
	\phiab(\Aab_t)=\displaystyle \bigcup\limits_{{\tab}\in [0,t]} \omegaab_{\tab}
	\,.
\end{equation}
It should be noted that $\phiac$ and $\phiab$ need not be injective. 
Next, we define the following two-point tensors
\begin{equation}
	\Fac(X):=\mathbf{F}(X,\tac(X))\,,  \qquad 	\Fab(X):=\mathbf{F}(X,\tab(X)) \,,
\end{equation}
and call them the accretion and ablation \emph{frozen deformation gradients}, respectively.
The derivative maps $T\phiac$ and $T\phiab$ are neither injective nor invertible, in general. They are related to $\Fac$ and $\Fab$ as
\begin{equation} \label{Frozen-F}
\begin{aligned}
	(T\phiac)^i{}_J &=\frac{\partial \varphi^i (X,\tac(X))}{\partial X^J} 
	= F^i{}_J(X,\tac(X))+V^i(X,\tac(X)) \frac{\partial \tac(X)}{\partial X^J} 
	=\accentset{+}{F}^i{}_J+  \accentset{+}{V}^i \frac{\partial \tac}{\partial X^J} \,, \\
	(T\phiab)^i{}_J &=\accentset{-}{F}^i{}_J+\accentset{-}{V}^i \frac{\partial {\tab}}{\partial X^J} \,,
\end{aligned}
\end{equation}
where $\Vac(X):=\mathbf{V}_{\tac(X)}(X)$, and $\Vab(X):=\mathbf{V}_{\tab(X)}(X)$. 
The coordinate-free forms of \eqref{Frozen-F} read
\begin{equation}
	T\phiac= \Fac+ \Vac \otimes \text{d}\tac\,, \qquad
	T\phiab= \Fab+ \Vab \otimes \text{d}{\tab} \,.
\end{equation}
For $t\in[0,T]$, on each layer, one has $\Fac|_{T\Omegaac_t}=T\phiac|_{T\Omegaac_t}=\mathbf{F}_t|_{T\Omegaac_t}$, and $\Fab|_{T\Omegaab_t}=T\phiac|_{T\Omegaab_t}=\mathbf{F}_t|_{T\Omegaab_t}$.

\subsection{The Growth velocity and the material metric}

The material metric of an accreting-ablating body at a given material point depends on the state of stress of the accreting particles, the growth velocity, and the state of deformation of the body at the time of attachment of the new material point. One can define an \emph{accretion tensor} via the frozen deformation gradient and a material growth velocity. Material metric is then defined to be the pull-back of the ambient space metric by the accretion tensor \citep{Sozio2017}. The time dependence of an accreting body can be modeled as a material motion in the material ambient space. It turns out that given a time of attachment map such material motions are not unique; instead one can define an equivalence class of material motions. Using any member of the equivalence class results in the same material metric. In the presence of ablation one has two material motions. Again, these are not unique. For calculating the material metric only the accretion material motion is needed as we are not concerned with the material metric of a particle after it has left the body. In the following we briefly review the construction of the material metric.

In an accreting-ablating body, what is given is the rate at which material points are added to the body on the accretion boundary and the rate at which material points are removed from the ablation boundary. These two vector fields may be unknown fields in the presence of phase change, for example. In this paper, we assume that they are given.
It should be emphasized that the material analogues of the accretion and ablation velocities are not defined uniquely.

Recall that we have assumed that for $t\in(0,T]$, $\Omegaac_t$ is diffeomorphic to $\Omegaac_0$ and $\Omegaab_t$ is diffeomorphic to $\Omegaab_0$. This guarantees the existence of $1$-parameter families of diffeomorphisms $\Phiac:\Omegaac_0 \times [0,T] \to \mathcal{M}$, and $\Phiab:\Omegaab_0 \times [0,T] \to \mathcal{M}$, such that $\Phiac(\Omegaac_0,t)=\Omegaac_t$, and $\Phiab(\Omegaab_0,t)=\Omegaab_t,~\forall t\in[0,T]$. 
However, these diffeomorphisms are not unique.
Let us fix two diffeomorphisms $\Phiac$ and $\Phiab$.
For $X\in \Aac_T$, there is a unique $\Xac \in \Omegaac_0$ such that $\Phiac(\Xac,\tac(X))=\Phiac_{\tac(X)}(\Xac)=X$. 
Similarly, for $X\in \Aab_T$ there is a unique $\Xab \in \Omegaab_0$ such that $\Phiab(\Xab,{\tab}(X))=\Phiab_{{\tab}(X)}(\Xab)=X$.\footnote{We use the notation $\accentset{\pm}{\Phi}_t(X)=\accentset{\pm}{\Phi}(X,t)$.}
The maps $\Phiac$ and $\Phiab$ define a pair of motions of layers in the material manifold, with which we associate the following velocities
\begin{equation}
	\Uac(\Xac,t)=\frac{\partial}{\partial t} \Phiac(\Xac,t)\,, \qquad
	\Uab(\Xab,t)= \frac{\partial}{\partial t} \Phiab(\Xab,t) \,.
\end{equation} 
Differentiating the relations $\tac(\Phiac(\Xac,t))=t$, and ${\tab}(\Phiab(\Xab,t))=t$ with respect to time, one obtains the following constraints
\begin{equation} \label{Material-Motion-Constraint}
	\big\langle \text{d}\tac (X), \Uac(\Xac,t) \big\rangle=1\,, \qquad 
	\big\langle \text{d}{\tab}(X), \Uab(\Xab,t) \big\rangle=1 \,,
\end{equation}
where $\Phiac(\Xac,t)=X=\Phiab(\Xab,t)$.
It should be emphasized that for a given pair of time of attachment and time of detachment maps, material motion is not uniquely defined; there are infinitely many material motions corresponding to a given pair $(\tac,\tab)$. These equivalent material motions satisfy the constraint \eqref{Material-Motion-Constraint}$_1$ \citep{Sozio2019} and correspond to isometric material metrics. 

One can define a material motion $(\Phiac_t,\Phiab_t)$ using foliation charts $(\Xi^1,\Xi^2,\tac)$ on $\Aac_T$, and $(\Theta^1,\Theta^2,{\tab})$ on $\Aab_T$, where $(\Xi^1,\Xi^2)$ and $(\Theta^1,\Theta^2)$ are the in-layer coordinates on $\Omegaac_{\tac}$ and $\Omegaab_{\tab}$, respectively, while $\tac$ and ${\tab}$ are globally defined on $\Aac_T$ and $\Aab_T$, respectively. 
The material motion preserves the in-layer coordinates $(\Xi^1,\Xi^2)$ and $(\Theta^1,\Theta^2)$.
This means that if $X_0 \in\Omegaac_0$ has coordinates $(\Xi^1,\Xi^2,0)$, then $\Phiac_t$ maps it to the point $X_t$ with coordinates $(\Xi^1,\Xi^2,t)$. Similarly, if $X_0 \in\Omegaab_0$ has coordinates $(\Theta^1,\Theta^2)$, then $\Phiab_t$ maps it to the point $X_t$ with coordinates $(\Theta^1,\Theta^2,t)$.
Trajectories of the material motion are the curves with constant layer coordinates, i.e., $\tac$-lines and $\tab$-lines. The growth and ablation velocities corresponding to $\Phiac_t$ and $\Phiab_t$, respectively, are defined as
\begin{equation}
	\Uac=\left. \frac{\partial}{\partial \tac} \right\rvert_{\Xi^1,\Xi^2} \,,\qquad
	\Uab=\left. \frac{\partial}{\partial \tab} \right\rvert_{\Theta^1,\Theta^2} \,.
\end{equation}
Notice that $\left\langle \text{d}\tac, \frac{\partial}{\partial \tac} \right\rangle=1$, and $\left\langle \text{d}{\tab},\frac{\partial}{\partial{\tab}}\right\rangle=1$. 

Let us next calculate the total velocities of $\omegaac_t$ and $\omegaab_t$. The map $\phiac\circ \Phiac: \Omegaac_0\times [0,T] \to \mathcal{S}$ tracks the deformed accretion surface $\omegaac_t$, and $\phiab\circ \Phiab: \Omegaab_0\times[0,T] \to \mathcal{S}$ tracks the deformed ablation surface $\omegaab_t$. The total velocities are calculated as
\begin{equation}
\begin{aligned}
	\Wac(X) &=\frac{\text{d}}{\text{d}t}\phiac\big( \Phiac(\Xac,t)\big)
	=[\Fac(X)+ \Vac(X) \otimes \text{d}\tac(X)]\Uac(\Xac,t)
	=\Fac(X)\Uac(\Xac,t)+\Vac(X) \,,\\
	\Wab(X) &=\frac{\text{d}}{\text{d}t}\phiab\big(\Phiab(\Xab,t)\big)
	=[\Fab(X)
	+ \Vab(X) \otimes \text{d}{\tab}(X)]\Uab(\Xab,t)
	=\Fab(X)\Uab(\Xab,t)+\Vab(X) \,.
\end{aligned}
\end{equation}
The spatial velocities are defined as $\wac\circ \varphi=\Wac$ and $\wab\circ \varphi=\Wab$.
Let $\uac_t$ be a vector field describing the velocity at which material is being added onto $\omegaac_t$, i.e., $-\uac_t$ is its relative velocity with respect to $\omegaac_t$ just before attachment, and let $-\uab_t$ be a vector field describing the velocity at which material is being removed from $\omegaab_t$, i.e., $\uab_t$ is its relative velocity with respect to $\omegaab_t$ just after detachment. If $\mathbf{n}_t$ denotes the unit normal to $\varphi_t(\partial\mathcal{B}_t)$, then we must have $(-\uac_t)\cdot\mathbf{n}_t<0$ for accretion, i.e.,  $\uac_t\cdot\mathbf{n}_t>0$, as the material would be moving towards the body. Similarly, in the case of ablation we need $\uab_t\cdot\mathbf{n}_t>0$ as the material removed would be moving away from the body.

For $X\in \mathcal{M}$, let us define the following time-independent tensor field
\begin{equation}
	\mathbf{Q}(X)=\Fac(X)+ \Big[\uac_{\tac(X)}\big(\phiac(X)\big)
	-\Fac(X)\Uac\big(\Phiac_{\tac(X)}^{-1}(X),\tac(X)\big)\Big]\otimes  \text{d}\tac (X) \,,
\end{equation}
where $\Phiac_{\tac(X)}^{-1}(X)$ can be understood as the pull back of $X\in \mathcal{M}$ to $\Omegaac_0$ along the trajectory induced by $\Phiac$. 
$\mathbf{Q}$ is called the \emph{accretion tensor} \citep{Sozio2019}.
Note that $\mathbf{Q}\big(\Phiac(\Xac,t)\big)\Uac\big(\Xac,t\big)=\uac\big((\phiac\circ \Phiac)(\Xac,t),t\big)$, $\forall~\Xac \in \Omegaac_0$.
Let $\mathbf{g}$ be the metric of the Euclidean ambient space $\mathcal{S}$. 
Assuming that the new material points are stress-free at their time of attachment the material metric on $\Aac_T$ is defined to be the pull-back of $\mathbf{g}$ using $\mathbf{Q}$, i.e.,\footnote{Generalizing the analysis to pre-stressed added material is straightforward \citep{Sozio2017}.}
\begin{equation}
	\mathbf{G}(X) 
	=\mathbf{Q}^{\star}(X)\,\mathbf{g}\big(\phiac(X)\big)\mathbf{Q}(X) \,,
\end{equation}
or in components, $G_{IJ}(X)=Q^I{}_J(X)\,g_{ij}(\phiac(X))\,Q^j{}_J(X)$.
Therefore, the material metric is written as
\begin{equation}
	\mathbf{G}(X)=
	\begin{cases}
	\mathbf{G}_0(X) & \text{for } X\in \mathcal{B}_0\,, \\
	\mathbf{Q}^{\star}(X)\,\mathbf{g}\big(\phiac(X)\big)\,\mathbf{Q}(X)
	& \text{for } X\in \Aac_T\,.
\end{cases}
\end{equation}
The natural distances in the accreting-ablating body are measured using the material metric $\mathbf{G}$.
Note that, for the layers that are joining the body, the material metric depends on the state of deformation of the body at the time of attachment. 
However, we are not interested in the state of deformation and stress of those layers that have already left the body. The material metric of those points after leaving the body has no effect on the rest of the body and does not appear in our accretion-ablation theory.

\begin{remark}
The map $\tac$ partitions the accreted manifold $(\Aac_T,\mathbf{G})$ into a collection of submanifolds $\big\{(\Omegaac_{\tac},\tilde{\mathbf{G}}_{\tac}):\tac\in [0,T]\big\}$. 
Similarly, ${\tab}$ partitions the ablated manifold $(\Aab_T,\mathbf{G})$ into another collection of submanifolds $\big\{(\Omegaac_{\tab},\tilde{\mathbf{G}}_{\tab}):{\tab}\in [0,T]\big\}$. 
Here, $\tilde{\mathbf{G}}_{\tac}$ and $\tilde{\mathbf{G}}_{\tab}$ are the first fundamental forms inherited from $\mathbf{G}$ by $\Omegaac_{\tac}$ and $\Omegaab_{\tab}$, respectively. 
Let $\Nac_{\tac}$ and $\Nab_{\tab}$ be the unit normals to $\Omegaac_{\tac}$ and $\Omegaab_{\tab}$, respectively, where orthonormality is with respect to the metric $\mathbf{G}$. Let $\nac_{\tac}$ and $\nab_{\tab}$ be the unit normals to $\omegaac_{\tac}$ and $\omegaab_{\tab}$, respectively (here orthonormality is with respect to the ambient metric $\mathbf{g}$). Let us decompose the accretion and ablation velocities as
\begin{equation}
	\Uac= \Uac^\parallel+\accentset{+}{U}^{\accentset{+}{N}}  \,\Nac\,, \qquad
	\Uab= \Uab^\parallel+ \accentset{-}{U}^{\accentset{-}{N}} \,\Nab \,,
\end{equation}
in the material manifold and
\begin{equation}
	\uac= \uac^\parallel+ \accentset{+}{u}^{\accentset{+}{n}}  \,\nac\,, \qquad
	\uab= \uab^\parallel+ \accentset{-}{u}^{\accentset{-}{n}}  \,\nab\,,
\end{equation}
in the deformed configuration, where we have $\accentset{+}{u}^{\accentset{+}{n}},\,\accentset{-}{u}^{\accentset{-}{n}}>0$. Note that, if $\mathbf{Y}$ is tangent to $\Omegaac$, then $\mathbf{Q}\mathbf{Y}$ is tangent to $\omegaac$. Notice that $\llangle \mathbf{Y},\mathbf{Q}^{-1}\nac \rrangle_{\mathbf{G}}=\llangle \mathbf{Q}\mathbf{Y},\nac \rrangle_{\mathbf{g}}=0$, and $\llangle \mathbf{Q}^{-1}\nac,\mathbf{Q}^{-1}\nac \rrangle_{\mathbf{G}}=\llangle \nac,\nac \rrangle_{\mathbf{g}}=1$.
Thus, it can be deduced that $\mathbf{Q}\Uac^\parallel=\uac^\parallel$, $\mathbf{Q}\Nac=\nac$, and hence $\accentset{+}{U}^{\accentset{+}{N}}=\accentset{+}{u}^{\accentset{+}{n}}\circ\varphi$.
\end{remark}

\subsection{Stress, strain, and constitutive equations}

\paragraph{Strain.}
A few different but related measures of strain are usually used in nonlinear elasticity and anelasticity \citep{MaHu1983,Ogden1984,Goriely2017,YavariSozio2023}.
The right Cauchy-Green strain is defined as $\mathbf{C}^{\flat}=\varphi^*\mathbf{g}=\mathbf{F}^{\star}\mathbf{g}\mathbf{F}$, which is the pulled-back of the spatial metric to the reference configuration. 
Here, $\flat:T_x\mathcal{C} \to T^*_x\mathcal{C}$ is the flat operator that maps a vector to its corresponding co-vector ($1$-form): $\mathbf{w}=w^a\frac{\partial}{\partial x^a} \mapsto \mathbf{w}^{\flat}=g_{ab}w^b\,dx^a$. In components, $C_{AB}=F^a{}_A\,g_{ab}F^b{}_B$.
The familiar definition of the right Cauchy-Green stain is $\mathbf{C}=\mathbf{F}^{\mathsf{T}}\mathbf{F}:T_X\mathcal{B}\to\mathcal{B}$, which has components $C^A{}_B=G^{AM}C_{MB}=(G^{AM}F^a{}_M\,g_{ab})F^b{}_B=\big(F^{\mathsf{T}}\big)^A{}_b\,F^b{}_B$.
The spatial analogue of the right Cauchy-Green strain is defined as $\mathbf{c}^{\flat}=\varphi_*\mathbf{G}=\mathbf{F}^{-\star}\mathbf{G}\mathbf{F}^{-1}$.
The left Cauchy-Green strain is defined as $\mathbf{B}^{\sharp}=\varphi^*\mathbf{g}^{\sharp}$, where
$\sharp:T^*_x\mathcal{C} \to T_x\mathcal{C}$ is the sharp operator that maps a co-vector ($1$-form) to its corresponding vector: $\boldsymbol{\omega}=\omega_a\,dx^a \mapsto  
  \boldsymbol{\omega}^{\sharp}=g^{ab}\omega_b\,\frac{\partial}{\partial x^a}$.
The left Cauchy-Green strain has components $B^{AB}=F^{-A}{}_a\,F^{-B}{}_b\,g^{ab}$, where $g^{ab}$ are components of the inverse spatial metric such that $g^{ac}g_{cb}=\delta^a_b$.
The spatial analogue of $\mathbf{B}^{\sharp}$ is defined as $\mathbf{b}^{\sharp}=\varphi_*\mathbf{G}^{\sharp}=\mathbf{F}\mathbf{G}^{\sharp}\mathbf{F}^{\star}$, which has components $b^{ab}=F^a{}_AF^b{}_B\,G^{AB}$.
The tensor $\mathbf{b}:T_x\mathcal{C}\to T_x\mathcal{C}$ is defined as $\mathbf{b}=\mathbf{b}^{\sharp}\mathbf{g}$. Similarly, $\mathbf{c}=\mathbf{g}^{\sharp}\mathbf{c}^{\flat}$. 
Notice that $\mathbf{c}\mathbf{b}=\mathbf{g}^{\sharp}\mathbf{c}^{\flat}\mathbf{b}^{\sharp}\mathbf{g}=\mathbf{g}^{\sharp}\mathbf{F}^{-\star}\mathbf{G}\mathbf{F}^{-1}  
\mathbf{F}\mathbf{G}^{\sharp}\mathbf{F}^{\star}\mathbf{g}=\mathbf{g}^{\sharp}\mathbf{F}^{-\star}\mathbf{G}\mathbf{G}^{\sharp}\mathbf{F}^{\star}\mathbf{g}=\mathbf{g}^{\sharp}\mathbf{F}^{-\star}\mathbf{F}^{\star}\mathbf{g}=\mathbf{g}^{\sharp}\mathbf{g}=\mathrm{id}_{\mathcal{S}}$, and hence $\mathbf{b} = \mathbf{c}^{-1}$. Similarly, $\mathbf{B} = \mathbf{C}^{-1}$.

\paragraph{Stress.}
For a hyper-elastic solid there exists an energy function $W=W(X,\mathbf{F},\mathbf{G},\mathbf{g})$. Material-frame-indifference (objectivity) implies that $W=\hat{W}(X,\mathbf{C}^{\flat},\mathbf{G})$.
The Cauchy $\boldsymbol{\sigma}$, the first Piola-Kirchhoff $\mathbf{P}$, and the second Piola-Kirchhoff $\mathbf{S}$ stress tensors are related to the energy function as 
\begin{equation} 
	\mathbf{P}= \mathbf{g}^\sharp \frac{\partial W}{\partial \mathbf{F}}\,,\qquad
	\boldsymbol{\sigma}	= \frac{2}{J}\frac{\partial W}{\partial \mathbf{g}}\,,\qquad
	\mathbf{S}= 2\frac{\partial W}{\partial \mathbf{C}^\flat} \,.
\end{equation}
They are also related as $\mathbf{S} = \mathbf{F}^{-1} \mathbf{P}$, and $\mathbf{P} = J \boldsymbol{\sigma} \mathbf{F}^{-\star}$.
Let us consider a surface element $dA$ in the reference configuration with unit normal vector $\mathbf{N}$. This surface element is mapped to its deformed surface element $da$ with unit normal $\mathbf{n}$. Traction $\mathbf{t}$ is related to the Cauchy stress as $\mathbf{t}=\boldsymbol{\sigma}\cdot\mathbf{n}=\llangle \boldsymbol{\sigma},\mathbf{n}\rrangle_{\mathbf{g}}$. In components, $t^a=\sigma^{ab}g_{bc}\,n^c$. The force acting on the deformed area element is $\mathbf{f}=\mathbf{t}da$. From the Piola identity $\mathbf{n}^{\flat}da=J\mathbf{F}^{-\star}\mathbf{N}^{\flat}dA$, or in components, $n_ada=J\,F^{-A}{}_aN_AdA$---Nanson's formula.
Thus, $\mathbf{f}=\mathbf{t}_0dA=\mathbf{P}\cdot\mathbf{N}dA$.

\paragraph{Material Symmetry.}
For an elastic body the material symmetry group $\mathcal{G}_X$ at $X\in\mathcal{B}$ with respect to the reference configuration $(\mathcal{B},\mathbf{G})$ is defined as 
\begin{equation} \label{Material-Sym0}
	W(X,\mathbf{K}^*\mathbf{F},\mathbf{K}^*\mathbf{G},\mathbf{g})
	=W(X,\mathbf{F}\mathbf{K},\mathbf{G},\mathbf{g})
	=W(X,\mathbf{F},\mathbf{G},\mathbf{g})\,, \qquad \forall\,\,
	\mathbf{K}\in \mathcal{G}_X\leqslant \mathrm{Orth}(\mathbf{G})\,,
\end{equation}
for any deformation gradient $\mathbf{F}$, where $\mathbf{K}:T_X\mathcal{B}\rightarrow T_X\mathcal{B}$ is an invertible linear transformation, and $\mathrm{Orth}(\mathbf{G})=\{\mathbf{Q}: T_X\mathcal{B}\rightarrow T_X\mathcal{B}~|~ \mathbf{Q}^{\star}\mathbf{G}\mathbf{Q}=\mathbf{G} \}$, and $\mathcal{G}_X\leqslant \mathrm{Orth}(\mathbf{G})$ means that $\mathcal{G}_X$ is a subgroup of $\mathrm{Orth}(\mathbf{G})$. Equivalently,
\begin{equation} 
	\hat{W}(X,\mathbf{K}^*\mathbf{C}^{\flat},\mathbf{K}^*\mathbf{G})
	=\hat{W}(X,\mathbf{K}^{\star}\mathbf{F}\mathbf{K},\mathbf{G})
	=\hat{W}(X,\mathbf{C}^{\flat},\mathbf{G})\,, \qquad \forall\,\,
	\mathbf{K}\in \mathcal{G}_X\leqslant \mathrm{Orth}(\mathbf{G})\,.
\end{equation}
In this paper, for the sake of simplicity, we restrict our calculations to isotropic solids. However, it should be emphasized that our accretion-ablation theory is not restricted to isotropic solids.

\paragraph{Isotropic solids.}
For an isotropic solid the symmetry group is the orthogonal group, i.e., the energy function is invariant under rotations in the reference configuration.
For an isotropic solid, $W$ depends only on the principal invariants of $\mathbf{C}^{\flat}$, i.e., $W=\overline{W}(X,I_1,I_2,I_3)$, where
\begin{equation}
\begin{aligned}
	I_1 &=\operatorname{tr}_{\mathbf{G}}\mathbf{C}=C^A{}_A=C_{AB}\,G^{AB}\,, \\
	I_2 &=\frac{1}{2}\left(I_1^2-\operatorname{tr}_{\mathbf{G}}\mathbf{C}^2\right)
	=\frac{1}{2}\left(I_1^2-C^A{}_B\,C^B{}_A\right)
	=\frac{1}{2}\left(I_1^2-C_{MB}\,C_{NA}\,G^{AM}G^{BN}\right)\,, \\
	I_3 &=\det\mathbf{C}=\frac{\det\mathbf{C}^{\flat}}{\det\mathbf{G}}\,.
\end{aligned}
\end{equation}
For an isotropic solid the Cauchy stress has the following representation \citep{DoyleEricksen1956,SimoMarsden1983}
\begin{equation}
	\boldsymbol{\sigma} = \frac{2}{\sqrt{I_3}} \left[ \left(I_2\,\overline{W}_2+I_3\,\overline{W}_3\right)
	\mathbf{g}^{\sharp} 
	+\overline{W}_1\,\mathbf{b}^{\sharp}-I_3\,\overline{W}_2\,\mathbf{c}^{\sharp} \right]\,,
\end{equation}
where $\overline{W}_i=\frac{\partial \overline{W}}{\partial I_i}$, $i=1,2,3$.
For an incompressible isotropic solid $I_3=1$, and hence
\begin{equation}
	\boldsymbol{\sigma} = -p\,\mathbf{g}^{\sharp} 
	+2\overline{W}_1\,\mathbf{b}^{\sharp}-2\,\overline{W}_2\,\mathbf{c}^{\sharp}\,,
\end{equation}
where $p$ is the Lagrange multiplier associated with the incompressibility constraint $J=\sqrt{I_3}=1$.

\subsection{The balance laws}

\subsubsection{Balance of mass}

Let $\rho_0(X)$ be the mass density field in the material configuration $\mathcal{M}$ and $\rho(X,t)$ be the mass density field in the deformed configuration $\varphi_t(\mathcal{B}_t)$. For a body undergoing accretion and ablation, mass is not conserved globally. However, local mass conservation still holds for $X \in \mathcal{B}_t$, away from $\Omegaac_t$ and $\Omegaab_t$, i.e.,
\begin{equation}
	\rho_0(X)=\rho\big(\varphi(X,t),t\big)\, J(X,t)\,.
\end{equation}
The total mass of the body at time $t$ is given by
\begin{equation}
	m(t)=\int_{\varphi_t(\mathcal{B}_t)}\rho\big(x,t\big) \text{d}v 
	= \int_{\mathcal{B}_t}\rho_0\big(X\big) \text{d}V\,.
\end{equation}
In terms of the material foliations one has
\begin{equation}
	m(t)=\int_{\mathcal{B}_0}\rho_0\text{d}V
	+\int_0^t\bigg(\int_{\Omegaac_{\tac}}\rho_0\, \accentset{+}{U}^{\accentset{+}{N}} \text{d}A \bigg)
	\text{d}{\tac} 
	-\int_0^t\bigg(\int_{\Omegaab_{\tab}}\rho_0\, \accentset{-}{U}^{\accentset{-}{N}} \text{d}A \bigg) 
	\text{d}{\tab} \,.
\end{equation}
The rate of change of mass can be expressed as
\begin{equation}
	\dot{m}(t)=\int_{\Omegaac_t}\rho_0\, \accentset{+}{U}^{\accentset{+}{N}}\, \text{d}A  
	-\int_{\Omegaab_t}\rho_0\, \accentset{-}{U}^{\accentset{-}{N}} \,\text{d}A
	=\int_{\omegaac_t}\rho_0\, \accentset{+}{u}^{\accentset{+}{n}} \,\text{d}a  
	-\int_{\omegaab_t}\rho_0\, \accentset{-}{u}^{\accentset{-}{n}}\, \text{d}a \,.
\end{equation}

\subsubsection{Balance of linear and angular momenta}

The local form of the balance of linear momentum  in terms of the Cauchy stress reads: 
\begin{equation}
	\operatorname{div}_\mathbf{g}\bm{\sigma}+\rho\mathbf{b}=\rho\mathbf{a} \,,
\end{equation}
where $\mathrm{div}_{\mathbf{g}}$ is divergence with respect to the spatial metric.
In components, $\left(\mathrm{div}_\mathbf{g}\boldsymbol \sigma\right)^a=\sigma^{ab}{}_{|b}=\frac{\partial \sigma^{ab}}{\partial x^b}+\sigma^{ac}\gamma^b{}_{cb} +\sigma^{cb}\gamma^a{}_{cb}$, where $\gamma^a{}_{bc}$ is the Christoffel symbol of the Levi-Civita connection $\nabla^{\mathbf{g}}$. 
In a local coordinate chart $\{x^a\}\,$, ${\nabla^{\mathbf{g}}}_ {\partial_b}\partial_c=\gamma^a{}_{bc}\,\partial_a$, where $\gamma^a{}_{bc}=\frac{1}{2}g^{ak}\left(g_{kb,c}+g_{kc,b}-g_{bc,k}\right)$.
$\mathbf{b}$ is the body force, and $\mathbf{a}$ is the spatial acceleration. 
The local form of the balance of angular momentum is the symmetry of the Cauchy stress, i.e., $\sigma^{ab}=\sigma^{ba}$.

The rate of change of linear momentum for the whole body is written as
\begin{equation}
\begin{aligned}
	\frac{\text{d}}{\text{d}t}\int_{\varphi_t(\mathcal{B}_t)}\rho\mathbf{v} \text{d}v
	=&\frac{\text{d}}{\text{d}t}\int_{\mathcal{B}_t}\rho_0\,\mathbf{V} \text{d}V \\
	=& \frac{\text{d}}{\text{d}t}\bigg[\int_{\mathcal{B}_0}\rho_0\,\mathbf{V} \text{d}V
	+\int_0^t\bigg(\int_{\Omegaac_{\tac}}\rho_0\, \accentset{+}{U}^{\accentset{+}{N}}
	\mathbf{V} \text{d}A\bigg) \text{d}{\tac} 
	-\int_0^t\bigg(\int_{\Omegaab_{\tab}}\rho_0\, \accentset{-}{U}^{\accentset{-}{N}}\mathbf{V} 
	\text{d}A\bigg) \text{d}{\tab}\bigg] \text{d}V \\
	=& \int_{\Aac_t}\rho_0\,\mathbf{A} \text{d}V
	+\int_{\Omegaac_t}\rho_0 \,\accentset{+}{U}^{\accentset{+}{N}}\mathbf{V} \text{d}A 
	-\int_{\Aab_t}\rho_0\,\mathbf{A} \text{d}V  
	-\int_{\Omegaab_t}\rho_0 \,\accentset{-}{U}^{\accentset{-}{N}}\mathbf{V} \text{d}A \\
	=& \int_{\mathcal{B}_t}\rho_0\mathbf{A} \text{d}V
	+\int_{\Omegaac_t}\rho_0 \,\accentset{+}{U}^{\accentset{+}{N}}\mathbf{V} \text{d}A  
	-\int_{\Omegaab_t}\rho_0\, \accentset{-}{U}^{\accentset{-}{N}}\mathbf{V} \text{d}A \,.
\end{aligned}
\end{equation}
Thus
\begin{equation}
	\frac{\text{d}}{\text{d}t}\int_{\varphi_t(\mathcal{B}_t)}\rho\,\mathbf{v} \text{d}v
	=\int_{\varphi_t(\mathcal{B}_t)}\rho\,\mathbf{a} \text{d}v
	+\int_{\omegaac_t}(\rho_0\circ \varphi) \accentset{+}{u}^{\accentset{+}{n}}\,\mathbf{v} \text{d}a 
	-\int_{\omegaab_t} (\rho_0\circ \varphi) \accentset{-}{u}^{\accentset{-}{n}} \,\mathbf{v} \text{d}a \,.
\end{equation}
Since $\operatorname{div}_\mathbf{g}\bm{\sigma}+\rho\mathbf{b}=\rho\mathbf{a}$, one writes
\begin{equation}
	\frac{\text{d}}{\text{d}t}\int_{\varphi_t(\mathcal{B}_t)}\rho\,\mathbf{v} \text{d}v
	=\int_{\varphi_t(\mathcal{B}_t)}\rho\,\mathbf{b} \text{d}v
	+\int_{\partial\varphi_t(\mathcal{B}_t)}\mathbf{t} \text{d}v
	+\int_{\omegaac_t}(\rho_0\circ \varphi) \accentset{+}{u}^{\accentset{+}{n}}\,\mathbf{v} \text{d}a  
	-\int_{\omegaab_t}(\rho_0\circ \varphi) \accentset{-}{u}^{\accentset{-}{n}}\, \mathbf{v} \text{d}a \,,
\end{equation}
where $\mathbf{b}$ is the body force and $\mathbf{t}$ is traction.

Let us decompose the traction vector as $\mathbf{t}=\mathbf{t}^\text{e}+\mathbf{t}^\pm$, where $\mathbf{t}^\text{e}$ is due to external loads and constraints, $\mathbf{t}^+$ is the effect of new particles being added and $\mathbf{t}^-$ is the effect of particles leaving the body. Since accretion and ablation cannot take place at the same point and at the same time, the traction on $\omegaac$ is $\mathbf{t}^\text{e}+\mathbf{t}^+$ and that on $\omegaab$ is $\mathbf{t}^\text{e}+\mathbf{t}^-$.
For a new particle that is being added to the body, the balance of linear momentum in the time interval $[t, t+\Delta t]$ implies that
\begin{equation}
	  \left[(\rho_0\circ \varphi)\,\accentset{+}{u}^{\accentset{+}{n}}\text{d}a \wedge \text{d}t\right]
	  (\mathbf{v}-\uac)
	  +(-\mathbf{t}^+ \text{d}a)\wedge \text{d}t
	= \left[(\rho_0\circ \varphi)\, \accentset{+}{u}^{\accentset{+}{n}}\text{d}a \wedge \text{d}t\right] \mathbf{v} \,.
\end{equation}
Similarly, for a particle that is leaving the body, the balance of linear momentum in the interval $[t, t+\Delta t]$ implies that
\begin{equation}
	 \left[(\rho_0\circ \varphi)\, \accentset{-}{u}^{\accentset{-}{n}}\text{d}a \wedge \text{d}t\right] \mathbf{v}
	= \left[(\rho_0\circ \varphi) \,\accentset{-}{u}^{\accentset{-}{n}}  \text{d}a \wedge \text{d}t\right]
	(\mathbf{v}+\uab)
	+(-\mathbf{t}^- \text{d}a)\wedge \text{d}t \,.
\end{equation}
The joining particles exert the force $\mathbf{t}^+ \text{d}a=-(\rho_0\circ \varphi)\, \accentset{+}{u}^{\accentset{+}{n}}\, \uac\,\text{d}a$ on the body, while the leaving particles exert the force $\mathbf{t}^- \text{d}a=(\rho_0\circ \varphi)\, \accentset{-}{u}^{\accentset{-}{n}}\, \uab\,\text{d}a$ on the body. 
Thus, the rate of change of the linear momentum of the body is written as 
\begin{equation}
	\frac{\text{d}}{\text{d}t}\int_{\varphi_t(\mathcal{B}_t)}\rho\,\mathbf{v}\, \text{d}v
	=\int_{\varphi_t(\mathcal{B}_t)}\rho\,\mathbf{b}\, \text{d}v
	+\int_{\partial\varphi_t(\mathcal{B}_t)}\mathbf{t}^\text{e} \text{d}v
	+\int_{\omegaac_t}(\rho_0\circ \varphi)\, \accentset{+}{u}^{\accentset{+}{n}}
	\,[\mathbf{v}-\uac] \,\text{d}a  
	-\int_{\omegaab_t}(\rho_0\circ \varphi)\, \accentset{-}{u}^{\accentset{-}{n}} 
	\,[\mathbf{v}-\uab] \,\text{d}a\,.
\end{equation}
Notice that the traction due to accretion/ablation is $\mathbf{t}^\pm=\mp(\rho_0\circ \varphi)\, \accentset{\pm}{u}^{\accentset{\pm}{n}}\, \accentset{\pm}{\mathbf{u}}$.

\section{Analysis of an accreting-ablating hollow cylindrical bar under finite extension} \label{Sec:Example}

In this section we present a detailed analysis of a thick hollow cylinder that while is under a time-dependent finite extension undergoes accretion on its outer cylinder boundary and ablation on its inner cylinder boundary.

\begin{figure}[t!]
\centering
\vskip 0.0in
\includegraphics[width=0.65\textwidth]{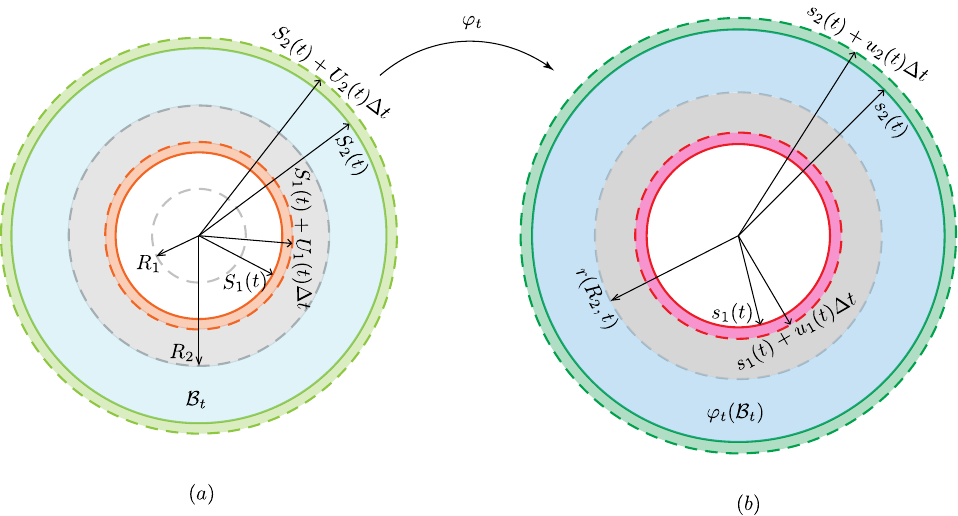}
\vskip 0.0in
\caption{Cross sections of the reference and current configurations of an accreting-ablating thick hollow cylinder. (a) The material manifold $(\mathcal{B}_t,\mathbf{G})$ at time $t$ has inner radial coordinate $S_1(t)$ and outer radial coordinate $S_2(t)$. Accretion is occurring on the outer surface and ablation on the inner surface. At time $t+\Delta t$, the inner radial coordinate is $S_1(t)+U_1(t)\Delta t$ and the outer radial coordinate is $S_2(t)+U_2(t)\Delta t$. (b) The deformed bar at time $t$ has inner radius $s_1(t)$ and outer radius $s_2(t)$. At time $t+\Delta t$, the inner radius is $s_1(t)+u_1(t)\Delta t$ and the outer radius is $s_2(t)+u_2(t)\Delta t$. 
}
\label{fig:CylinderMotion}
\end{figure}

\subsection{Kinematics}

Let us consider a hollow circular cylindrical bar with initial length $L$, inner radius $R_1$ and outer radius $R_2>R_1$. 
We assume a homogeneous isotropic and incompressible material with an energy function $W=W(I_1,I_2)$, and use the cylindrical coordinates $(R,\Theta,Z)$ in the reference configuration, and cylindrical coordinates $(r,\theta,z)$ in the current configuration. The metric of the current configuration has the following representation 
\begin{equation} \label{Metrics}
    \mathbf{g}=\begin{bmatrix}
  1 & 0  & 0  \\
  0 & r^2  & 0  \\
  0 & 0  & 1
\end{bmatrix}\,.
\end{equation}
Accretion is assumed to occur on the outer cylindrical boundary in the current configuration. 
The outer radius of the deformed body is denoted by $s_2(t)$ (Fig.~\ref{fig:CylinderMotion}).
Ablation is assumed to occur on the inner cylindrical boundary of the current configuration that has the radius $s_1(t)$. 
Let us assign a time of attachment $\tac(R)$ and a time of detachment ${\tab}(R)$ to each layer with the radial coordinate $R>R_1$ in the reference configuration. 
Notice that for $R_1\leq R \leq R_2$, $\tac(R)=0$. Hence, $\tac(R)$ is invertible for $R\geq R_2$, while ${\tab}(R)$ is invertible for $R\geq R_1$. We assume that $\tab$ and $\tac$ are diffeomorphisms with non-vanishing derivatives.
Their inverses are denoted by $S_1=\tab^{-1}$, and $S_2=\tac^{-1}$, so that 
\begin{equation}
	s_1(t)=r(S_1(t),t)\,,\qquad s_2(t)=r(S_2(t),t) \,.
\end{equation}
We also assume that accretion and ablation take place continuously in the time interval $t\in (0,T]$.  
It is assumed that the accreting-ablating body has non-vanishing volume at all times, i.e., $S_1(t)<S_2(t)$, $\forall~ t\in (0,T]$.
Let us consider a time-dependent finite extension of the bar such that it is slow enough for the inertial effects to be negligible and assume the following deformation mapping
\begin{equation} \label{Deformation}
   r=r(R,t)\,,\quad \theta=\Theta\,,\quad z=\lambda^2(t) Z\,,
\end{equation}
where $\lambda^2(t)$ is the axial stretch and is an unknown function to be determined.\footnote{In a displacement-control loading $\lambda(t)$ is given.} The deformation gradient reads
\begin{equation}
   \mathbf{F}=\mathbf{F}(R,t)=\begin{bmatrix}
  r_{,R}(R,t) & 0  & 0  \\
  0 & 1  & 0  \\
  0 & 0  & \lambda^2(t)
\end{bmatrix}\,.
\end{equation}

\subsection{The material metric}

Let us define the following functions
\begin{equation}
	\rab(R)=r(R,{\tab}(R))\,,\qquad
	\rac(R)=r(R,\tac(R))\,,\qquad 
	\lambdab(R)=\lambda({\tab}(R))\,,\qquad 
	\lambdac(R)=\lambda(\tac(R))\,.
\end{equation}
We assume that the accreted cylindrical layer at any instant of time $t$ is stress-free. In other words, stress-free cylindrical layers are continuously added to the outer cylindrical boundary of the bar.\footnote{It is straightforward to extend this analysis to pre-stressed accreting cylindrical layers \citep{Sozio2017}.} This implies that the material metric at $R=S_2(t)$ is the pull-back of the metric of the (Euclidean) ambient space, i.e.,
\begin{equation}
	\mathbf{G}(S_2(t))=\varphi_t^*\mathbf{g}(r(S_2(t),t))\,,~~\text{or}\qquad 
	\mathbf{G}(R)=\varphi_{\tac(R)}^*\,\mathbf{g}(r(R,\tac(R)))\,.
\end{equation}
In components, $G_{AB}(S_2(t))=G_{AB}(R)=F^a{}_A(R,\tac(R))\,F^b{}_B(R,\tac(R))\,g_{ab}(r(R,\tac(R)))$. For this accretion-ablation problem, the material manifold (the natural configuration of the body) is an evolving Riemannian manifold $(\mathcal{B}_t,\mathbf{G})$ with
\begin{equation}
	\mathcal{B}_t=\left\{ (R,\Theta,Z): 0\leq \Theta <2\pi\,, S_1(t) \leq R \leq S_2(t)\,,~
	0\leq Z \leq L \right\}\,.
\end{equation}
The reference configuration is equipped with the following material metric:
\begin{equation}
\begin{aligned}
 \mathbf{G}(R)=
\begin{cases} 
     \; 
\begin{bmatrix}
  1 & 0  & 0  \\
  0 & R^2  & 0  \\
  0 & 0  & 1
\end{bmatrix}\,, 
     & \quad S_1(t) \leq R < R_2 \,, \\
     & \\
\begin{bmatrix}
   [r_{,R}(R,\tac(R))]^2 & 0  & 0  \\
  0 & \rac^2(R)  & 0  \\
  0 & 0  & \lambda^4(\tac(R))
\end{bmatrix}\,,
 & \quad R_2 \leq R \leq  S_2(t)\,.
\end{cases} 
\end{aligned} 
\end{equation}
Observe that $\timeablation:=\tab(R_2)$ is the time when the initial body is completely ablated.\footnote{Note that $S_1(t) < R_2$ is equivalent to $t<\timeablation$. }

\subsection{The incompressibility constraint}
The Jacobian is calculated as
\begin{equation} 
	J(R,t)=\sqrt{\frac{\det\mathbf{g}}{\det\mathbf{G}}}\det\mathbf{F}
	=\begin{cases} 
	 \displaystyle    \frac{\lambda^2(t)\, r(R,t)\, r_{,R}(R,t)}{R}\,,     & S_1(t) < R \leq R_2 	\,, \\[10pt]
	\displaystyle    \frac{\lambda^2(t) \,r(R,t) \,r_{,R}(R,t)}{\lambdac^2(R)\, \rac(R) \,r_{,R}(R,\tac(R))} \,,
	& R_2 \leq R <  S_2(t)\,.
\end{cases} 
\end{equation}
When $t<\timeablation$, incompressibility in the region $S_1(t)\leq R \leq  R_2$ gives us
\begin{equation} \label{Kinematics-Initial}
	r(R,t)\,r_{,R}(R,t)=\frac{R}{\lambda^2(t)}\,,
\end{equation}
thus implying that
\begin{equation}
	r^2(R,t)=s_1^2(t)+\frac{R^2-S_1^2(t)}{\lambda^2(t)} \,. \label{rPrim}
\end{equation}
Similarly, for the region $ R_2 \leq R \leq S_2(t)$, incompressibility requires that
\begin{equation} \label{Kinematics-Accreted}
	\lambda^2(t)\, r(R,t)\, r_{,R}(R,t)=\lambdac^2(R)\, \rac(R) \,r_{,R}(R,\tac(R))\,,
\end{equation}
which can be integrated to obtain
\begin{equation}\label{rSec}
	r^2(R,t) =	r^2(R_2,t)+\frac{2}{\lambda^2(t)}\int_{R_2}^R \lambdac^2(\xi)\, \rac(\xi) 
	\,r_{,R}(\xi,\tac(\xi)) \text{d}\xi\,,
\end{equation}
for $t<\timeablation$.
Equivalently, one may integrate \eqref{Kinematics-Accreted} from $R$ to $S_2(t)$ to obtain:
\begin{equation}\label{raccretedincomp}
r^2(R,t) =s_2^2(t)-\frac{2}{\lambda^2(t)}\int_{R}^{S_2(t)} \lambdac^2(\xi)\, \rac(\xi) \,r_{,R}(\xi,\tac(\xi))\text{d}\xi \,,
\end{equation}
for $R \geq R_2$. Note that \eqref{raccretedincomp} holds for all $t>0$.

\subsection{The accretion and ablation velocities}

The accretion surfaces in the reference and the current configurations have the following representations:
\begin{equation} 
\begin{aligned}
	\Omegaac_t &=\left\{(S_1(t),\Theta,Z): 0\leq \Theta <2\pi\,, 0\leq Z \leq L \right\}\,, \\
	\omegaac_t &=\left\{(s_1(t),\Theta,\lambda^2(t) Z): 0\leq \Theta <2\pi\,, 0\leq Z \leq L \right\}\,.
\end{aligned}
\end{equation}
The ablation surfaces in the reference and the current configurations are represented as:
\begin{equation} 
\begin{aligned}
	\Omegaab_t &=\left\{(S_2(t),\Theta,Z): 0\leq \Theta <2\pi\,, 0\leq Z \leq L \right\}\,, \\
	\omegaab_t &=\left\{(s_2(t),\Theta,\lambda^2(t) Z): 0\leq \Theta <2\pi\,, 0\leq Z \leq L \right\}\,.
\end{aligned}
\end{equation}
Thus
\begin{equation} 
\begin{aligned} \label{accrlaw}
    \dot{s_1}(t)=r_{,R}(S_1(t),t)~U_1(t)+V(S_1(t),t)\,, \\
    \dot{s_2}(t)=r_{,R}(S_2(t),t)~U_2(t)+V(S_2(t),t) \,,
\end{aligned}
\end{equation}
where $U_1(t)=\dot{S_1}(t)>0$, and $U_2(t)=\dot{S_2}(t)>0$, i.e., both accretion and ablation surfaces are moving radially outward. Here, $V=\displaystyle\frac{\partial r}{\partial t}$ is the radial component of the material velocity on the accretion/ablation surface. 
We denote the ablation and accretion velocities by $u_1(t)$ and $u_2(t)$, respectively, which are defined as
\begin{equation} \label{u}
	u_1(t)=r_{,R}(S_1(t),t)\,U_1(t)\,,\qquad  u_2(t)=r_{,R}(S_2(t),t)\,U_2(t)\,.
\end{equation}
The choices $U_1(t)=u_1(t)$, and $U_2(t)=u_2(t)$ impose the following constraints on $r(R,t)$:\footnote{Other choices will lead to isometric material manifolds, and hence identical stresses \citep{Sozio2017,Yavari2022Torsion,Yavari2023Accretion}.}
\begin{equation} \label{r-constraint}
\begin{aligned} 
	r_{,R}(S_1(t),t) &=1\,,\quad \text{or} \qquad r_{,R}(R,\tac(R))=1\,, \\
	r_{,R}(S_2(t),t) &=1\,,\quad \text{or} \qquad r_{,R}(R,{\tab}(R))=1\,. 
\end{aligned}
\end{equation}
In particular, the choice \eqref{r-constraint}$_1$ makes the material metric $\mathbf{G}(R,t)$ and the Jacobian $J(R,t)$ continuous at $R=R_2$.  Now, \eqref{r-constraint}$_1$ and \eqref{rSec} imply that for $t<\timeablation$
\begin{equation} \label{rSecwconstrnt}
r^2(R,t) =	r^2(R_2,t)+\frac{2}{\lambda^2(t)}\int_{R_2}^R \lambdac^2(\xi)\, \rac(\xi) \,\text{d}\xi\,.
\end{equation}
Since $r(R,t)$ is continuous at $R=R_2$, \eqref{rPrim} and \eqref{rSecwconstrnt} imply that
\begin{equation} \label{rsecusings1}
	 r^2(R,t) =s_1^2(t)+\frac{R_2^2-S_1^2(t)}{\lambda^2(t)}
	+\frac{2}{\lambda^2(t)}\int_{R_2}^R \lambdac^2(\xi) \,\rac(\xi) \,\text{d}\xi \,,
\end{equation}
for $R_2\leq R\leq S_2(t)$ with $t< \timeablation$. Substituting $R=S_2(t)$ in \eqref{rsecusings1}, one obtains
\begin{equation} \label{s1squared}
	 s_1^2(t)=s_2^2(t)-\frac{R_2^2-S_1^2(t)}{\lambda^2(t)}
	-\frac{2}{\lambda^2(t)}\int_{R_2}^{S_2(t)} \lambdac^2(\xi) \,\rac(\xi) \,\text{d}\xi \,.
\end{equation}
Now, one may substitute \eqref{s1squared} into \eqref{rPrim} and combine with \eqref{raccretedincomp} to write the kinematics in terms of $\rac$ and $\lambdac$. As $s_2(t)=r(S_2(t),t)=\rac(S_2(t))$, when $t <\timeablation$:
\begin{equation}\label{kinrplusandlambda1}
r^2(R,t) =
\begin{dcases}
	\rac^2(S_2(t))-\frac{R_2^2-R^2}{\lambda^2(t)}
	-\frac{2}{\lambda^2(t)}\int_{R_2}^{S_2(t)} \lambdac^2(\xi) \,\rac(\xi) \,\text{d}\xi \,, 
	&  R \leq R_2\,, \\
	\rac^2(S_2(t))-\frac{2}{\lambda^2(t)}\int_{R}^{S_2(t)} \lambdac^2(\xi)\, \rac(\xi) \,\text{d}\xi\,, 
	& R\geq R_2 \,.
\end{dcases}
\end{equation}
For $t\geq\timeablation$:
\begin{equation}\label{kinrplusandlambda2}
	r^2(R,t) =	\rac^2(S_2(t))-\frac{2}{\lambda^2(t)}\int_{R}^{S_2(t)} \lambdac^2(\xi)\, 
	\rac(\xi) \,\text{d}\xi\,. 
\end{equation}
In a force-control problem, the functions $r(R,t)$ and $\lambda(t)$ are not known. However, it can be inferred from \eqref{kinrplusandlambda1}-\eqref{kinrplusandlambda2} that the knowledge of $\rac(R)$ and $\lambda(t)$ is sufficient to calculate $r(R,t)$. Since $\tac(R)$ is assumed to be given, $\lambdac(R)=\lambda ( \tac(R))$ is not an independent function. Thus, $\rac(R)$ and $\lambda(t)$ are the only independent functions.
Since $\lambda(t)$ is given in a displacement-control problem, $\rac(R)$ is the only independent unknown function in displacement-control problems.

\subsection{Stress calculation}

In this section, we compute the stresses for  both $t<\timeablation$ and $t>\timeablation$. The stresses are calculated separately for the initial and the accreted parts of the body. The radial equilibrium equation reads $\frac{\partial \sigma^{rr}}{\partial r}+\frac{1}{r}\sigma^{rr}-r\sigma^{\theta\theta}=0$.\footnote{The other two equilibrium equations imply that $p=p(R,t)$.} In terms of reference coordinates, $\frac{1}{r_{,R}}\frac{\partial \sigma^{rr}}{\partial R}+ \frac{\sigma^{rr}}{r}-r\sigma^{\theta\theta} =0$. 
Thus, we have
\begin{equation} \label{sigmarrEq}
	\sigma^{rr}_{,R}(R,t)=\left[r^2(R,t) \sigma^{\theta\theta}(R,t)-\sigma^{rr}(R,t)\right]\frac{r_{,R}(R,t)}{r(R,t)}\,.
\end{equation}
Recall that $\mathbf{b}^{\sharp}$ has components $b^{ab}=F^a{}_{A}\,F^b{}_{B}\,G^{AB}$ and $\mathbf{c}^{\sharp}$ has components $c^{ab}=g^{am}g^{bn}c_{mn}$, where $c_{ab}=F^{-A}{}_{a}\,F^{-B}{}_{b}\,G_{AB}$. For $S_1(t) \leq R \leq R_2$, and $t<\timeablation$:
\begin{equation}\label{bcInitialBod}
  \mathbf{b}^{\sharp}(R,t)=\begin{bmatrix}
r_{,R}^2(R,t) & 0 & 0 \\
 0 & \displaystyle\frac{1}{R^2} &0 \\
 0 & 0 & \lambda^4(t) \\
\end{bmatrix} \,,\qquad 
\mathbf{c}^{\sharp}(R,t)=\begin{bmatrix}
\displaystyle\frac{1}{r_{,R}^2(R,t)} & 0 & 0 \\
 0 & \displaystyle\frac{R^2}{r^4(R,t)} & 0 \\
 0 & 0& \displaystyle\frac{1}{\lambda^4(t)} \\
\end{bmatrix}\,.
\end{equation}
The principal invariants of $\mathbf{b}$ read
\begin{equation}
\begin{aligned}
   I_1(R,t) &=r_{,R}^2(R,t)+\frac{r^2(R,t)}{R^2} +\lambda^4(t) \,,\\
   I_2(R,t) &=r_{,R}^2(R,t)\,\lambda^4(t)+ \frac{r_{,R}^2(R,t)\,r^2(R,t)}{R^2}
   +\frac{r^2(R,t)\, \lambda^4(t)}{R^2} \,.
\end{aligned}
\end{equation}
The Cauchy stress has the following non-zero components\footnote{Notice that $\sigma^{\theta\theta}$ does not have the dimension of stress. Its physical component is $\hat{\sigma}^{\theta\theta}=r^2 \sigma^{\theta\theta}$.}
\begin{equation} \label{cauchystressInitialbody}
\begin{aligned}
   \sigma^{rr}(R,t) &= -p(R,t)+\alpha(R,t)\,r_{,R}^2(R,t)-\frac{\beta(R,t)}{r_{,R}^2(R,t)}  \,,  \\
   \sigma^{\theta\theta}(R,t) &= -\frac{p(R,t)}{r^2(R,t)}
   +\frac{\alpha(R,t)}{R^2} -\frac{\beta(R,t)\,R^2}{r^4(R,t)}  \,, \\
   \sigma^{zz}(R,t) &= -p(R,t)+\alpha(R,t)\, \lambda^4(t)   -\,\frac{\beta(R,t) }{\lambda^4(t)}  \,,
\end{aligned}
\end{equation}
where $\alpha=2\frac{\partial W}{\partial I_1}$ and $\beta=2\frac{\partial W}{\partial I_2}$. 
Using the incompressibility constraint \eqref{Kinematics-Initial} each of the components of $\bm{\sigma}(R,t)$ can be expressed solely in terms of the kinematic quanities $r(R,t)$ and $\lambda(t)$, i.e., one can eliminate $r_{,R}(R,t)$ in \eqref{cauchystressInitialbody}$_1$ to obtain 
\begin{equation} \label{sigmarrInbodsimplified}
   \sigma^{rr}(R,t) = -p(R,t)+ \frac{R^2\alpha(R,t)}{r^2(R,t)\lambda^4(t)}- \frac{\beta(R,t)r^2(R,t)\lambda^4(t)}{R^2}  \,.
\end{equation}
Substituting \eqref{sigmarrInbodsimplified} and \eqref{cauchystressInitialbody}$_2$ in \eqref{sigmarrEq}, one obtains
\begin{equation} \label{sigmarrsubRInital}
	\sigma^{rr}_{,R}(R,t)=\frac{r^4(R,t)\lambda^4(t)-R^4}{R\, r^4(R,t)\lambda^2(t)}\left[\frac{\alpha(R,t)}{\lambda^4(t)}
	+\beta(R,t)\right]\,.
\end{equation}
Since $\sigma^{rr}(S_1(t),t)=0$, it is implied that
\begin{equation}
	\sigma^{rr}(R,t)=\int_{S_1(t)}^R \frac{r^4(\xi,t)\lambda^4(t)-\xi^4}{\xi\, r^4(\xi,t)\lambda^2(t)}
	\left[\frac{\alpha(\xi,t)}{\lambda^4(t)}+\beta(\xi,t)\right] d\xi \,,
\end{equation}
for $S_1(t) \leq R \leq R_2$ and $t<\timeablation$.\footnote{Alternatively, one may integrate \eqref{sigmarrsubRInital} from $R$ to $R_2$ to obtain 
\begin{equation}\label{sigmarralternate}
\sigma^{rr}(R,t)=\sigma^{rr}(R_2,t)-\int_{R}^{R_2} \frac{r^4(\xi,t)\lambda^4(t)-\xi^4}{\xi\, r^4(\xi,t)\lambda^2(t)}\left[\frac{\alpha(\xi,t)}{\lambda^4(t)}+\beta(\xi,t)\right] d\xi \,,
\end{equation}
for $S_1(t) \leq R \leq R_2$, and $t<\timeablation$.} Thus, \eqref{sigmarrInbodsimplified} gives the following expression for pressure
\begin{align}\label{innerpressurev1}
	p(R,t) =  \frac{R^2\alpha(R,t)}{r^2(R,t)\lambda^4(t)}- \frac{\beta(R,t)r^2(R,t)\lambda^4(t)}{R^2} 
	- \int_{S_1(t)}^R \frac{r^4(\xi,t)\lambda^4(t)-\xi^4}{\xi\, r^4(\xi,t)\lambda^2(t)}
	\left[\frac{\alpha(\xi,t)}{\lambda^4(t)}+\beta(\xi,t)\right] d\xi \,.
\end{align}
Substituting \eqref{innerpressurev1} into \eqref{cauchystressInitialbody}$_{2-3}$, one obtains
\begin{equation} \label{Innerstress}
\begin{aligned}
  	\hat{\sigma}^{\theta\theta}(R,t) =& \frac{r^4(R,t)\lambda^4(t)-R^4}{R^2\,r^2(R,t)}
	\left[\frac{\alpha(R,t)}{\lambda^4(t)}
	+\beta(R,t)\right] + \int_{S_1(t)}^R \frac{r^4(\xi,t)\lambda^4(t)-\xi^4}{\xi\, r^4(\xi,t)\lambda^2(t)}
	\left[\frac{\alpha(\xi,t)}{\lambda^4(t)}+\beta(\xi,t)\right] d\xi\,, \\
	\sigma^{zz}(R,t) =& \alpha(R,t)\left[\lambda^4(t)-\frac{R^2}{r^2(R,t)\lambda^4(t)}\right]
	+\beta(R,t)\left[\frac{r^2(R,t)\lambda^4(t)}{R^2}-\frac{1}{\lambda^4(t)}\right] \\
	&+\int_{S_1(t)}^R \frac{r^4(\xi,t)\lambda^4(t)-\xi^4}{\xi\, r^4(\xi,t)\lambda^2(t)}\left[\frac{\alpha(\xi,t)}{\lambda^4(t)}
	+\beta(\xi,t)\right] d\xi\,.
\end{aligned}
\end{equation}
For $\big\{R_2\leq R \leq S_2(t),\,t<\timeablation\big\}$, or $\big\{t\geq \timeablation,\,S_1(t)\leq R\leq S_2(t)\big\}$:
\begin{equation}\label{bcAccretedBody}
\begin{aligned}
  \mathbf{b}^{\sharp}(R,t) =\begin{bmatrix} \displaystyle
 \left[\frac{r_{,R}(R,t)}{r_{,R}(R,\tac(R))}\right]^2 & 0 & 0 \\
 0 & \displaystyle\frac{1}{\rac^2(R)} 
 & 0 \\
 0 & 0 & \displaystyle\frac{\lambda^4(t)}{\lambdac^4(R)} \\
\end{bmatrix} \,,\quad
\mathbf{c}^{\sharp}(R,t) =\begin{bmatrix}
\displaystyle
 \left[\frac{r_{,R}(R,\tac(R))}{r_{,R}(R,t)}\right]^2 & 0 & 0 \\
 0 & \displaystyle\frac{\rac^2(R)}{r^4(R,t) } & 0 \\
 0 & 0 & \displaystyle\frac{\lambdac^4(R)
 }{\lambda^4(t)} \\
   \end{bmatrix}\,.
\end{aligned}
\end{equation}
The principal invariants of $\mathbf{b}$ read
\begin{equation}
\begin{aligned}
   I_1(R,t) &=r_{,R}^2(R,t)+\frac{r^2(R,t)}{\rac^2(R)}+\frac{\lambda^4(t)}{\lambdac^4(R)}  \;,\\
   I_2(R,t) &=\frac{r_{,R}^2(R,t)\,r^2(R,t)}{\rac^2(R)}+\frac{r^2(R,t)\,\lambda^4(t)}{\rac^2(R)\lambdac^4(R)}+\frac{r_{,R}^2(R,t)\lambda^4(t)}{\lambdac^4(R)}\,.
\end{aligned}
\end{equation}
The non-zero components of the Cauchy stress are
\begin{equation}\label{cauchystressAccretedbody}
\begin{aligned}
   \sigma^{rr}(R,t) &= -p(R,t)+\alpha(R,t)r_{,R}^2(R,t)-\frac{\beta(R,t)}{r_{,R}^2(R,t)} \,,  \\
   \sigma^{\theta\theta}(R,t) &= -\frac{p(R,t)}{r^2(R,t)}
   +\frac{\alpha(R,t)}{\rac^2(R)}-\frac{\beta(R,t)\,\rac^2(R)}{r^4(R,t)}     \,, \\
   \sigma^{zz}(R,t) &= -p(R,t)+\frac{\alpha(R,t)\, \lambda^4(t)}{\lambdac^4(R)}
   -\frac{\beta(R,t)\, \lambdac^4(R)}{\lambda^4(t)} \,.
\end{aligned}
\end{equation}
Using the constraints \eqref{Kinematics-Accreted} and \eqref{r-constraint}, each of the components of $\bm{\sigma}(R,t)$ can be expressed solely in terms of the kinematic quanities $r(R,t)$ and $\lambda(t)$, i.e., one can eliminate $r_{,R}(R,t)$ in \eqref{cauchystressAccretedbody}$_1$ to obtain
\begin{equation} \label{sigmarrOutbodsimplified}
	\sigma^{rr}(R,t) = -p(R,t)
	+\frac{\alpha(R,t)\,\rac^2(R)\lambdac^4(R)}{r^2(R,t)\lambda^4(t)}
	-\frac{\beta(R,t)r^2(R,t)\,\lambda^4(t)}{\rac^2(R)\,\lambdac^4(R)}  \,.
\end{equation}
Substituting \eqref{cauchystressAccretedbody}$_{1-2}$ in \eqref{sigmarrEq}, one obtains
\begin{equation}
	\sigma^{rr}_{,R}(R,t)=\frac{\lambdac^2(R)\left[r^4(R,t)\lambda^4(t)-\rac^4(R)\lambdac^4(R)\right]}
	{r^4(R,t)\lambda^2(t)\,\rac(R)}\left[\frac{\alpha(R,t)}{\lambda^4(t)}+\frac{\beta(R,t)}{\lambdac^4(R)}\right].
\end{equation}
Since $\sigma^{rr}(S_2(t),t)=0$, it is implied that
\begin{equation}\label{sigmarraccretedBody}
	\sigma^{rr}(R,t)=-\int_R^{S_2(t)} \frac{\lambdac^2(\xi)\left[r^4(\xi,t)\lambda^4(t)-\rac^4(\xi)\lambdac^4(\xi)\right]}
	{r^4(\xi,t)\lambda^2(t)\,\rac(\xi)}\left[\frac{\alpha(\xi,t)}{\lambda^4(t)}
	+\frac{\beta(\xi,t)}{\lambdac^4(\xi)}\right] d\xi\,,
\end{equation}
whenever $\big\{R_2\leq R \leq S_2(t),\,t<\timeablation\big\}$, or $\big\{t\geq \timeablation,\,S_1(t)\leq R\leq S_2(t)\big\}$. Thus, \eqref{sigmarrOutbodsimplified} gives the following expression for the pressure field
\begin{equation}\label{pressureOuter}
\begin{aligned}
	p(R,t) &=  \frac{\alpha(R,t)\rac^2(R)\lambdac^4(R)}{r^2(R,t)\lambda^4(t)} 
	-\frac{\beta(R,t)r^2(R,t)\lambda^4(t)}{\rac^2(R)\lambdac^4(R)}\\  
	& \quad+\int_R^{S_2(t)} \frac{\lambdac^2(\xi)\left[r^4(\xi,t)\lambda^4(t)-\rac^4(\xi)\lambdac^4(\xi)\right]}
	{r^4(\xi,t)\lambda^2(t)\,\rac(\xi)}\left[\frac{\alpha(\xi,t)}{\lambda^4(t)}
	+\frac{\beta(\xi,t)}{\lambdac^4(\xi)}\right] d\xi \,.
\end{aligned}
\end{equation}
Substituting \eqref{pressureOuter} into \eqref{cauchystressAccretedbody}$_{2-3}$, one obtains
\begin{equation} \label{Outerstress}
\begin{aligned}
	\hat{\sigma}^{\theta\theta}(R,t) =& \frac{r^4(R,t)\lambda^4(t)-\rac^4(R)\lambdac^4(R)}{r^2(R,t)\rac^2(R)}
	\left[\frac{\alpha(R,t)}{\lambda^4(t)}+\frac{\beta(R,t)}{\lambdac^4(R)}\right] \\
	&-\int_R^{S_2(t)} \frac{\lambdac^2(\xi)\left[r^4(\xi,t)\lambda^4(t)
	-\rac^4(\xi)\lambdac^4(\xi)\right]}{r^4(\xi,t)\lambda^2(t)\rac(\xi)}\left[\frac{\alpha(\xi,t)}{\lambda^4(t)}
	+\frac{\beta(\xi,t)}{\lambdac^4(\xi)}\right] d\xi\,, \\
	\sigma^{zz}(R,t) =& \alpha(R,t)\left[\frac{\lambda^4(t)}{\lambdac^4(R)}
	-\frac{\lambdac^4(R)\rac^2(R)}{\lambda^4(t)r^2(R,t)}\right]
	+\beta(R,t)\left[\frac{\lambda^4(t)r^2(R,t)}{\lambdac^4(R)\rac^2(R)}
	-\frac{\lambdac^4(R)}{\lambda^4(t)}\right] \\
	&-\int_R^{S_2(t)} \frac{\lambdac^2(\xi)\left[r^4(\xi,t)\lambda^4(t)-\rac^4(\xi)\lambdac^4(\xi)\right]}
	{r^4(\xi,t)\lambda^2(t)\rac(\xi)}\left[\frac{\alpha(\xi,t)}{\lambda^4(t)}
	+\frac{\beta(\xi,t)}{\lambdac^4(\xi)}\right] d\xi \,,
\end{aligned}
\end{equation}
for  $\big\{R_2\leq R \leq S_2(t),\,t<\timeablation\big\}$, and $\big\{t\geq \timeablation,\,S_1(t)\leq R\leq S_2(t)\big\}$. Note that $\sigma^{\theta\theta}(S_2(t),t)=0$, and $\sigma^{zz}(S_2(t),t)=0$, i.e., $\bm{\sigma}(S_2(t),t)=\mathbf{0}$. 
Since $\sigma^{rr}(R,t)$ has to be continuous in $R$ at $R_2$ at any time $t<\timeablation$, we must have
\begin{equation}\label{sigmarrcontatR2} 
\begin{aligned}
&\int_{S_1(t)}^{R_2} \frac{r^4(\xi,t)\lambda^4(t)-\xi^4}{\xi\, r^4(\xi,t)\lambda^2(t)}\left[\frac{\alpha(\xi,t)}{\lambda^4(t)}+\beta(\xi,t)\right] d\xi \\
&\qquad +\int_{R_2}^{S_2(t)} \frac{\lambdac^2(\xi)\left[r^4(\xi,t)\lambda^4(t)-\rac^4(\xi)\lambdac^4(\xi)\right]}{r^4(\xi,t)\lambda^2(t)\rac(\xi)}\left[\frac{\alpha(\xi,t)}{\lambda^4(t)}+\frac{\beta(\xi,t)}{\lambdac^4(\xi)}\right] d\xi=0\,,
\end{aligned}
\end{equation}
for all $0<t<\timeablation$.\footnote{For $t\geq\timeablation$, the boundary condition $\sigma^{rr}(S_1(t),t)=0$ yields a similar constraint
\begin{equation}\label{sigmarrzeroatS1}
\int_{S_1(t)}^{S_2(t)} \frac{\lambdac^2(\xi)\left[r^4(\xi,t)\lambda^4(t)-\rac^4(\xi)\lambdac^4(\xi)\right]}
	{r^4(\xi,t)\lambda^2(t)\,\rac(\xi)}\left[\frac{\alpha(\xi,t)}{\lambda^4(t)}
	+\frac{\beta(\xi,t)}{\lambdac^4(\xi)}\right] d\xi=0\,,
\end{equation}
in view of \eqref{sigmarraccretedBody}.
}

\begin{remark}
Using \eqref{sigmarraccretedBody}, one obtains
\begin{equation}
	\sigma^{rr}(R_2,t)=-\int_{R_2}^{S_2(t)} \frac{\lambdac^2(\xi)\left[r^4(\xi,t)\lambda^4(t)
	-\rac^4(\xi)\lambdac^4(\xi)\right]}{r^4(\xi,t)\lambda^2(t)\rac(\xi)}\left[\frac{\alpha(\xi,t)}{\lambda^4(t)}
	+\frac{\beta(\xi,t)}{\lambdac^4(\xi)}\right] d\xi\,,
\end{equation}
which can be substituted into \eqref{sigmarralternate} to give
\begin{equation}\label{sigmarrcontversion}
\begin{aligned}
	\sigma^{rr}(R,t)=&-\int_{R}^{R_2} \frac{r^4(\xi,t)\lambda^4(t)-\xi^4}{\xi\, r^4(\xi,t)\lambda^2(t)}
	\left[\frac{\alpha(\xi,t)}{\lambda^4(t)}+\beta(\xi,t)\right] d\xi  \\
	&-\int_{R_2}^{S_2(t)} \frac{\lambdac^2(\xi)\left[r^4(\xi,t)\lambda^4(t)
	-\rac^4(\xi)\lambdac^4(\xi)\right]}{r^4(\xi,t)\lambda^2(t)\rac(\xi)}\left[\frac{\alpha(\xi,t)}{\lambda^4(t)}
	+\frac{\beta(\xi,t)}{\lambdac^4(\xi)}\right] d\xi \,,
\end{aligned}
\end{equation}
where $S_1(t) \leq R \leq R_2$, and $t<\timeablation$.\footnote{Setting $\sigma^{rr}(S_1(t),t)=0$ in \eqref{sigmarrcontversion} provides an alternative approach to recover \eqref{sigmarrcontatR2}.} Thus, \eqref{sigmarrInbodsimplified} gives the following expression for the pressure field
\begin{equation} \label{Newpeq}
\begin{aligned}
	p(R,t) & =  \frac{R^2\alpha(R,t)}{r^2(R,t)\lambda^4(t)}- \frac{\beta(R,t)r^2(R,t)\lambda^4(t)}{R^2} 
	+\int_{R}^{R_2} \frac{r^4(\xi,t)\lambda^4(t)-\xi^4}{\xi\, r^4(\xi,t)\lambda^2(t)}
	\left[\frac{\alpha(\xi,t)}{\lambda^4(t)}+\beta(\xi,t)\right] d\xi  \\
	& \quad +\int_{R_2}^{S_2(t)} \frac{\lambdac^2(\xi)\left[r^4(\xi,t)\lambda^4(t)
	-\rac^4(\xi)\lambdac^4(\xi)\right]}{r^4(\xi,t)\lambda^2(t)\rac(\xi)}\left[\frac{\alpha(\xi,t)}{\lambda^4(t)}
	+\frac{\beta(\xi,t)}{\lambdac^4(\xi)}\right] d\xi\,,
\end{aligned}
\end{equation}
where $S_1(t) \leq R \leq R_2$. Substituting \eqref{Newpeq} into \eqref{cauchystressInitialbody}$_{2-3}$ one obtains
\begin{equation} \label{InitialBodStressv2}
\begin{aligned}
	\hat{\sigma}^{\theta\theta}(R,t) & = \frac{r^4(R,t)\lambda^4(t)-R^4}{R^2\, r^2(R,t)}
	\left[\frac{\alpha(R,t)}{\lambda^4(t)}+\beta(R,t)\right]- \int_{R}^{R_2} \frac{r^4(\xi,t)\lambda^4(t)
	-\xi^4}{\xi\, r^4(\xi,t)\lambda^2(t)}\left[\frac{\alpha(\xi,t)}{\lambda^4(t)}+\beta(\xi,t)\right] d\xi  \\
	& \quad -\int_{R_2}^{S_2(t)} \frac{\lambdac^2(\xi)\left[r^4(\xi,t)\lambda^4(t)
	-\rac^4(\xi)\lambdac^4(\xi)\right]}{r^4(\xi,t)\lambda^2(t)\rac(\xi)}\left[\frac{\alpha(\xi,t)}{\lambda^4(t)}
	+\frac{\beta(\xi,t)}{\lambdac^4(\xi)}\right] d\xi, \\
	\sigma^{zz}(R,t) &= \alpha(R,t)\left[\lambda^4(t)-\frac{R^2}{r^2(R,t)\lambda^4(t)}\right]
	+\beta(R,t)\left[\frac{r^2(R,t)\lambda^4(t)}{R^2}-\frac{1}{\lambda^4(t)}\right] \\
	& \quad - \int_{R}^{R_2} \frac{r^4(\xi,t)\lambda^4(t)-\xi^4}{\xi\, r^4(\xi,t)\lambda^2(t)}
	\left[\frac{\alpha(\xi,t)}{\lambda^4(t)}+\beta(\xi,t)\right] d\xi  \\
	& \quad -\int_{R_2}^{S_2(t)} \frac{\lambdac^2(\xi)\left[r^4(\xi,t)\lambda^4(t)
	-\rac^4(\xi)\lambdac^4(\xi)\right]}{r^4(\xi,t)\lambda^2(t)\rac(\xi)}\left[\frac{\alpha(\xi,t)}{\lambda^4(t)}
	+\frac{\beta(\xi,t)}{\lambdac^4(\xi)}\right] d\xi\,,
\end{aligned}
\end{equation}
for $S_1(t) \leq R \leq R_2$.\footnote{It is clear from \eqref{InitialBodStressv2} and \eqref{Outerstress} that $\sigma^{\theta\theta}$ and $\sigma^{zz}$ are continuous at $R_2$.} On the ablation boundary
\begin{equation} 
\begin{aligned}
	\hat{\sigma}^{\theta\theta}(S_1(t),t) 
	&=\begin{cases}
	\displaystyle \left[\frac{s_1^2(t)\lambda^4(t)}{S_1^2(t)}-\frac{S_1^2(t)}{s_1^2(t)}\right]
	\left[\frac{\alpha(S_1(t),t)}{\lambda^4(t)}+\beta(S_1(t),t)\right]
	\,, & t<\timeablation\,, \\[10 pt]
	\displaystyle	\left[\frac{s_1^2(t)\lambda^4(t)}{\rac^2(S_1(t))}
	-\frac{\rac^2(S_1(t))\lambdac^4(S_1(t))}{s_1^2(t)}\right]
	\left[\frac{\alpha(S_1(t),t)}{\lambda^4(t)}+\frac{\beta(S_1(t),t)}{\lambdac^4(S_1(t))}\right] 
	\,, &  t\geq \timeablation\,, \\ 
	\end{cases}
	\\[16 pt]
	\sigma^{zz}(S_1(t),t) 
	&=\begin{cases}
	\displaystyle \left[\frac{s_1^2(t)\lambda^4(t)}{S_1^2(t)}-\frac{1}{\lambda^4(t)}\right]
	\left[\frac{\alpha(S_1(t),t) S_1^2(t)}{s_1^2(t)}+\beta(S_1(t),t)\right]
	\,, & t<\timeablation\,, \\[10 pt]
	\displaystyle \alpha(S_1(t),t)\left[\frac{\lambda^4(t)}{\lambdac^4(S_1(t))}
	-\frac{\lambdac^4(S_1(t))\rac^2(S_1(t))}{\lambda^4(t)s_1^2(t)}\right]& \\
	\qquad\qquad +\beta(S_1(t),t)\left[\frac{\lambda^4(t)s_1^2(t)}{\lambdac^4(S_1(t))\rac^2(S_1(t))}	
	-\frac{\lambdac^4(S_1(t))}{\lambda^4(t)}\right]	
	\,, &  t\geq \timeablation\,. \\ 
	\end{cases}
\end{aligned}
\end{equation}
Notice that $\sigma^{\theta\theta}$ and $\sigma^{zz}$ do not vanish on the ablation boundary.\footnote{
In a similar problem, \citet{naghibzadeh2022accretion} observed non-zero $\sigma^{\theta\theta}$ and $\sigma^{\phi\phi}$ on the ablation boundary of a hollow spherical body undergoing accretion through its fixed inner boundary while ablation takes place on its traction-free outer boundary.}  
\end{remark}

\begin{remark}
In \citep{Yavari2023Accretion} it was shown that the finite extension of an accreting circular cylindrical bar made of an arbitrary incompressible isotropic solid is a universal deformation even in the presence of radially-symmetric accretion. We have observed here that this result holds even when there is simultaneous radially-symmetric accretion and ablation on the outer and inner boundaries, respectively.
\end{remark}

\paragraph{The applied axial force.}
The axial force at the two ends of the bar is given by
\begin{equation}
	F(t) =2\pi \int_{S_1(t)}^{S_2(t)}P^{zZ}(R,t)R\,\text{d}R\,,
\end{equation}
where $P^{zZ}(R,t)= \displaystyle\frac{\sigma^{zz}(R,t)}{\lambda^2(t)}$ is the $zZ$-component of first Piola-Kirchhoff stress. For $S_1(t) \leq R \leq R_2$ and $t<\timeablation$:
\begin{equation} 
\begin{aligned}
P^{zZ}(R,t) &= \alpha(R,t)\left[\lambda^2(t)-\frac{R^2}{r^2(R,t)\lambda^6(t)}\right]+\beta(R,t)\left[\frac{r^2(R,t)\lambda^2(t)}{R^2}-\frac{1}{\lambda^6(t)}\right] \\
 & \quad - \int_{R}^{R_2} \frac{r^4(\xi,t)\lambda^4(t)-\xi^4}{\xi\, r^4(\xi,t)\lambda^4(t)}\left[\frac{\alpha(\xi,t)}{\lambda^4(t)}+\beta(\xi,t)\right] d\xi  \\
& \quad -\int_{R_2}^{S_2(t)} \frac{\lambdac^2(\xi)\left[r^4(\xi,t)\lambda^4(t)-\rac^4(\xi)\lambdac^4(\xi)\right]}{r^4(\xi,t)\lambda^4(t)\rac(\xi)}\left[\frac{\alpha(\xi,t)}{\lambda^4(t)}+\frac{\beta(\xi,t)}{\lambdac^4(\xi)}\right] d\xi\,.
\end{aligned}
\end{equation}
For $\big\{R_2\leq R \leq S_2(t),t<\timeablation\big\}$ and $\big\{t\geq \timeablation,S_1(t)\leq R\leq S_2(t)\big\}$:
\begin{equation} 
\begin{aligned}
   P^{zZ}(R,t) &= \alpha(R,t)\left[\frac{\lambda^2(t)}{\lambdac^4(R)}
   -\frac{\lambdac^4(R)\,\rac^2(R)}{\lambda^6(t)\,r^2(R,t)}\right]
   + \beta(R,t)\left[\frac{\lambda^2(t)\,r^2(R,t)}{\lambdac^4(R)\,\rac^2(R)}
   -\frac{\lambdac^4(R)}{\lambda^6(t)}\right] \\
   & \quad -\int_R^{S_2(t)} \frac{\lambdac^2(\xi)\left[r^4(\xi,t)\,\lambda^4(t)
   -\rac^4(\xi)\,\lambdac^4(\xi)\right]}{r^4(\xi,t)\,\lambda^4(t)\,\rac(\xi)}
   \left[\frac{\alpha(\xi,t)}{\lambda^4(t)}+\frac{\beta(\xi,t)}{\lambdac^4(\xi)}\right] d\xi\, \,.
\end{aligned}
\end{equation}

\begin{remark}
Observe that \eqref{sigmarrcontatR2} and \eqref{sigmarrzeroatS1} can be combined to define the following function\footnote{$\Upsilon$ is continuous at $\timeablation$. Equivalently, $\Upsilon \circ \tac$ is continuous at $R_2$. }  
\begin{equation}\label{sigmarrcontatR2}
\begin{aligned}
\Upsilon(t):=
\begin{cases}
\displaystyle \int_{S_1(t)}^{R_2} \frac{r^4(\xi,t)\lambda^4(t)-\xi^4}{\xi\, r^4(\xi,t)\lambda^2(t)}\left[\frac{\alpha(\xi,t)}{\lambda^4(t)}+\beta(\xi,t)\right] d\xi &\\
\quad\quad  \displaystyle +\int_{R_2}^{S_2(t)} \frac{\lambdac^2(\xi)\left[r^4(\xi,t)\lambda^4(t)-\rac^4(\xi)\lambdac^4(\xi)\right]}{r^4(\xi,t)\lambda^2(t)\rac(\xi)}\left[\frac{\alpha(\xi,t)}{\lambda^4(t)}+\frac{\beta(\xi,t)}{\lambdac^4(\xi)}\right] d\xi\,, & t \leq \timeablation\,, \\[20pt]
\displaystyle \int_{S_1(t)}^{S_2(t)} \frac{\lambdac^2(\xi)\left[r^4(\xi,t)\lambda^4(t)-\rac^4(\xi)\lambdac^4(\xi)\right]}{r^4(\xi,t)\lambda^2(t)\,\rac(\xi)}\left[\frac{\alpha(\xi,t)}{\lambda^4(t)} +\frac{\beta(\xi,t)}{\lambdac^4(\xi)}\right] d\xi \,,  & t \geq \timeablation\,.
\end{cases}
\end{aligned}
\end{equation}
In a displacement-control problem, where $\lambda(t)$ is given, $\Upsilon(\tac(R))=0$ is to be solved with $\rac(R_2)=R_2$ to find the unknown function $\rac(R)$ for $R>R_2$.
\end{remark}

\subsection{The accretion-ablation initial-boundary-value problem for a neo-Hookean solid}

Let $T \leq \timeablation$. Consider a homogeneous neo-Hookean material for which $\alpha(R,t)={\mu}(R)=\mu_0>0$ and $\beta(R,t)=0$. 
In order to simplify the calculations, assume that the spatial accretion/ablation velocities are constant, i.e., $u_1(t)=\velab>0$ and $u_2(t)=\velac >0$. The signs of $\velac$ and $\velab$ indicate that both accretion and ablation interfaces are moving radially outward. Thus,
\begin{equation} 
\begin{aligned} 
	S_1(t) &=R_1+\velab t\,,\quad \text{or}\qquad {\tab}(R)=\frac{R-R_1}{\velab}\,, \\
	S_2(t) &=R_2+\velac t\,,\quad \text{or}\qquad \tac(R)=\frac{R-R_2}{\velac }\,.
\end{aligned}
\end{equation}
The nonzero physical components of the Cauchy stress for this problem are listed as follows:\footnote{
On the ablation boundary,
\begin{equation} 
	\hat{\sigma}^{\theta\theta}(S_1(t),t) = \mu_0\left[\frac{s_1^2(t)}{S_1^2(t)}
	-\frac{S_1^2(t)}{\lambda^4(t) s_1^2(t)}\right]\,, \qquad
	\sigma^{zz}(S_1(t),t) = \mu_0\left[\lambda^4(t)-\frac{S_1^2(t)}{\lambda^4(t) s_1^2(t)}\right]
	\,.
\end{equation}
}
\begin{equation}\label{Strescompsimplified}
\begin{aligned}
\sigma^{rr}(R,t)&= 
\begin{dcases}
	\frac{\mu_0}{\lambda^2(t)}\int_{S_1(t)}^R \frac{\text{d}\xi}{\xi\, }-\frac{\mu_0}{\lambda^6(t)}\int_{S_1(t)}^R \frac{\xi^3 \text{d}\xi}{ r^4(\xi,t)}
	  \,, &  R < R_2\,, \\
	-\frac{\mu_0}{\lambda^2(t)}\int_R^{S_2(t)} \frac{\lambdac^2(\xi)\text{d}\xi}
	{\rac(\xi)}  +\frac{\mu_0}{\lambda^6(t)}\int_R^{S_2(t)} \frac{\rac^3(\xi)\lambdac^6(\xi)\text{d}\xi}{r^4(\xi,t)} \,, & R\geq R_2 \,,
\end{dcases} \\
\hat{\sigma}^{\theta\theta}(R,t)&= 
\begin{dcases}
	 \mu_0 \left[\frac{r^2(R,t)}{R^2}-\frac{ R^2}{\lambda^4(t) \,r^2(R,t)}\right]+\sigma^{rr}(R,t) \,, &  R < R_2\,, \\
	\mu_0\left[\frac{r^2(R,t)}{\rac^2(R)}
	-\frac{\lambdac^4(R)\rac^2(R)}{\lambda^4(t)r^2(R,t)}\right]
	+\sigma^{rr}(R,t)\,, & R\geq R_2 \,,
\end{dcases} \\
\sigma^{zz}(R,t)&=
\begin{dcases}
	\mu_0\left[\lambda^4(t)-\frac{R^2}{\lambda^4(t) r^2(R,t)}\right] +\sigma^{rr}(R,t) \,, &  R < R_2\,, \\
	\mu_0\left[\frac{\lambda^4(t)}{\lambdac^4(R)}
	-\frac{\lambdac^4(R)\rac^2(R)}{\lambda^4(t)r^2(R,t)}\right]
	+\sigma^{rr}(R,t)\,, & R\geq R_2 \,.
\end{dcases} 
\end{aligned}
\end{equation}
Traction continuity equation \eqref{sigmarrcontatR2} simplifies to read
\begin{equation} \label{tractcont}
\lambda^4(t)\left[\int_{S_1(t)}^{R_2}\frac{\text{d}\xi}{\xi}+\int_{R_2}^{S_2(t)}\frac{\lambdac^2(\xi)\text{d}\xi}{\rac(\xi)}\right]= \int_{S_1(t)}^{R_2}\frac{\xi^3 \text{d}\xi}{r^4(\xi,t)}+\int_{R_2}^{S_2(t)}\frac{\rac^3(\xi)\lambdac^6(\xi) \text{d}\xi}{r^4(\xi,t)}\,.
\end{equation}
Now, one can differentiate \eqref{tractcont} with respect to $t$ and use \eqref{rPrim} and \eqref{rsecusings1} to deduce that (the detailed calculations are given in Appendix \ref{AppendixA})
\begin{equation}
 \rac(R)=\frac{R_2+ \int_{R_2}^R \lambdac(\xi)\text{d}\xi}{\lambdac(R)}\,,
\end{equation}
and
\begin{equation}\label{AnalytSolGuess}
r(R,t) =
\begin{cases} 
	\displaystyle \frac{R}{\lambda(t)} \,, & S_1(t) \leq R \leq R_2\,, \\[10pt]
	\displaystyle \frac{\lambdac(R) \rac(R) }{\lambda(t)}\,,  & R_2 \leq R \leq S_2(t) \,,
\end{cases}
\end{equation}
which when substituted into \eqref{Strescompsimplified} implies that $\sigma^{rr}(R,t)=\sigma^{\theta \theta}(R,t)=0$. Morever, 
\begin{equation}
\begin{aligned}
P^{zZ}(R,t)=
\begin{cases}
\displaystyle  \mu_0\left[\lambda^2(t)-\frac{1}{\lambda^4(t)}\right] \,,& S_1(t) \leq R \leq R_2\,, \\[10pt]
\displaystyle  \mu_0 \left[\frac{\lambda^2(t)}{\lambdac^4(R)}
   -\frac{\lambdac^2(R)}{\lambda^4(t)}\right]  \,, & R_2\leq R \leq S_2(t)\,,
\end{cases}
\end{aligned}
\end{equation}
using which the axial force can be expressed as
\begin{equation} \label{axialforceintermsofLambda}
	\frac{F(t)}{2\pi\mu_0}  =  \frac{R_2^2-S_1^2(t)}{2}\left[\lambda^2(t)-\frac{1}{\lambda^4(t)}\right] + \lambda^2(t)\int_{R_2}^{S_2(t)}\frac{R\,\text{d}R}{\lambdac^4(R)} - \frac{1}{\lambda^4(t)} \int_{R_2}^{S_2(t)}R \lambdac^2(R) \text{d}R   \,.
\end{equation}
In a force-control problem $F(t)$ is given (with $F(0)=0$), and $S_1(t)$, $S_2(t)$ are both known. All one needs to find is the unknown function $\lambda(t)$ that satisfies \eqref{axialforceintermsofLambda}.

\begin{example} \label{Ex1}
Consider a displacement-control problem with $R_2=2R_1$, $\velab=2 \frac{R_1}{T}$, $\velac=3 \frac{R_1}{T}$, and $\lambda(t)= 1+ a \left(\frac{t}{T}\right)^n$, where $n \geq 0$, and $a \in \mathbb{R}$. Then $\timeablation=\frac{T}{2}$ and $S_2(\timeablation)=3.5 R_1$. 
We solve this problem for $a \in \big\{\pm 1\big\}$ and $n\in\big\{\frac{1}{2}, 1, 2\big\}$ assuming the numerical values $R_1=1$ and $T=1$. 
At any given $t<\timeablation$, $\sigma^{zz}(R,t)$ is constant in the initial body ($R\leq R_2$), as we have assumed the material to be homogeneous and the deformation to be uniform (Figs.~\ref{fig:DispControlLambdaInc} and \ref{fig:DispControlLambdaDec}). As foreseen, $\sigma^{zz}\geq 0$ for $\lambda\geq 1$ (Fig.~\ref{fig:DispControlLambdaInc}) and $\sigma^{zz}\leq 0$ for $\lambda \leq 1$ (Fig.~\ref{fig:DispControlLambdaDec}). Further, $\sigma^{zz}$ is nonzero on the ablation boundary (Fig.~\ref{fig:DispControlsigmazzAblBd}). Note that $\dot{\lambda}(t)$ taken in this example blows up near $t=0$ for $n<1$ and $a>0$. This is why when the bar is subjected to the elongation $\lambda(t)=1+\sqrt{\frac{t}{T}}$\,, we first observe a reduction in the outer diameter $\rac(R)$, or equivalently $s_2(t)$, see Fig.~\ref{fig:DispControlLambdaInc}. Later, as the rate of elongation reduces, this effect is overshadowed by accretion and $\rac(R)$ increases monotonically. 
\begin{figure}[t!]
\centering
\includegraphics[width=0.9\textwidth]{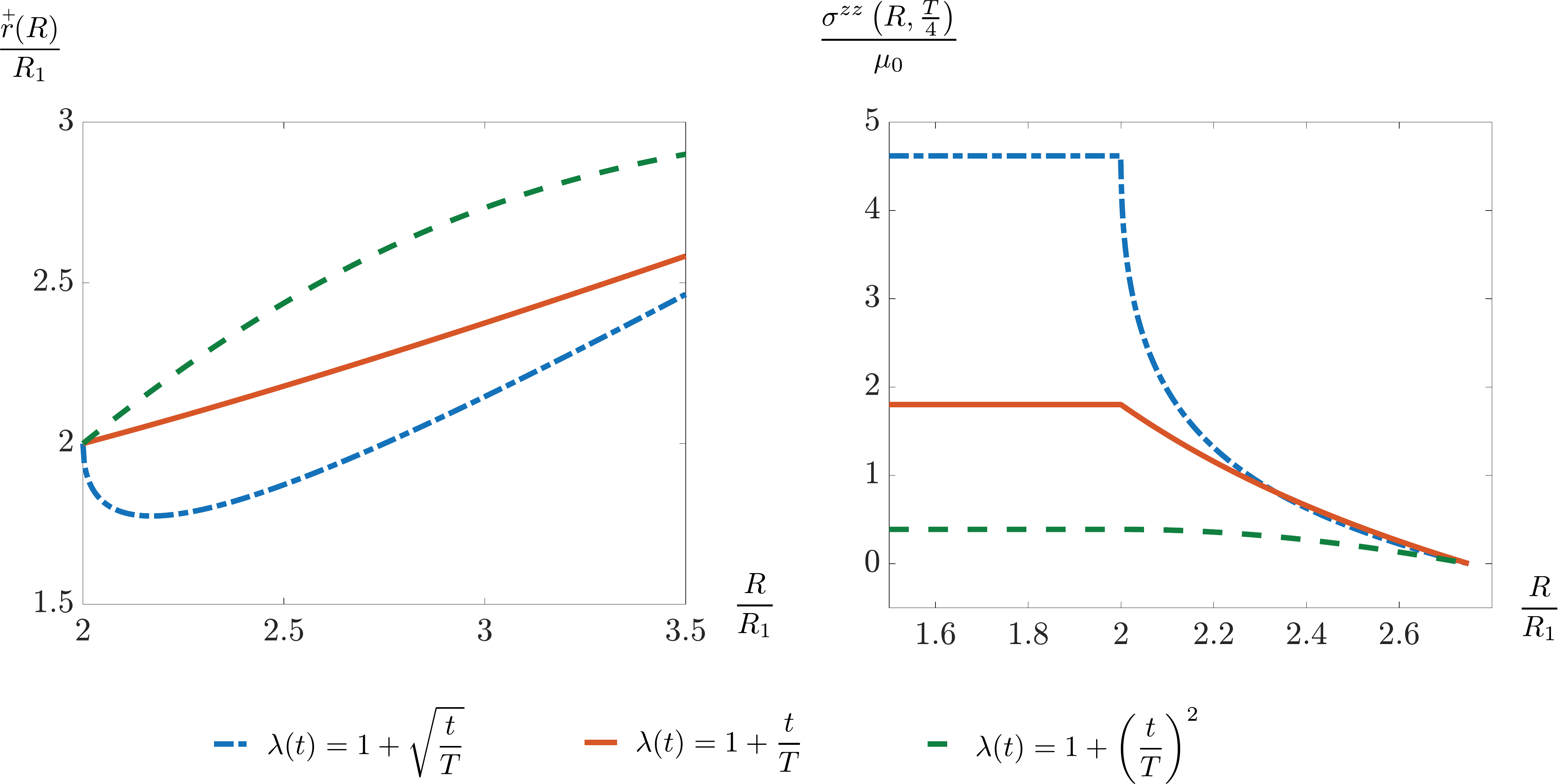}
\vspace*{0.0in}
\caption{Solution to the displacement-control problem described in Example \ref{Ex1} for $a=1$. For $\lambda(t)= 1+\left(\frac{t}{T}\right)^n$ (an increasing function) the variation of the axial stress $\sigma^{zz}(R,t)$ with $R$ at $t=\frac{T}{4}$ for the cases $n\in\left\{\frac{1}{2},1,2\right\}$ is shown.}
\label{fig:DispControlLambdaInc}
\end{figure}
\begin{figure}[t!]
\centering
\vskip 0.3in
\includegraphics[width=0.9\textwidth]{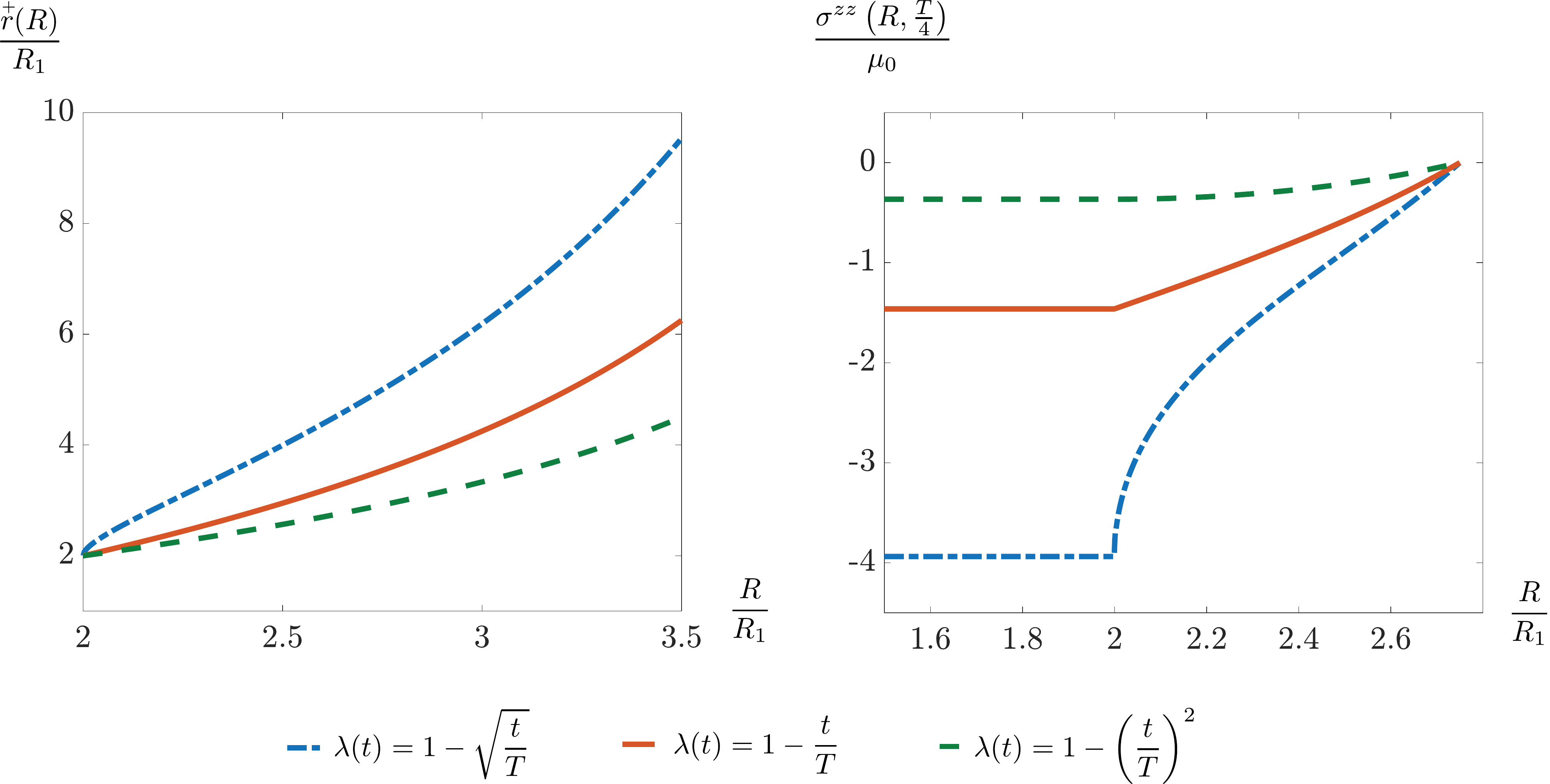}
\vspace*{0.0in}
\caption{Solution to the displacement-control problem described in Example \ref{Ex1} for $a=-1$. For $\lambda(t)= 1-\left(\frac{t}{T}\right)^n$ (a decreasing function) the variation of the axial stress $\sigma^{zz}(R,t)$ with $R$ at $t=\frac{T}{4}$ for the cases $n\in\left\{\frac{1}{2},1,2\right\}$ is shown.}
\label{fig:DispControlLambdaDec}
\end{figure}
\begin{figure}[t!]
\centering
\vskip 0.3in
\includegraphics[width=0.9\textwidth]{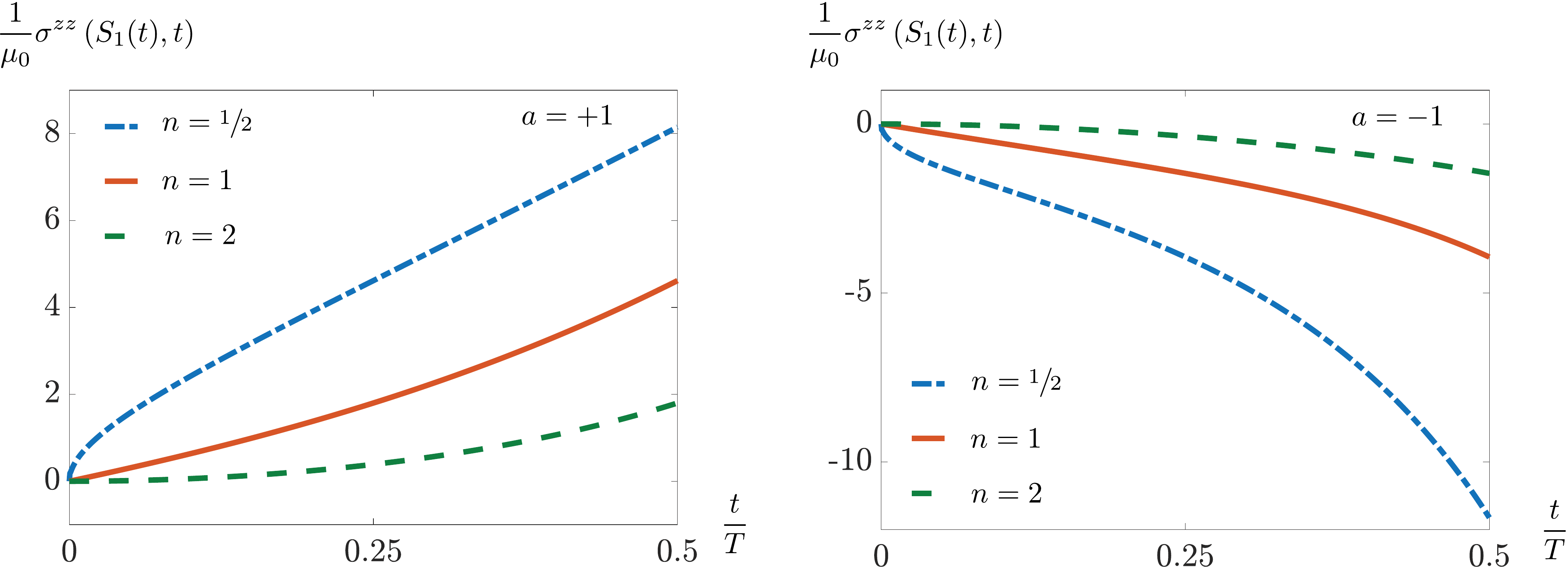}
\vspace*{0.0in}
\caption{The axial stress on the ablation boundary for the displacement-control problem described in Example \ref{Ex1}. The variation of $\sigma^{zz}(S_1(t),t)$ with $t(<\timeablation)$ for $\lambda(t)= 1+ a \left(\frac{t}{T}\right)^n$, where $a \in \big\{\pm 1\big\}$ and $n\in\left\{\frac{1}{2},1,2\right\}$ is shown.}
\label{fig:DispControlsigmazzAblBd}
\end{figure}
\end{example}

\begin{example} \label{Ex2}
Consider a force-control problem with $R_2=2R_1$, $\velab=\frac{R_1}{T}$, $\velac=2\frac{R_1}{T}$. Then $\timeablation=T$ and $S_2(T)=4 R_1$. To find solutions of the form \eqref{AnalytSolGuess}, first define $h(t):=\int_{R_2}^{S_2(t)}\frac{R\,\text{d}R}{\lambdac^4(R)}$, and $k(t):= \int_{R_2}^{S_2(t)}R \lambdac^2(R) \text{d}R$, so that $\dot{h}(t)=\frac{\velac S_2(t)}{\lambda^4(t)}$ and $\dot{k}(t)=\velac S_2(t)\lambda^2(t)$. Differentiating \eqref{axialforceintermsofLambda} with respect to $t$, one obtains 
\begin{equation}
\Bigg[2\lambda(t)\left[\frac{R_2^2-S_1^2(t)}{2}+h(t)\right]+\frac{4}{\lambda^5(t)}\left[\frac{R_2^2-S_1^2(t)}{2}+k(t)\right]\Bigg]\dot{\lambda}(t)+\velab S_1(t)\left[\frac{1}{\lambda^4(t)}-\lambda^2(t)\right]=\frac{\dot{F}(t)}{2\pi \mu_0}\,.
\end{equation}
Thus, we need to solve the following system of nonlinear ODEs:
\begin{equation}
\begin{aligned}
\begin{cases}
\displaystyle \dot{\lambda}(t)=\frac{f(t)+\velab S_1(t)\left[\lambda^2(t)-\frac{1}{\lambda^4(t)}\right]}{2\lambda(t)\left[\frac{R_2^2-S_1^2(t)}{2}+h(t)\right]+\frac{4}{\lambda^5(t)}\left[\frac{R_2^2-S_1^2(t)}{2}+k(t)\right]}\,,\\
\displaystyle \dot{h}(t)=\frac{\velac S_2(t)}{\lambda^4(t)} \,,\\
\displaystyle  \dot{k}(t)=\velac S_2(t)\lambda^2(t) \,,\\
\displaystyle \lambda(0)=1\,, ~ h(0)=0 \,, ~ k(0)=0 \,,
\end{cases}
\end{aligned}
\end{equation}
where $f(t):=\frac{\dot{F}(t)}{2\pi \mu_0}$. Assume the numerical values $R_1=1$ and $T=1$. The time-dependent force $F(t)$ is taken as a polynomial function of $t$ in Fig.~\ref{fig:ForceControlPoly}, and an error function of $t$ in Fig.~\ref{fig:ForceControlErf}, and a sinusoidal function of $t$ in Fig.~\ref{fig:ForceControlTrig}. 
The sign of $\sigma^{zz}$ is the same as that of $F$ for monotonic loads (Fig.~\ref{fig:ForceControlPoly}). In Fig.~\ref{fig:ForceControlErf}, as $F(t)$ increases from $0$ until the asymptotic value $\mu_0 \pi R_1^2$ is reached, $\dot{\lambda}(t)$ (and hence the axial strain rate) decreases until an asymptotic limit is reached. In Fig.~\ref{fig:ForceControlTrig} we look at three different time-dependent loads with $F(0)=F(T)=0$, but all of them have different stretches at $t=T$ because their loading histories are different. Similarly,  $F(t)=\mu_0 \pi R_1^2\sin(\frac{\pi t}{T})$ and $F(t)=\mu_0 \pi R_1^2\sin^2(\frac{\pi t}{T})$ have the same load at $t=\frac{T}{2}$, but different radial variation of $\sigma^{zz}$ at that instant because of their different loading histories.  
\begin{figure}[t!]
\centering
\vskip 0.3in
\includegraphics[width=0.9\textwidth]{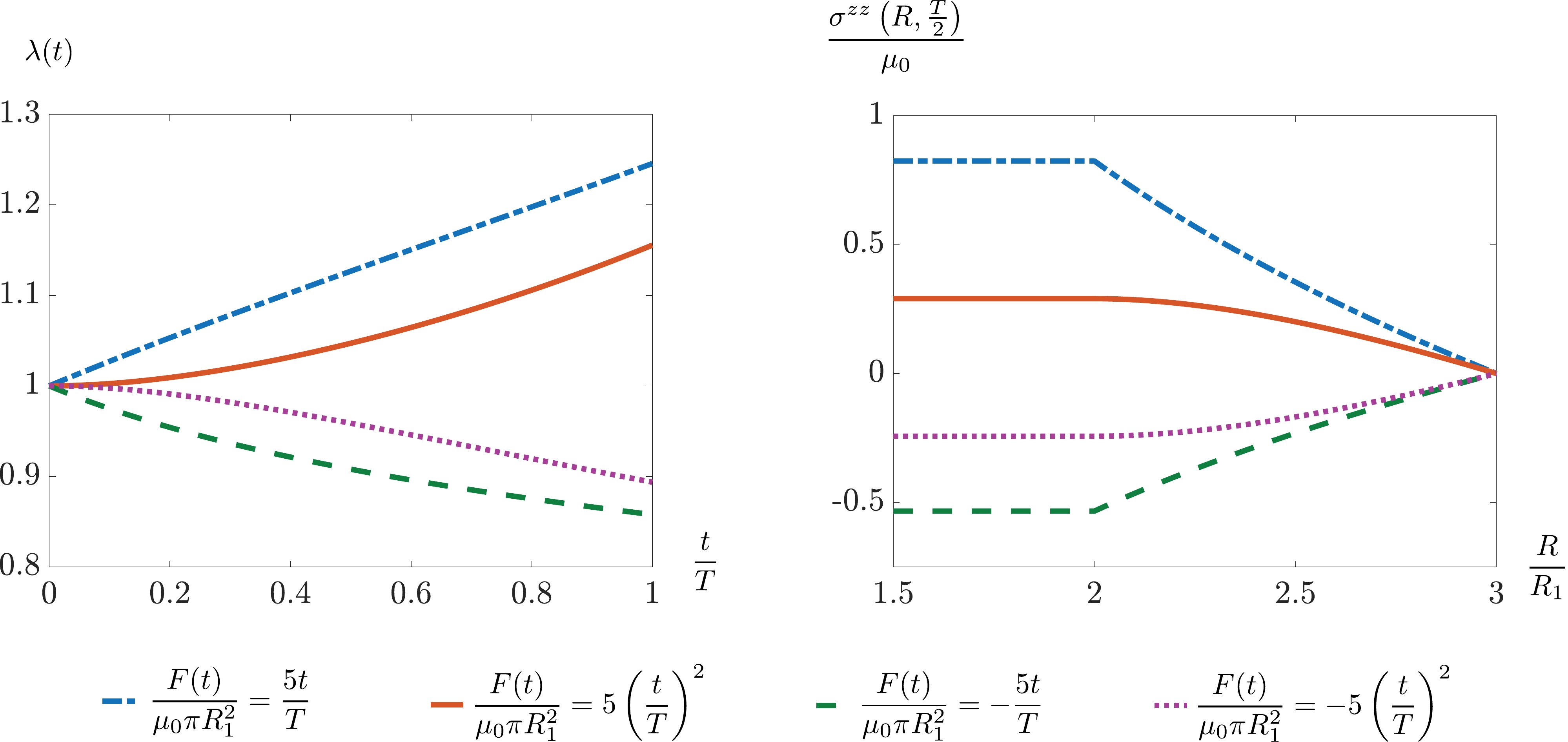}
\vspace*{0.0in}
\caption{Solution to the force-control problem described in Example \ref{Ex2} with $\displaystyle F(t)=\pm 5 \mu_0 \pi R_1^2 \left(\frac{t}{T}\right)^n $, where $n \in \{1,2\}$. The function $\lambda(t)$ reported here is the solution to the integral equation \eqref{axialforceintermsofLambda}. The variation of the axial stress $\sigma^{zz}$ with $R$ at $t=\frac{T}{2}$ is shown.}
\label{fig:ForceControlPoly}
\end{figure}
\begin{figure}[t!]
\centering
\vskip 0.3in
\includegraphics[width=0.9\textwidth]{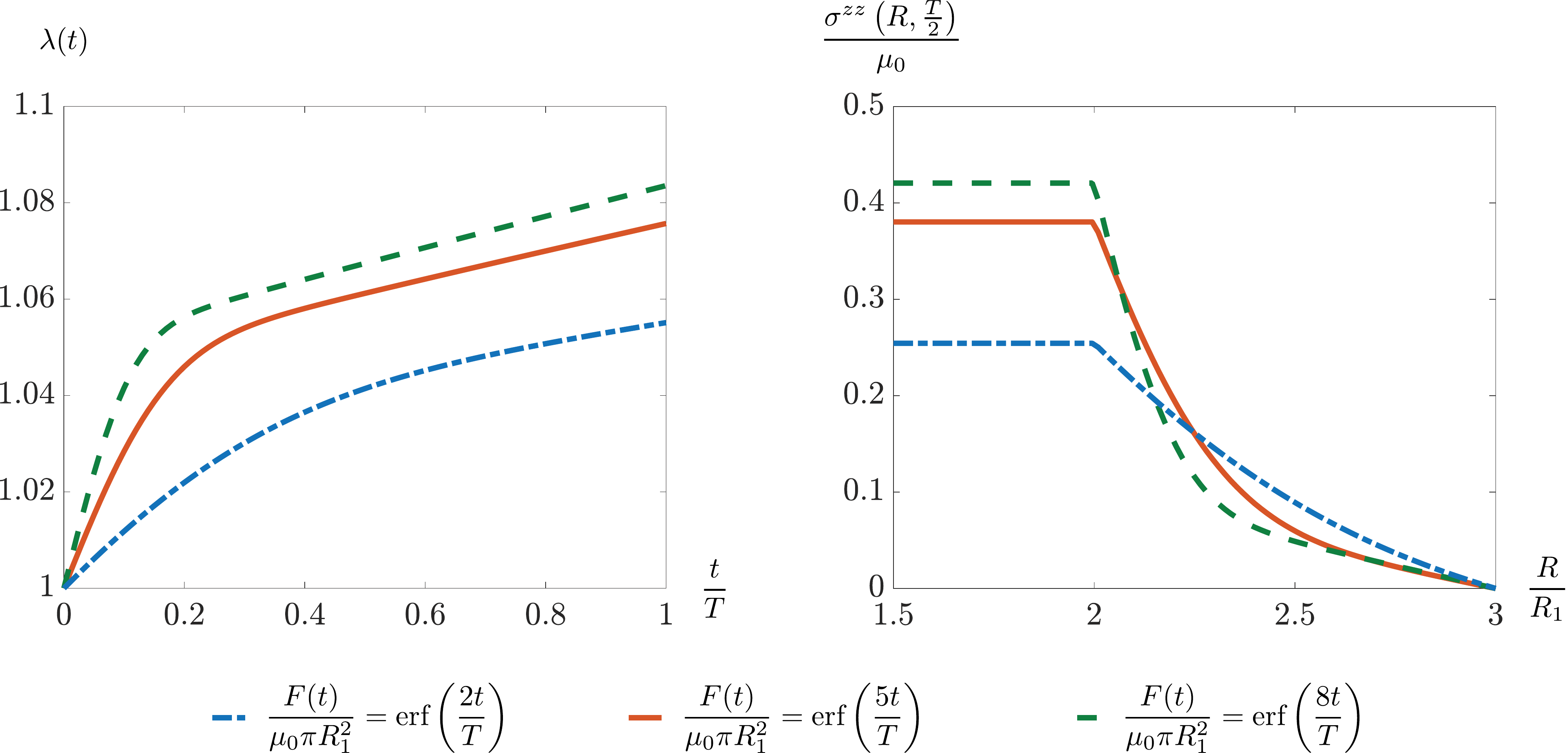}
\vspace*{0.2in}
\caption{ Solution to the force-control problem described in Example \ref{Ex2} with $\displaystyle F(t)= \mu_0 \pi R_1^2\, \mathrm{erf}\left(\frac{At}{T}\right) $, where $A \in \{2,5,8\}$. The function $\lambda(t)$ reported here is the solution to the integral equation \eqref{axialforceintermsofLambda}. The variation of the axial stress $\sigma^{zz}$ with $R$ at $t=\frac{T}{2}$ is shown.}
\label{fig:ForceControlErf}
\end{figure}
\begin{figure}[t!]
\centering
\includegraphics[width=0.9\textwidth]{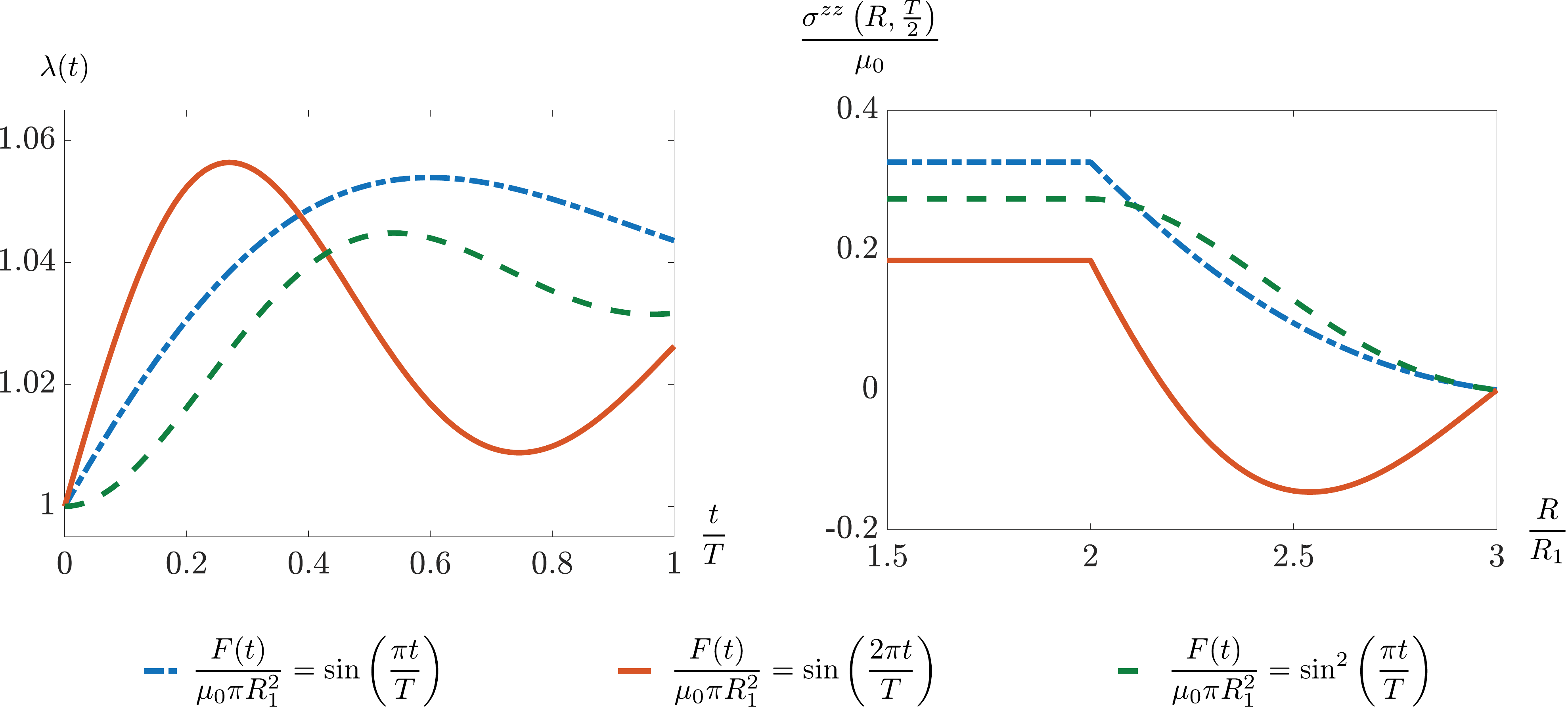}
\vspace*{0.2in}
\caption{Solution to the force-control problem described in Example \ref{Ex2}. The function $\lambda(t)$ reported here is the solution to the integral equation \eqref{axialforceintermsofLambda} with $F(t)=\mu_0\pi R_1^2\sin^m\left(\frac{\omega t}{T}\right)$, where $m\in\{1,2\}$ and $\omega\in\{\pi, 2\pi\}$. Further, the variation of the axial stress $\sigma^{zz}$ with $R$ at $t=\frac{T}{2}$ is shown.} 
\label{fig:ForceControlTrig}
\end{figure}
\end{example}

\subsection{Residual stress}
We assume that the body is unloaded after the accretion and ablation processes end. Let $T<\timeablation$. For $t> T$, $\lambda(t)=1$, and $F(t)=0$.  The material metric of the resulting body has the following representation\footnote{The constraint \eqref{r-constraint}$_1$ has been used in calculating the metric of the accreted portion of the body.}
\begin{equation}
\begin{aligned}
 \mathbf{G}=
\begin{cases} 
     \; 
\begin{bmatrix}
  1 & 0  & 0  \\
  0 & R^2  & 0  \\
  0 & 0  & 1
\end{bmatrix}\,,
     & \quad S_1(T) \leq R < R_2 \,, \\
\begin{bmatrix}
   1 & 0  & 0  \\
  0 & \rac^2(R)  & 0  \\
  0 & 0  & \lambdac^4(R)
\end{bmatrix}\,,
 & \quad R_2 \leq R \leq  S_2(T)\,,
\end{cases} 
\end{aligned} 
\end{equation}
wherein the case $R < R_2$ is present only if $T<\tab(R_2)$. Let $\tilde{\varphi}: \mathcal{B}_T \rightarrow \mathcal{S}$ map the  material manifold to the residually-stressed configuration. Let us consider the map $\tilde{\varphi}(R, \Theta,Z)=(\tilde{r},\tilde{\theta},\tilde{z})$ with $\tilde{r}=\tilde{r}(R)$, $\tilde{\theta}=\Theta$ and $\tilde{z}=\tilde{\lambda}^2 Z$. Incompressibilty constraint can be expressed as
\begin{equation}\label{IncompRes}
\begin{dcases}
\frac{\tilde{\lambda}^2\tilde{r}(R)\tilde{r}'(R)}{R} =1\,, & R \leq R_2\,, \\
\frac{\tilde{\lambda}^2\tilde{r}(R)\tilde{r}'(R)}{\lambdac^2(R)\,\rac(R)}=1\,, & R_2 \leq R\,,
\end{dcases}
\end{equation}
which implies that
\begin{equation}
	\tilde{r}^2(R) =
	\begin{cases} 
	\displaystyle \tilde{r}^2(S_1(T))+\frac{R^2-S_1^2(T)}{\tilde{\lambda}^2} \,, & R \leq R_2\,, \\[10pt]
	\displaystyle \tilde{r}^2(S_2(T)) -\frac{2}{\tilde{\lambda}^2}\int_R^{S_2(T)} \lambdac^2(\xi) \rac(\xi) \text{d}\xi\,,  & R_2 \leq R\,.
\end{cases}
\end{equation}
The continuity of $ \tilde{r}(R)$ at $R_2$ requires that
\begin{equation} \label{respostcont}
\tilde{r}^2(S_2(T))-\tilde{r}^2(S_1(T))= \frac{1}{\tilde{\lambda}^2}\left[R_2^2-S_1^2(T)+2\int_{R_2}^{S_2(T)} \lambdac^2(\xi) \rac(\xi) \text{d}\xi\right]\,.
\end{equation}
Observe that the knowledge of either $\tilde{r}(S_1(T))$ or $\tilde{r}(S_2(T))$ is sufficient to calculate $\tilde{r}(R)$. Let us take $\tilde{r}(S_1(T))$ as the only independent variable (other than $\tilde{\lambda}$), in terms of which
\begin{equation}
\tilde{r}^2(R) =
\begin{cases} 
	\displaystyle \tilde{r}^2(S_1(T))+\frac{R^2-S_1^2(T)}{\tilde{\lambda}^2} \,, & R \leq R_2\,, \\[10pt]
	\displaystyle \tilde{r}^2(S_1(T))+\frac{1}{\tilde{\lambda}^2}\left[R_2^2-S_1^2(T)+2\int_{R_2}^R \lambdac^2(\xi) \rac(\xi) \text{d}\xi\right]\,,  & R_2 \leq R\,.
\end{cases}
\end{equation}
The deformation gradient reads
\begin{equation}
   \tilde{\mathbf{F}}(R)=\begin{bmatrix}
  \tilde{r}'(R) & 0  & 0  \\
  0 & 1  & 0  \\
  0 & 0  & \tilde{\lambda}^2
\end{bmatrix}\,.
\end{equation}
For $ R \leq R_2$:
\begin{equation}\label{bcInitialBod}
  \mathbf{b}^{\sharp}(R)=\begin{bmatrix}
[\tilde{r}'(R) ]^2 & 0 & 0 \\
 0 & \displaystyle\frac{1}{R^2} &0 \\
 0 & 0 & \tilde{\lambda}^4 \\
\end{bmatrix} \,,\qquad 
\mathbf{c}^{\sharp}(R)=\begin{bmatrix}
\displaystyle\frac{1}{[\tilde{r}'(R)]^2} & 0 & 0 \\
 0 & \displaystyle\frac{R^2}{\tilde{r}^4(R)} & 0 \\
 0 & 0& \displaystyle\frac{1}{\tilde{\lambda}^4} \\
\end{bmatrix}\,.
\end{equation}
The principal invariants of $\mathbf{b}$ read
\begin{equation}
\begin{aligned}
   I_1(R) =[\tilde{r}'(R)]^2+\frac{\tilde{r}^2(R)}{R^2} +\tilde{\lambda}^4 \,,\qquad
   I_2(R) =[\tilde{r}'(R)]^2\,\tilde{\lambda}^4+ \frac{[\tilde{r}'(R)]^2\,\tilde{r}^2(R)}{R^2}
   +\frac{\tilde{r}^2(R)\, \tilde{\lambda}^4}{R^2} \,.
\end{aligned}
\end{equation}
Using \eqref{IncompRes}$_1$, the nonzero physical components of the residual Cauchy stress can be expressed as\footnote{Note that $\hat{\tilde{\sigma}}^{\theta\theta}=\tilde{r}^2 \tilde{\sigma}^{\theta\theta}$.}
\begin{equation} 
\begin{aligned}
\tilde{\sigma}^{rr}(R) &= -p(R)+ \frac{\alpha(R) \, R^2}{\tilde{\lambda}^4\tilde{r}^2(R)}-\frac{\beta(R) \tilde{\lambda}^4\tilde{r}^2(R)}{ R^2}  \,,  \\
\hat{\tilde{\sigma}}^{\theta\theta}(R) &= -p(R) +\frac{\alpha(R)\tilde{r}^2(R)}{R^2} -\frac{\beta(R)\,R^2}{\tilde{r}^2(R)}  \,, \\
 \tilde{\sigma}^{zz}(R) &= -p(R)+\alpha(R)\, \tilde{\lambda}^4   -\,\frac{\beta(R) }{\tilde{\lambda}^4}  \,.
\end{aligned}
\end{equation}
For $R< R_2$, the radial equilibrium equation and the traction boundary condition $\tilde{\sigma}^{rr}(S_1(T))=0$ imply that
\begin{equation}
\begin{aligned}
\tilde{\sigma}^{rr}(R)  =& \int_{S_1(T)}^R \left[\frac{\alpha(\xi)}{\tilde{\lambda}^4}+\beta(\xi)\right]  \left[ \frac{\tilde{\lambda}^2}{\xi}-\frac{\xi^3}{\tilde{\lambda}^2 \,\tilde{r}^4(\xi)}\right] d\xi \,,\\
  	\hat{\tilde{\sigma}}^{\theta\theta}(R) =& \left[\frac{\alpha(R)}{\tilde{\lambda}^4}+\beta(R)\right] \left[\frac{ \tilde{\lambda}^4 \tilde{r}^2(R)}{R^2}-\frac{R^2}{\tilde{r}^2(R)}\right]+	 \tilde{\sigma}^{rr}(R)\,, \\
\tilde{\sigma}^{zz}(R) =& \alpha(R)\left[\tilde{\lambda}^4-\frac{R^2}{\tilde{\lambda}^4\tilde{r}^2(R)}\right]	+\beta(R)\left[\frac{\tilde{\lambda}^4 \tilde{r}^2(R)}{R^2}-\frac{1}{\tilde{\lambda}^4}\right]+ \tilde{\sigma}^{rr}(R)\,.
\end{aligned}
\end{equation}
Now for $R\geq R_2$:
\begin{equation}
\begin{aligned}
  \mathbf{b}^{\sharp}(R) =\begin{bmatrix} \displaystyle
[\tilde{r}'(R) ]^2 & 0 & 0 \\
 0 & \displaystyle\frac{1}{\rac^2(R)} 
 & 0 \\
 0 & 0 & \displaystyle\frac{\tilde{\lambda}^4}{\lambdac^4(R)} \\
\end{bmatrix} \,,\qquad
\mathbf{c}^{\sharp}(R) =\begin{bmatrix}
\displaystyle
 \frac{1}{[\tilde{r}'(R) ]^2} & 0 & 0 \\
 0 & \displaystyle\frac{\rac^2(R)}{\tilde{r}^4(R) } & 0 \\
 0 & 0 & \displaystyle\frac{\lambdac^4(R)
 }{\tilde{\lambda}^4} \\
   \end{bmatrix}\,.
\end{aligned}
\end{equation}
The principal invariants of $\mathbf{b}$ read

\begin{equation}
\begin{aligned}
   I_1(R) =[\tilde{r}'(R) ]^2+\frac{\tilde{r}^2(R)}{\rac^2(R)}+\frac{\tilde{\lambda}^4}{\lambdac^4(R)} \,,
   \qquad
   I_2(R) =\frac{[\tilde{r}'(R) ]^2\,\tilde{r}^2(R)}{\rac^2(R)}
   +\frac{\tilde{\lambda}^4 \tilde{r}^2(R)}{\rac^2(R)\lambdac^4(R)}
   +\frac{\tilde{\lambda}^4 [\tilde{r}'(R) ]^2}{\lambdac^4(R)}\,.
\end{aligned}
\end{equation}
Using \eqref{IncompRes}$_2$, the nonzero physical components of the residual Cauchy stress can be expressed as
\begin{equation}
\begin{aligned}
   \tilde{\sigma}^{rr}(R) &= -p(R)+\alpha(R,t)\tilde{r}_{,R}^2(R)-\frac{\beta(R)}{\tilde{r}_{,R}^2(R)} \,,  \\
   \hat{\tilde{\sigma}}^{\theta\theta}(R) &= -p(R)
   +\frac{\alpha(R)\tilde{r}^2(R)}{\rac^2(R)}-\frac{\beta(R)\,\rac^2(R)}{\tilde{r}^2(R)}     \,, \\
   \tilde{\sigma}^{zz}(R) &= -p(R)+\frac{\alpha(R)\, \tilde{\lambda}^4}{\lambdac^4(R)}
   -\frac{\beta(R)\, \lambdac^4(R)}{\tilde{\lambda}^4} \,.
\end{aligned}
\end{equation}
For $R \geq R_2$, the radial equilibrium equation and the traction boundary condition $\tilde{\sigma}^{rr}(S_2(T))=0$ give us
\begin{equation} 
\begin{aligned}
\tilde{\sigma}^{rr}(R)=& -\int_R^{S_2(T)} \lambdac^2(\xi) \left[\frac{\alpha(\xi)}{\tilde{\lambda}^4}	+\frac{\beta(\xi)}{\lambdac^4(\xi)}\right] \left[ \frac{\tilde{\lambda}^2 }{\rac(\xi)}-\frac{\rac^3(\xi)\lambdac^4(\xi)}{\tilde{\lambda}^2\,\tilde{r}^4(\xi)}\right] d\xi\,, \\
	\hat{\tilde{\sigma}}^{\theta\theta}(R) =& \left[\frac{\alpha(R)}{\tilde{\lambda}^4 }+\frac{\beta(R)}{\lambdac^4(R)}\right]\left[\frac{\tilde{\lambda}^4  \tilde{r}^2(R)}{\rac^2(R)}-\frac{\rac^2(R)\lambdac^4(R)}{\tilde{r}^2(R)}\right]
	+\tilde{\sigma}^{rr}(R), \\
	\tilde{\sigma}^{zz}(R) =& \alpha(R)\left[\frac{\tilde{\lambda}^4 }{\lambdac^4(R)}
	-\frac{\lambdac^4(R)\rac^2(R)}{\tilde{\lambda}^4 \tilde{r}^2(R)}\right]
	+\beta(R)\left[\frac{\tilde{\lambda}^4  \tilde{r}^2(R)}{\lambdac^4(R)\rac^2(R)}
	-\frac{\lambdac^4(R)}{\tilde{\lambda}^4 }\right] +\tilde{\sigma}^{rr}(R).
\end{aligned}
\end{equation}
The continuity of $\tilde{\sigma}^{rr}(R)$ at $R_2$ requires that\footnote{
Alternatively, $\tilde{\sigma}^{rr}$ for $ R \leq R_2$  can be expressed as:
\begin{equation} 
\tilde{\sigma}^{rr}(R) = -\int_{R}^{R_2} \frac{\tilde{\lambda}^4 \tilde{r}^4(\xi)-\xi^4}{\tilde{\lambda}^2 \xi\, \tilde{r}^4(\xi)}\left[\frac{\alpha(\xi)}{\tilde{\lambda}^4}+\beta(\xi)\right] d\xi  -\int_{R_2}^{S_2(T)} \frac{\lambdac^2(\xi)\left[\tilde{\lambda}^4 \tilde{r}^4(\xi)-\rac^4(\xi)\lambdac^4(\xi)\right]}{\tilde{\lambda}^2 \tilde{r}^4(\xi)\rac(\xi)}\left[\frac{\alpha(\xi)}{\tilde{\lambda}^4}+\frac{\beta(\xi)}{\lambdac^4(\xi)}\right] d\xi\,,
\end{equation}
in which case the condition \eqref{residualtractioncont} is recovered from the traction boundary condition $\tilde{\sigma}^{rr}(S_1(T))=0$.
}
\begin{equation} \label{residualtractioncont}
\begin{aligned}
\tilde{\Upsilon}:=\int_{S_1(T)}^{R_2} \left[\frac{\alpha(\xi)}{\tilde{\lambda}^4}+\beta(\xi)\right]  \left[ \frac{\tilde{\lambda}^2}{\xi}-\frac{\xi^3}{\tilde{\lambda}^2 \,\tilde{r}^4(\xi)}\right] d\xi 
+\int_{R_2}^{S_2(T)} \lambdac^2(\xi) \left[\frac{\alpha(\xi)}{\tilde{\lambda}^4}	+\frac{\beta(\xi)}{\lambdac^4(\xi)}\right] \left[ \frac{\tilde{\lambda}^2 }{\rac(\xi)}-\frac{\rac^3(\xi)\lambdac^4(\xi)}{\tilde{\lambda}^2\,\tilde{r}^4(\xi)}\right] d\xi =0 \,.
\end{aligned}
\end{equation}
Absence of an axial force in the residually-stressed state implies that
\begin{equation}\label{axialRes_v1}
	\int_{S_1(T)}^{S_2(T)}2\pi R\, \tilde{P}^{zZ}(R) \text{d}R=0\,,
\end{equation}
where $\tilde{P}^{zZ}(R)= \displaystyle\frac{\tilde{\sigma}^{zz}(R)}{\tilde{\lambda}^2}$ is the $zZ$- component of the residual Piola-Kirchhoff stress, calculated as
\begin{equation}
\begin{aligned}
\tilde{P}^{zZ}(R)=
\begin{cases}
 \alpha(R)\left[\tilde{\lambda}^2-\frac{R^2}{\tilde{\lambda}^6 \tilde{r}^2(R)}\right]+\beta(R)\left[\frac{\tilde{\lambda}^2 \tilde{r}^2(R)}{R^2}-\frac{1}{\tilde{\lambda}^6}\right] +\frac{\tilde{\sigma}^{rr}(R)}{\tilde{\lambda}^2}\,,& R \leq R_2\,, \\[10pt]
 \alpha(R)\left[\frac{\tilde{\lambda}^2}{\lambdac^4(R)}-\frac{\lambdac^4(R)\rac^2(R)}{\tilde{\lambda}^6\tilde{r}^2(R)}\right]+ \beta(R)\left[\frac{\tilde{\lambda}^2\tilde{r}^2(R)}{\lambdac^4(R)\rac^2(R)}-\frac{\lambdac^4(R)}{\tilde{\lambda}^6}\right] +\frac{\tilde{\sigma}^{rr}(R)}{\tilde{\lambda}^2} \,, &R_2 \leq R\,. 
\end{cases}
\end{aligned}
\end{equation}

\paragraph{Residual stress in the case of a neo-Hookean solid.}
Consider a homogeneous neo-Hookean material for which $\alpha(R)={\mu}(R)=\mu_0>0$ and $\beta(R)=0$ as in the previous section. The nonzero components of the residual Cauchy stress are written as
\begin{equation}
\begin{aligned}
\tilde{\sigma}^{rr}(R)&= 
\begin{dcases}
	\frac{\mu_0}{\tilde{\lambda}^2}\int_{S_1(T)}^R \frac{\text{d}\xi}{\xi }-\frac{\mu_0}{\tilde{\lambda}^6}\int_{S_1(T)}^R \frac{\xi^3 \text{d}\xi}{ \tilde{r}^4(\xi)}
	  \,, &  R < R_2\,, \\
	-\frac{\mu_0 }{\tilde{\lambda}^2} \int_R^{S_2(T)}    \frac{\lambdac^2(\xi) d\xi}{\rac(\xi)} +\frac{\mu_0 }{\tilde{\lambda}^6} \int_R^{S_2(T)}  \frac{\rac^3(\xi)\lambdac^6(\xi)d\xi}{\tilde{r}^4(\xi)} \,, & R\geq R_2 \,,
\end{dcases} \\
\hat{\tilde{\sigma}}^{\theta\theta}(R)&= 
\begin{dcases}
	 \mu_0 \left[\frac{\tilde{r}^2(R)}{R^2}-\frac{ R^2}{\tilde{\lambda}^4 \,\tilde{r}^2(R)}\right]+\tilde{\sigma}^{rr}(R) \,, &  R < R_2\,, \\
	\mu_0\left[\frac{\tilde{r}^2(R)}{\rac^2(R)}
	-\frac{\lambdac^4(R)\rac^2(R)}{\tilde{\lambda}^4 \tilde{r}^2(R)}\right]
	+\tilde{\sigma}^{rr}(R)\,, & R\geq R_2 \,,
\end{dcases} \\
\tilde{\sigma}^{zz}(R)&=
\begin{dcases}
	\mu_0\left[\tilde{\lambda}^4-\frac{R^2}{\tilde{\lambda}^4 \tilde{r}^2(R)}\right] +\tilde{\sigma}^{rr}(R) \,, &  R < R_2\,, \\
	\mu_0\left[\frac{\tilde{\lambda}^4}{\lambdac^4(R)}
	-\frac{\lambdac^4(R)\rac^2(R)}{\tilde{\lambda}^4 \tilde{r}^2(R)}\right]
	+\tilde{\sigma}^{rr}(R)\,, & R\geq R_2 \,.
\end{dcases} 
\end{aligned}
\end{equation}
Continuity of $\tilde{\sigma}^{rr}$, i.e., \eqref{residualtractioncont} simplifies to read
\begin{equation}\label{restractcont_v2} 
\tilde{\lambda}^4 \left[ \log\left({\frac{R_2}{S_1(T)}}\right) + \int_{R_2}^{S_2(T)} \frac{\lambdac^2(\xi)d\xi}{\rac(\xi)} \right]-\left[  \int_{S_1(T)}^{R_2} \frac{\xi^3 d\xi }{\tilde{r}^4(\xi)} +\int_{R_2}^{S_2(T)} \frac{\rac^3(\xi)\lambdac^6(\xi)d\xi}{\tilde{r}^4(\xi)} \right]=0 \,.
\end{equation}
Further,
\begin{equation}
\begin{aligned}
\tilde{P}^{zZ}(R)=
\begin{cases}
 \mu_0\left[\tilde{\lambda}^2-\frac{R^2}{\tilde{\lambda}^6 \tilde{r}^2(R)}\right]+\frac{\tilde{\sigma}^{rr}(R)}{\tilde{\lambda}^2}\,,& R \leq R_2\,, \\[10pt]
 \mu_0\left[\frac{\tilde{\lambda}^2}{\lambdac^4(R)}-\frac{\lambdac^4(R)\rac^2(R)}{\tilde{\lambda}^6\tilde{r}^2(R)}\right]+\frac{\tilde{\sigma}^{rr}(R)}{\tilde{\lambda}^2} \,, &R_2 \leq R\,, 
\end{cases}
\end{aligned}
\end{equation}
so that the zero axial force condition \eqref{axialRes_v1} implies that
\begin{equation}\label{ZerAxforce_v2}
	\mu_0\int_{S_1(T)}^{R_2} \left[\tilde{\lambda}^2-\frac{R^2}{\tilde{\lambda}^6 \tilde{r}^2(R)}\right]R\, \text{d}R + \mu_0\int_{R_2}^{S_2(T)} \left[\frac{\tilde{\lambda}^2}{\lambdac^4(R)}-\frac{\lambdac^4(R)\rac^2(R)}{\tilde{\lambda}^6\tilde{r}^2(R)}\right]R\, \text{d}R + \frac{1}{\tilde{\lambda}^2}\int_{S_1(T)}^{S_2(T)} R\, \tilde{\sigma}^{rr}(R)\, \text{d}R=0\,.
\end{equation}
Note that 
\begin{equation}
\begin{aligned}
	\int_{S_1(T)}^{S_2(T)} R\, \tilde{\sigma}^{rr}(R)\, \text{d}R & =  
	\frac{\mu_0 }{\tilde{\lambda}^2} \int_{S_1(T)}^{R_2} \left[\int_{S_1(T)}^R \frac{\text{d}\xi}{\xi }\right] R\, 
	\text{d}R -\frac{\mu_0 }{\tilde{\lambda}^6} \int_{S_1(T)}^{R_2} 
	\left[\int_{S_1(T)}^R \frac{\xi^3 \text{d}\xi}{ \tilde{r}^4(\xi)}\right] R\, \text{d}R  \\
	& \quad -\frac{\mu_0 }{ \tilde{\lambda}^2}  \int_{R_2}^{S_2(T)}  \left[\int_R^{S_2(T)}    
	\frac{\lambdac^2(\xi) d\xi}{\rac(\xi)}\right] R\, \text{d}R \\
	& \quad  +\frac{\mu_0 }{ \tilde{\lambda}^6}  \int_{R_2}^{S_2(T)} \left[\int_R^{S_2(T)}  
	 \frac{\rac^3(\xi)\lambdac^6(\xi)d\xi}{\tilde{r}^4(\xi)}\right]R\, \text{d}R\,,
\end{aligned}
\end{equation}
or equivalently,
\begin{equation}
\begin{aligned}
	\int_{S_1(T)}^{S_2(T)} R\, \tilde{\sigma}^{rr}(R)\, \text{d}R & = 
	\frac{\mu_0 }{2\tilde{\lambda}^2}\left[R_2^2 \log\left(\frac{R_2}{S_2(T)}\right)
	-\frac{R_2^2-S_1^2(T)}{2}\right] -\frac{\mu_0 }{2\tilde{\lambda}^6} \int_{S_1(T)}^R 
	\frac{ \left[R_2^2-\xi^2\right] \xi^3 \text{d}\xi}{ \tilde{r}^4(\xi)} \\
	& \quad -\frac{\mu_0 }{2\tilde{\lambda}^2}  \int_{R_2}^{S_2(T)}  
	\frac{\left[\xi^2-R_2^2\right] \lambdac^2(\xi) \, d\xi}{\rac(\xi)}  +\frac{\mu_0 }{2\tilde{\lambda}^6}  
	\int_{R_2}^{S_2(T)}  \frac{\left[\xi^2-R_2^2\right]\rac^3(\xi)\lambdac^6(\xi)d\xi}{\tilde{r}^4(\xi)}\,.
\end{aligned}
\end{equation}
This in view of \eqref{restractcont_v2} implies that
\begin{equation}
\begin{aligned}
\int_{S_1(T)}^{S_2(T)} R\, \tilde{\sigma}^{rr}(R)\, \text{d}R=&  -\frac{\mu_0 [R_2^2-S_1^2(T)]}{4\tilde{\lambda}^2} +\frac{\mu_0 }{2\tilde{\lambda}^6} \int_{S_1(T)}^R \frac{  \xi^5 \text{d}\xi}{ \tilde{r}^4(\xi)} \\
&-\frac{\mu_0 }{2\tilde{\lambda}^2}  \int_{R_2}^{S_2(T)}  \frac{\xi^2 \lambdac^2(\xi) \, d\xi}{\rac(\xi)} 
 +\frac{\mu_0 }{2\tilde{\lambda}^6}  \int_{R_2}^{S_2(T)}  \frac{\xi^2\rac^3(\xi)\lambdac^6(\xi)d\xi}{\tilde{r}^4(\xi)}\,.
\end{aligned}
\end{equation}
Thus, \eqref{ZerAxforce_v2} can be rewritten as
\begin{equation} \label{ZerAxforce_v3}
\begin{aligned}
\tilde{\lambda}^2\left[ \frac{R_2^2-S_1^2(T)}{2}+\int_{R_2}^{S_2(T)}\frac{  R\,\text{d}R}{\lambdac^4(R)}\right]  
 - \frac{1}{\tilde{\lambda}^6}\left[\int_{S_1(T)}^{R_2} \frac{R^3 \,\text{d}R}{ \tilde{r}^2(R)} + \int_{R_2}^{S_2(T)}\frac{R \,\lambdac^4(R)\rac^2(R)\,\text{d}R}{\tilde{r}^2(R)} \right]\\
- \frac{1}{2 \tilde{\lambda}^4}\left[ \frac{R_2^2-S_1^2(T)}{2}+  \int_{R_2}^{S_2(T)}  \frac{\xi^2\lambdac^2(\xi) \, d\xi }{ \rac(\xi)}    \right] + \frac{1}{2 \tilde{\lambda}^8}\left[ \int_{S_1(T)}^{R_2} \frac{\xi^5 d\xi  }{ \, \tilde{r}^4(\xi)}   +   \int_{R_2}^{S_2(T)}  \frac{ \xi^2 \rac^3(\xi)\lambdac^6(\xi)\, d\xi }{ \tilde{r}^4(\xi)}  \right]=0  \,.
\end{aligned}
\end{equation}
Determining the residual stretch and stress requires calculating the unknowns $\tilde{r}(S_1(T))$ and $\tilde{\lambda}$ that satisfy \eqref{restractcont_v2} and \eqref{ZerAxforce_v3}.
\begin{remark} \label{ResSolAnalyl}
The functions $\lambdac(R)$ and $\rac(R)$ are known from the deformation history of the bar prior to the removal of external loads, and hence, are treated as given quantities while solving for the residually-stressed state.
Observe that 
\begin{equation}
\tilde{r}(R) =
\begin{cases} 
	\displaystyle \frac{R}{\tilde{\lambda}} \,, & R \leq R_2\,, \\[10pt]
	\displaystyle \frac{\lambdac(R) \rac(R) }{\tilde{\lambda}}\,,  & R_2 \leq R\,,
\end{cases}
\end{equation}
is a solution that satisfies \eqref{IncompRes}, \eqref{restractcont_v2} and \eqref{ZerAxforce_v3}. The residual Cauchy stress components $\tilde{\sigma}^{rr}$ and $\hat{\tilde{\sigma}}^{\theta\theta}$ vanish for this solution. Moreover,
\begin{equation}
\begin{aligned}
\tilde{P}^{zZ}(R)=
\begin{cases}
 \displaystyle \mu_0 \left[\tilde{\lambda}^2-\frac{1}{\tilde{\lambda}^4 }\right] \,,& R \leq R_2\,, \\[10pt]
 \displaystyle  \mu_0\left[\frac{\tilde{\lambda}^2}{\lambdac^4(R)}-\frac{\lambdac^2(R)}{\tilde{\lambda}^4}\right] \,, &R_2 \leq R\,, 
\end{cases}
\end{aligned}
\end{equation}
so that the zero axial force condition \eqref{axialRes_v1} requires that
\begin{equation} \label{axialforceforResSimplified}
	\int_{S_1(T)}^{R_2} \mu_0 R  \left[\tilde{\lambda}^2-\frac{1}{\tilde{\lambda}^4 }\right] \text{d}R 
	+\int_{R_2}^{S_2(T)} \mu_0 R \left[\frac{\tilde{\lambda}^2}{\lambdac^4(R)}
	-\frac{\lambdac^2(R)}{\tilde{\lambda}^4}\right] \text{d}R=0\,.
\end{equation}
This implies that
\begin{equation}
	\tilde{\lambda}^6=\displaystyle\frac{R_2^2-S_1^2(T)+2 \displaystyle\int_{R_2}^{S_2(T)}R\,\lambdac^2(R)\text{d}R}{R_2^2-S_1^2(T)+2 \displaystyle\int_{R_2}^{S_2(T)}\frac{R\,\text{d}R}{\lambdac^4(R)}}\,.
\end{equation}
\end{remark}

\begin{example} \label{Ex3}
Consider a displacement-control problem with $R_2=2R_1\,$, $\velab=\frac{R_1}{2T}$, $\velac= \frac{R_1}{T}$, and $\lambda(t)= 1+ a \left(\frac{t}{T}\right)^n$, where $n \geq 0$, and $a \in \mathbb{R}$. Then $\timeablation=2T>T$, $S_1(T)=1.5 R_1$ and $S_2(T)=3 R_1$. First we need to find $\rac:[R_2,S_2(T)]\rightarrow \mathbb{R}$ as in Example \ref{Ex1}. With the knowledge of $\lambdac$ and $\rac$ (see Fig.~\ref{fig:rPDispControlRes}), we need to find $\tilde{r}(S_1(T))$ and $\tilde{\lambda}$ that satisfy the following nonlinear integral equations
\begin{equation} \label{ResProbSys}
\begin{aligned}
\begin{cases}
\displaystyle \tilde{\lambda}^2\left[ \frac{R_2^2-S_1^2(T)}{2}+\int_{R_2}^{S_2(T)}\frac{  R\,\text{d}R}{\lambdac^4(R)}\right]  
  - \frac{1}{\tilde{\lambda}^6}\left[\int_{S_1(T)}^{R_2} \frac{R^3 \,\text{d}R}{ \tilde{r}^2(R)} + \int_{R_2}^{S_2(T)}\frac{R \,\lambdac^4(R)\rac^2(R)\,\text{d}R}{\tilde{r}^2(R)} \right]\\[8pt]
\displaystyle \qquad - \frac{1}{2 \tilde{\lambda}^4}\left[ \frac{R_2^2-S_1^2(T)}{2}+  \int_{R_2}^{S_2(T)}  \frac{\xi^2\lambdac^2(\xi) \, d\xi }{ \rac(\xi)}    \right] + \frac{1}{2 \tilde{\lambda}^8}\left[ \int_{S_1(T)}^{R_2} \frac{\xi^5 d\xi  }{ \, \tilde{r}^4(\xi)}   +   \int_{R_2}^{S_2(T)}  \frac{ \xi^2 \rac^3(\xi)\lambdac^6(\xi)\, d\xi }{ \tilde{r}^4(\xi)}  \right]=0 \,,\\[20pt]
 \displaystyle \tilde{\lambda}^4 \left[ \log\left({\frac{R_2}{S_1(T)}}\right) + \int_{R_2}^{S_2(T)} \frac{\lambdac^2(\xi)d\xi}{\rac(\xi)} \right]-\left[  \int_{S_1(T)}^{R_2} \frac{\xi^3 d\xi }{\tilde{r}^4(\xi)} +\int_{R_2}^{S_2(T)} \frac{\rac^3(\xi)\lambdac^6(\xi)d\xi}{\tilde{r}^4(\xi)} \right]=0   \,,
\end{cases}
\end{aligned}
\end{equation}
where
\begin{equation} \label{ResProbKinedep}
\tilde{r}^2(R) =
\begin{cases} 
	\displaystyle \tilde{r}^2(S_1(T))+\frac{R^2-S_1^2(T)}{\tilde{\lambda}^2} \,, & R \leq R_2\,, \\[10pt]
	\displaystyle \tilde{r}^2(S_1(T))+\frac{1}{\tilde{\lambda}^2}\left[R_2^2-S_1^2(T)+2\int_{R_2}^R \lambdac^2(\xi) \rac(\xi) \text{d}\xi\right]\,,  & R_2 \leq R\,.
\end{cases}
\end{equation}
We solve this system numerically in Matlab for $a \in \big\{\pm \frac{1}{2}\big\}$ and $n\in\big\{\frac{1}{2}, 1, 2\big\}$ (with the numerical values $R_1=1$ and $T=1$). Since the numerical values of $\tilde{\sigma}^{rr}(R)$ and $\tilde{\sigma}^{\theta\theta}(R)$ are negligible, they are not reported here. The values of $\tilde{\lambda}^2$ (see Table \ref{table:dispControlResstretch}) and $\tilde{\sigma}^{zz}$ (see Fig.~\ref{fig:DispControlsigmazzRes}) obtained numerically agree with those described in Remark \ref{ResSolAnalyl}.
From Fig.~\ref{fig:DispControlsigmazzRes}, we observe that even if a bar is subjected only to elongation, the residual axial stresses in the initial portion of the final body are compressive due to stress redistribution after the bar is set free. Similarly, tensile residual axial stress is observed in the initial portion of a bar shortened during accretion-ablation. 
\begin{table} 
\centering
\begin{tabular}{ p{2cm}|p{1.5cm}|p{1.5cm} |p{1.5cm}}
& $n=\frac{1}{2}$ & $n=1$ & $n=2$  \\[2pt]
\hline
$a=\frac{1}{2}$ & $1.4733$  & $1.3547$ & $1.2349$\\  
$a=-\frac{1}{2}$ & $0.4727$  & $0.5514$ & $0.6477$
\end{tabular}
\caption{Residual stretch $\tilde{\lambda}^2$ in the body when it is set free after the displacement-control loading described in Example \ref{Ex3}.}
\label{table:dispControlResstretch}
\end{table}
\begin{figure}[t!]
\centering
\vskip 0.3in
\includegraphics[width=0.9\textwidth]{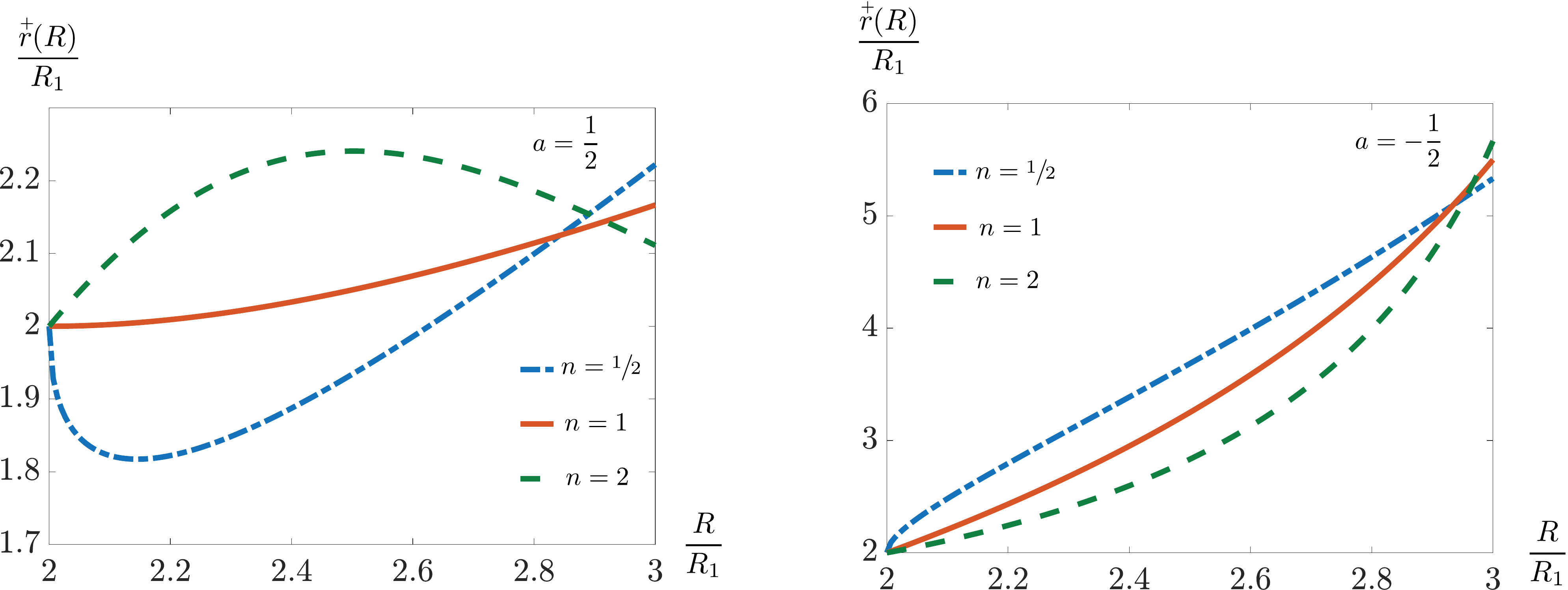}
\vspace*{0.10in}
\caption{Solution of the displacement-control problem described in Example \ref{Ex3} during the loading process.}
\label{fig:rPDispControlRes}
\end{figure}
\begin{figure}[t!]
\centering
\vskip 0.3in
\includegraphics[width=0.9\textwidth]{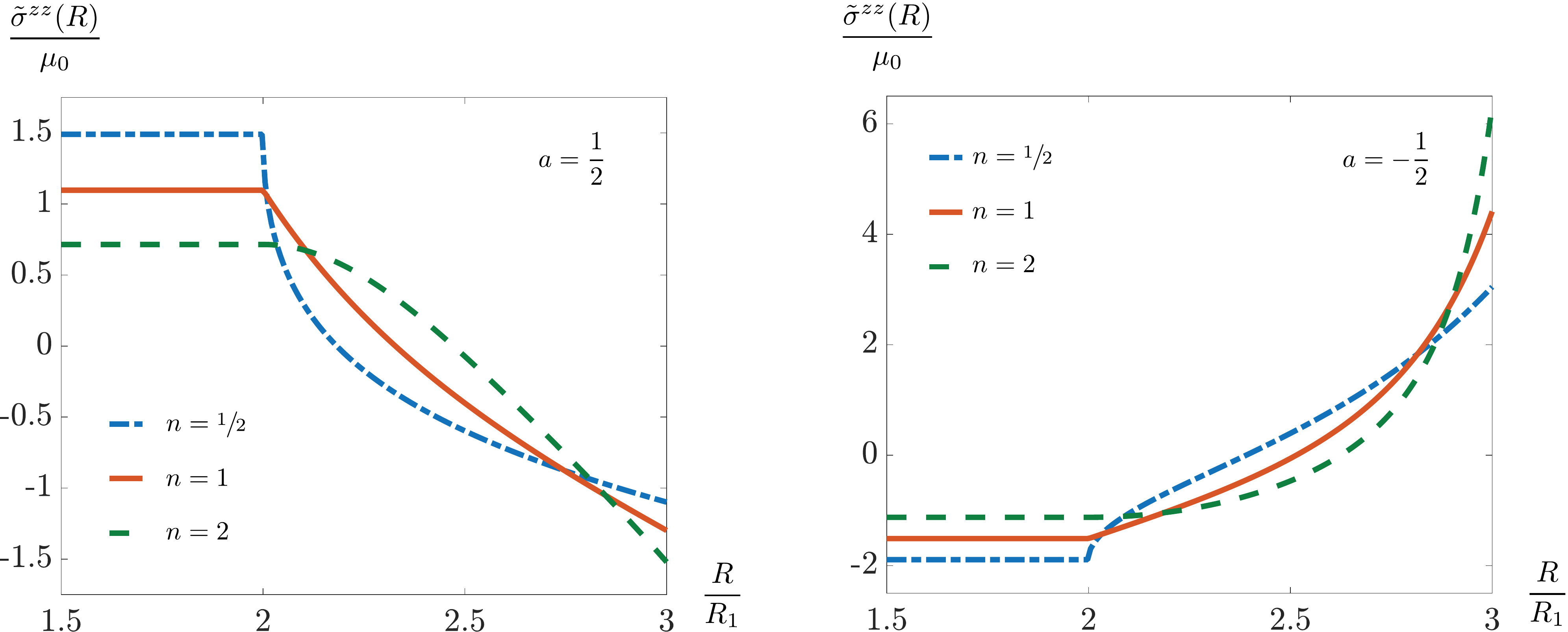}
\vspace*{0.10in}
\caption{Residual stress $\tilde{\sigma}^{zz}(R)$ in the body (Example \ref{Ex3}) after the removal of the external forces for the displacement-control loading $\lambda(t)= 1+ a \left( \frac{t}{T} \right)^n$,  where $a \in \left\{\pm \frac{1}{2}\right\}$ and $n\in\left\{\frac{1}{2},1,2\right\}$. }
\label{fig:DispControlsigmazzRes}
\end{figure}
\end{example}

\begin{example} \label{Ex4}
Consider a force-control problem with $R_2=2R_1$, $\velab=\frac{R_1}{2T}$, and $\velac= \frac{R_1}{T}$. Then $\timeablation=2T>T$, $S_1(T)=1.5 R_1$, and $S_2(T)=3 R_1$.  First we find the functions $\lambda(t)$ and $\rac(R)$ as in Example \ref{Ex2}, which are then used to solve \eqref{ResProbSys} and \eqref{ResProbKinedep} for $\tilde{r}(S_1(T))$ and $\tilde{\lambda}$. Assume the numerical values $R_1=1$ and $T=1$. We report $\tilde{\sigma}^{zz}$ as a function of $R$ in the residually-stressed configuration taking $F(t)$ as polynomial (Fig.~\ref{fig:ForceControlResPoly}), error (Fig.~\ref{fig:ForceControlResErf}), and sinusoidal (Fig.~\ref{fig:ForceControlResTrig}) functions of time $t$. The residual stretches $\tilde{\lambda}^2$ for the same choices of $F(t)$ are given in Table \ref{table:forceControlResstretch}.
In Fig.~\ref{fig:ForceControlResPoly} we compare the axial residual stress for loads varying monotonically as linear and quadratic functions of time. A monotonically increasing tensile $F(t)$ induces a compressive $\tilde{\sigma}^{zz}$ in the initial portion of the body, although $\tilde{\lambda}^2>1$. 
Similarly, a monotonically increasing compressive $F(t)$ leaves a tensile $\tilde{\sigma}^{zz}$ in the initial portion of the final body along with a residual contraction $\tilde{\lambda}^2<1$.
In Fig.~\ref{fig:ForceControlResErf}, $\tilde{\sigma}^{zz}$ is observed to be almost the same towards the outermost accreted layers for all the three loading paths. This is probably because the load when those outermost layers were accreted was very close to the asympotic limit of $F(t)$ in all the three cases. As a result, all those layers experience the same state of stress during loading as well as when they are set free.
In Fig.~\ref{fig:ForceControlResTrig}, we look at the residual stress print left after different sinusoidal loading cycles and observe that $\tilde{\sigma}^{zz}$ remains zero on the outer boundary even after the bar is unloaded.
\begin{table} 
\centering
\begin{tabular}{ ccccc } 
$\displaystyle\frac{F(t)}{\mu_0 \pi R_1^2}$ &  & $\tilde{\lambda}^2$ \\[8pt]
\hline
\multirow{3}{4em}{$\text{erf}\left(\frac{At}{T}\right)$} & $A=2$ & $1.0603$ \\ 
& $A=5$ & $1.0835$ \\ 
& $A=8$ & $1.0910$ \\ 
\hline
\multirow{3}{4em}{$a\left(\frac{t}{T}\right)^n$} & $a=5$, $n=1$ & $1.2158$ \\ 
& $a=5$, $n=2$ & $1.1166$ \\ 
& $a=-5$, $n=1$ & $0.8479$ \\ 
& $a=-5$, $n=2$ & $0.9051$ \\ 
\hline
\multirow{3}{4em}{$\sin^m\left(\frac{\omega t}{T}\right)$} & $m=1$, $\omega=\pi$ & $1.0617$ \\ 
& $m=1$, $\omega=2\pi$  & $1.0219$ \\ 
& $m=2$, $\omega=\pi$  & $1.0468$ \\ 
\hline
\end{tabular}
\caption{Residual stretch $\tilde{\lambda}^2$ in the body when it is set free after the force-control loading described in Example \ref{Ex4}.}
\label{table:forceControlResstretch}
\end{table}
\begin{figure}[t!]
\centering
\vskip 0.3in
\includegraphics[width=0.75\textwidth]{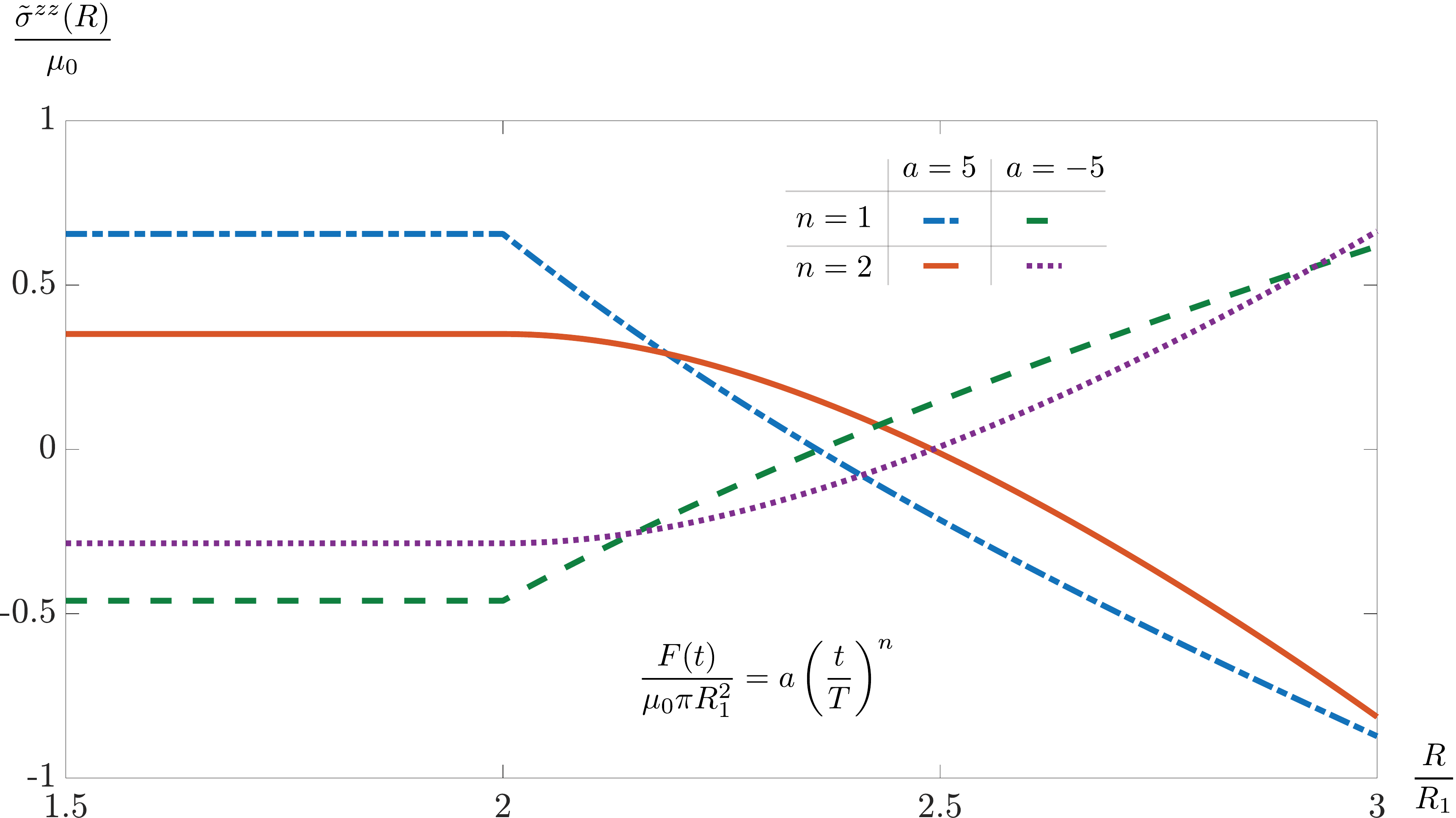}
\vspace*{0.2in}
\caption{Residual stress $\tilde{\sigma}^{zz}(R)$ in the body (Example \ref{Ex4}) when it is set free after the loading $F(t)= \pm 5\mu_0 \pi R_1^2 \left( \frac{t}{T} \right)^n$, where $n\in \{1,2\}$. }
\label{fig:ForceControlResPoly}
\end{figure}
\begin{figure}[t!]
\centering
\includegraphics[width=0.75\textwidth]{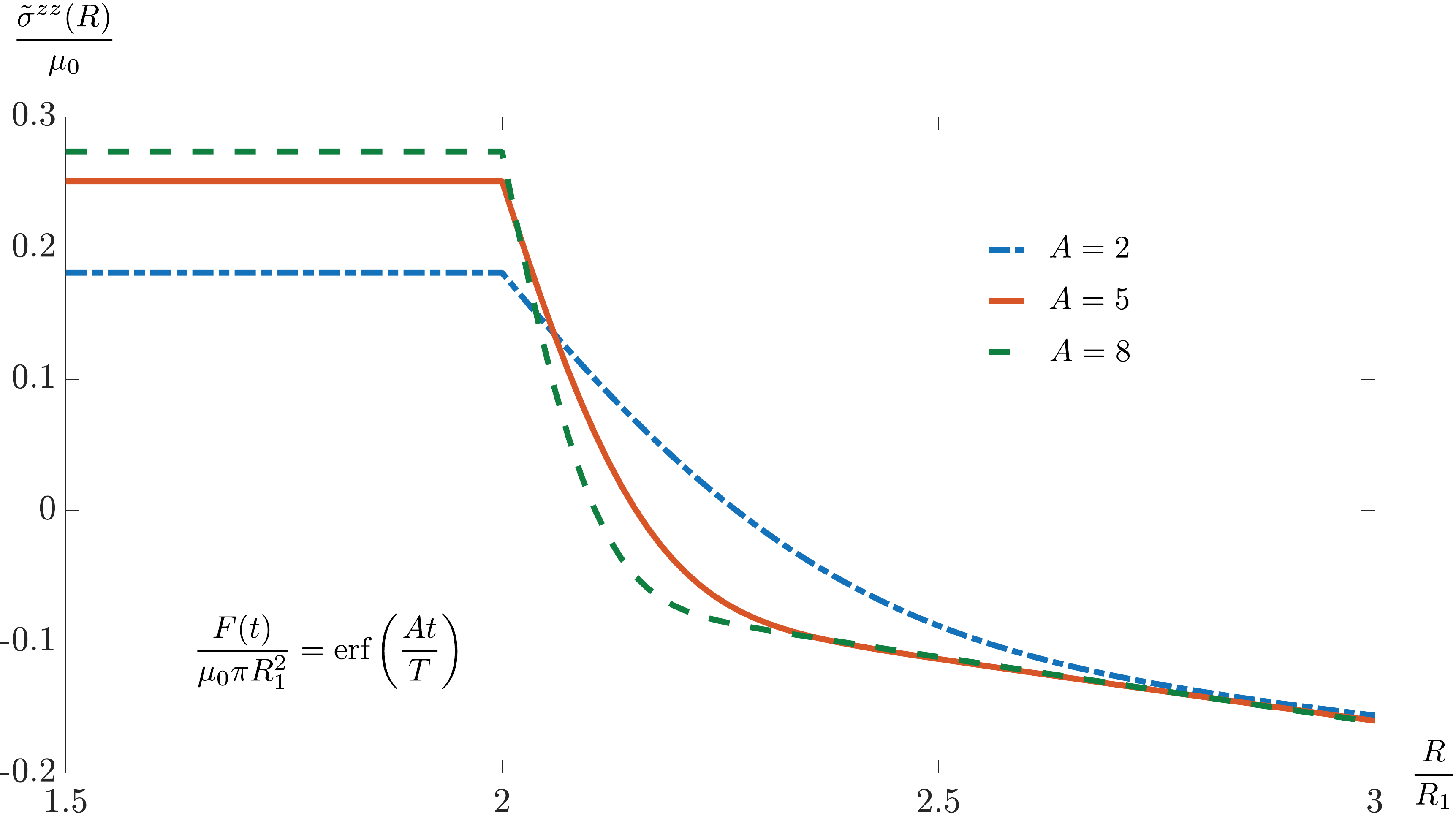}
\vspace*{0.2in}
\caption{Residual stress $\tilde{\sigma}^{zz}(R)$ in the body (Example \ref{Ex4}) when it is set free after the loading $F(t)= \mu_0 \pi R_1^2\, \mathsf{erf}\left( \frac{At}{T} \right)$, where $A\in \{2,5,8\}$. }
\label{fig:ForceControlResErf}
\end{figure}
\begin{figure}[t!]
\centering
\vskip 0.3in
\includegraphics[width=0.75\textwidth]{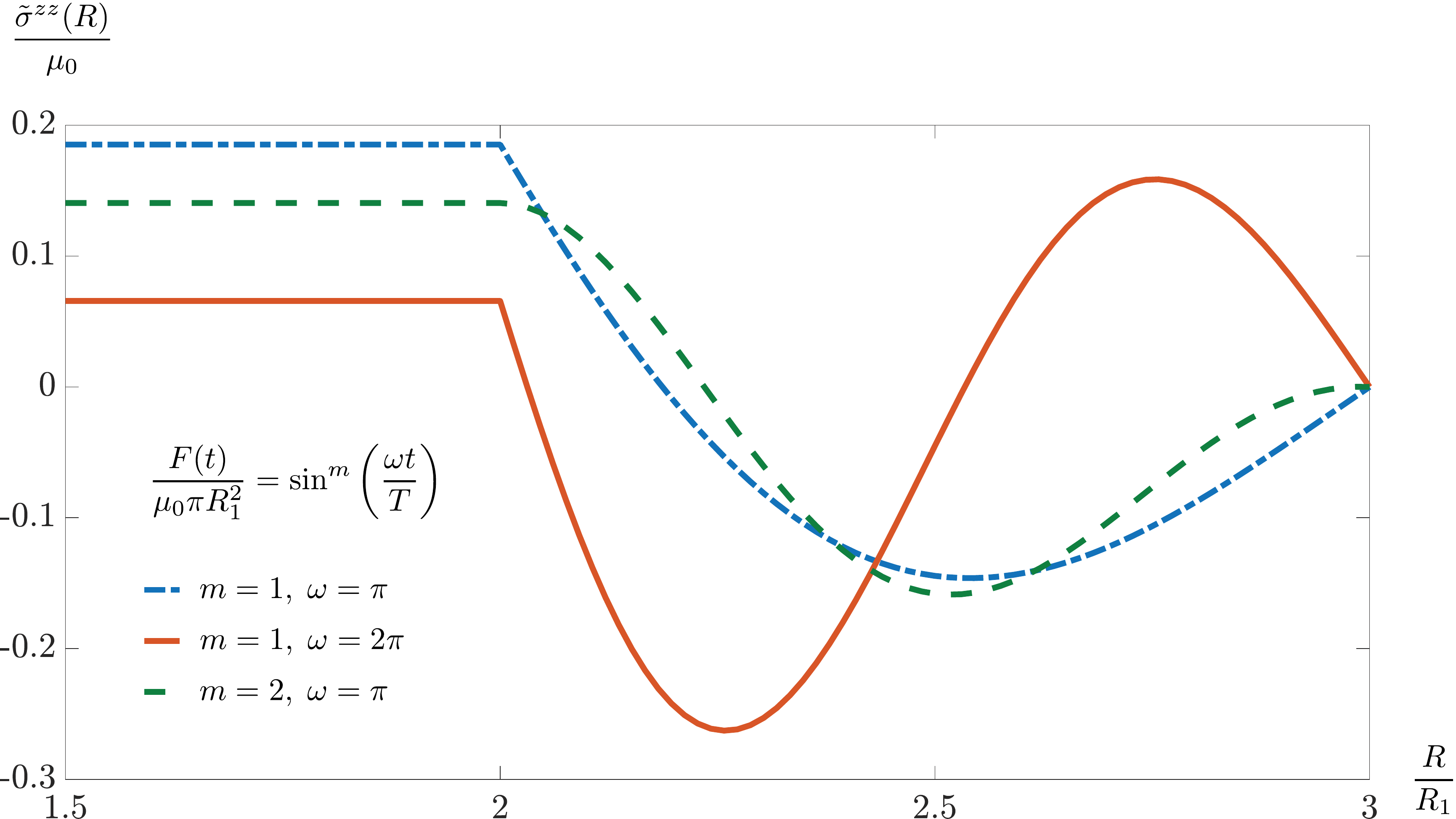}
\vspace*{0.2in}
\caption{Residual stress $\tilde{\sigma}^{zz}(R)$ in the final body (Example \ref{Ex4}). During the accretion-ablation process this body was under a sinusoidal axial force.}
\label{fig:ForceControlResTrig}
\end{figure}
\end{example}

\subsection{Response of a $3$D-printed bar under axial loads}

Let us assume that $T_1$ and $T_2$ are such that $0<T<T_1<T_2$, and consider a $3$D-printed thick hollow cylinder in the time interval $[0,T_2]$. Accretion/ablation occurs during the interval $[0,T]$ under some time-dependent axial load, after which the body is in an unloaded state until $t=T_1$, and service loads are applied during $[T_1,T_2]$. The motion map has the following representation 
\begin{equation}
\varphi_t(R,\Theta,Z)= 
\begin{cases}
\big(\, r(R,t), \Theta, \lambda^2(t)Z \,\big) \,,\quad &  t\leq T \,,\\
\big(\, \tilde{r}(R), \Theta, \tilde{\lambda}^2 Z\,\big) \,,\quad & T< t<T_1\,, \\
\big(\, r_s(R,t), \Theta, \lambda_s^2(t)\tilde{\lambda}^2 Z \,\big) \,,\quad &  T_1\leq t\leq T_2 \,,
\end{cases}
\end{equation}
where $S_1(t) \leq R\leq S_2(t)$, and $\lambda_s^2(t)$ is the axial stretch due to the service load. The functions $\lambda$, $\lambda_s$ are assumed to satisfy the following conditions: $\lambda(0)=1$, $\lambda_s(T_1)=1$. Further, notice that
\begin{equation}
S_1(t)= 
\begin{cases}
R_1+ \velab t \,,\quad &  t\leq T\,, \\
S_1(T)  \,,\quad &  t >T\,
\end{cases}
\;,\qquad
S_2(t)= 
\begin{cases}
R_2+ \velac t \,,\quad &  t\leq T\,, \\
S_2(T)  \,,\quad &  t >T\,,
\end{cases}
\end{equation}
so that they are invertible only in $[0,T]$, with their inverses being $\tab$ and $\tac$, respectively.
Axial force and the Cauchy stress are written as 
\begin{equation}
F(t)= 
\begin{cases}
\mu_0 \pi R_1^2 \; f_l(t) \,,\quad &  t\leq T\,, \\
0 \,,\quad &  T<t<T_1\,, \\
\mu_0 \pi R_1^2 \;  f_s(t) \,,\quad &  T_1\leq t\leq T_2 \,,
\end{cases}
\qquad
\bm{\sigma}(R,t)= 
\begin{cases}
\bm{\sigma}_l(R,t) \,,\quad &  t\leq T\,, \\
\tilde{\bm{\sigma}}(R)\,,\quad &  T<t<T_1\,, \\
\bm{\sigma}_s(R,t) \,,\quad &  T_1\leq t\leq T_2 \,,
\end{cases}
\end{equation}
where the functions $f_l$, $f_s$ are the dimensionless axial forces during the acrretion-ablation process and service loading, respectively, and are assumed to satisfy the following conditions: $f_l(0)=0$, $f_s(T_1)=0$. The $zz$-component of the Cauchy stress during service loading is written as
\begin{equation}
	\sigma_s^{zz}(R,t)= 
	\begin{cases}
	\displaystyle \lambda_s^4(t)\tilde{\lambda}^4-\frac{1}{\lambda_s^2(t)\tilde{\lambda}^2} \,,\quad 
	& S_1(T)\leq R\leq R_2\,, \\[12pt]
	\displaystyle\frac{\lambda_s^4(t)\tilde{\lambda}^4}{\lambdac^4(R)}
	-\frac{\lambdac^2(R)}{\lambda_s^2(t)\tilde{\lambda}^2} \,,\quad &  R_2\leq R \leq S_2(T)\, ,
\end{cases}
\end{equation}
which implies that
\begin{equation} \label{axforceresp}
	\frac{F(t)}{2\pi\mu_0}  =  \frac{R_2^2-S_1^2(T)}{2}\left[\lambda_s^2(t)\tilde{\lambda}^2-\frac{1}{\lambda_s^4(t)\tilde{\lambda}^4}\right] + \lambda_s^2(t)\tilde{\lambda}^2 \int_{R_2}^{S_2(T)}\frac{R\,\text{d}R}{\lambdac^4(R)} - \frac{1}{\lambda_s^4(t)\tilde{\lambda}^4} \int_{R_2}^{S_2(T)}R \lambdac^2(R) \text{d}R   \,,
\end{equation}
where $T_1\leq t\leq T_2 $. This can be rearranged as
\begin{equation} 
	\frac{F(t)}{2\pi\mu_0}  =  \lambda_s^2(t)\tilde{\lambda}^2\left[  \frac{R_2^2-S_1^2(T)}{2}+\int_{R_2}^{S_2(T)}\frac{R\,\text{d}R}{\lambdac^4(R)}\right] - \frac{1}{\lambda_s^4(t)\tilde{\lambda}^4} \left[  \frac{R_2^2-S_1^2(T)}{2}+ \int_{R_2}^{S_2(T)}R \lambdac^2(R) \text{d}R \right]   \,,
\end{equation}
with 
\begin{equation}
	\tilde{\lambda}^6= \frac{ R_2^2-S_1^2(T)
	+ 2\int_{R_2}^{S_2(T)}R \lambdac^2(R) \text{d}R }{ R_2^2-S_1^2(T)
	+ 2\int_{R_2}^{S_2(T)}\frac{R}{\lambdac^4(R)}\text{d}R}\,.
\end{equation}
This implies the following force-stretch relationship
\begin{equation} \label{FlsqManf}
	  f_s(t) = a\left[\lambda_s^2(t) - \frac{1}{\lambda_s^4(t)}   \right]\,,
\end{equation}
where
\begin{equation} \label{aVal}
	a= \frac{ 1}{R_1^2} \left[  R_2^2-S_1^2(T)
	+ 2\int_{R_2}^{S_2(T)}R \lambdac^2(R) \text{d}R \right]^\frac{1}{3} 
	\left[ R_2^2-S_1^2(T)+ 2\int_{R_2}^{S_2(T)}\frac{R}{\lambdac^4(R)}\text{d}R\right]^\frac{2}{3}\,.
\end{equation}

\paragraph*{A stress-free elastic body with the same size as the $3$D-printed body.}
First observe that in the absence of accretion/ablation (i.e., $\velac=\velab=0$) Eq.~\eqref{axforceresp} simplifies to read
\begin{equation} 
	\frac{F(t)}{2\pi\mu_0}  =  \frac{R_2^2-R_1^2}{2}\left[\lambda^2(t)-\frac{1}{\lambda^4(t)}\right]     \,.
\end{equation}
In this problem, we replace $R_1$ by $\tilde{r}(S_1(T))$, $R_2 $ by $ \tilde{r}(S_2(T))$, and $L$ by $\tilde{\lambda}^2 L$.
Consider a stress-free thick cylinder of inner radius $\tilde{r}(S_1(T))=\frac{S_1(T)}{\tilde{\lambda}}$, outer radius $\tilde{r}(S_2(T))=\frac{\lambda(T) s_2(T)}{\tilde{\lambda}}$, initial length $\tilde{\lambda}^2 L$ and subject it to the same service load during the time interval $[T_1, T_2]$. For this new problem, let the motion be denoted as
\begin{equation}
\accentset{\circ}{\varphi}_t(R,\Theta,Z)= 
\big(\, \accentset{\circ}{r}_s(R,t), \Theta, \accentset{\circ}{\lambda}_s^2(t) Z \,\big) \,,
\end{equation}
where $\frac{S_1(T)}{\tilde{\lambda}} \leq R \leq \frac{\lambda(T) \,s_2(T)}{\tilde{\lambda}}$, and $0 \leq Z\leq \tilde{\lambda}^2 L $. Let the axial force be $F_s(t)= \mu_0 \pi R_1^2 \;  f_s(t)$, and denote the Cauchy stress by $\accentset{\circ}{\bm{\sigma}}_s$. For $\tilde{r}(S_1(T))\leq \tilde{R}\leq \tilde{r}(S_2(T))$,
\begin{equation}
\accentset{\circ}{\sigma}_s^{zz}(\tilde{R},t)= \mu_0\left[\accentset{\circ}{\lambda}_s^4(t)	-\frac{1}{\accentset{\circ}{\lambda}_s^2(t)}\right] \,,
\end{equation}
which implies that the axial force can be expressed as
\begin{equation} 
	\frac{F_s(t)}{2\pi\mu_0}  =  \frac{\lambda^2(T)s_2^2(T)-S_1^2(T)}{2 \tilde{\lambda}^2}\left[\accentset{\circ}{\lambda}_s^2(t)-\frac{1}{\accentset{\circ}{\lambda}_s^4(t)}\right]     \,.
\end{equation}
Thus, we have the following force-stretch relationship
\begin{equation} \label{FlsqStFr}
	f_s(t)
	= \accentset{\circ}{a}\left[\accentset{\circ}{\lambda}_s^2(t)
	-\frac{1}{\accentset{\circ}{\lambda}_s^4(t)}\right]     \,,
\end{equation}
where 
\begin{equation} \label{aCircVal}
	\accentset{\circ}{a} 
	= \frac{\left[\lambda^2(T)s_2^2(T)-S_1^2(T)\right]\left[  R_2^2-S_1^2(T)
	+ 2\int_{R_2}^{S_2(T)}R \lambdac^2(R) \text{d}R \right]^\frac{1}{3}}
	{R_1^2 \left[  R_2^2-S_1^2(T)+ 2\int_{R_2}^{S_2(T)}\frac{R}{\lambdac^4(R)} \,\text{d}R \right]^\frac{1}{3} }   \,.
\end{equation}
Note that if $a> \accentset{\circ}{a}$, then the $3$D-printed body is stiffer in comparison to a stress-free body of the same dimensions, and vice-versa.

\begin{example} \label{Ex5}
Consider $R_2=2R_1$, $\velab=\frac{R_1}{2T}$, $\velac=\frac{R_1}{T}$, $T_1= 2T$, and $T_2=3T$.  Then, $\timeablation=2T>T$, $S_1(T)=1.5 R_1$, and $S_2(T)=3 R_1$, where we assume the numerical values $R_1=1$, and $T=1$ as in Example \ref{Ex4}.  
The values of $a$ and $\accentset{\circ}{a}$ for several loads (same as those from Example \ref{Ex4}) are given in Table \ref{table:serviceloadresponse}. 
The force-stretch relationship during the service loading is shown in Fig.~\ref{fig:serviceresponse} for two particular cases. It is observed that the $3$D-printed body is less stiff than a stress-free body of the same size and made of the same material provided that it was subjected to monotonic tensile loading during the accretion-ablation time interval. Similarly, a body which was under monotonic compressive loading during accretion-ablation time interval is stiffer (in tension) than a stress-free body of the same size and made of the same material.
\begin{table} 
\centering
\begin{tabular}{cccccccc} 
$f_l(t)$ &  & $a$  & $\accentset{\circ}{a}$\\[8pt]
\hline
\multirow{3}{4em}{$\text{erf}\left(\frac{At}{T}\right)$} & $A=2$ & $6.3782$ & $6.5900$\\ 
& $A=5$ & $6.2486$ & $6.5388$\\ 
& $A=8$ & $6.2076$ & $6.5233$\\ 
\hline
\multirow{3}{4em}{$a \left(\frac{t}{T}\right)^n$} & $a=5$, $n=1$ & $5.7295$ & $6.2902$\\ 
& $a=5$, $n=2$ & $6.1431$ & $6.4657$\\ 
& $a=-5$, $n=1$ & $8.0862$ & $7.3032$\\ 
& $a=-5$, $n=2$ & $7.5304$ & $7.0902$\\ 
\hline
\multirow{3}{4em}{$\sin^m\left(\frac{t}{T}\right)$} & $m=1$, $\omega=\pi$ & $6.3695$ & $6.5918$\\ 
& $m=1$, $\omega=2\pi$  & $6.6185$ & $6.7104$\\ 
& $m=2$, $\omega=\pi$  & $6.4570$ & $6.6276$\\ 
\hline
\end{tabular}
\caption{The coefficients of the force-stretch relations (see equations \eqref{FlsqManf} and \eqref{FlsqStFr}) for a $3$D-printed bar (see \eqref{aVal}) and a stress-free bar of the same size and made of the same material (see \eqref{aCircVal}).}
\label{table:serviceloadresponse}
\end{table}
\begin{figure}[t!]
\centering
\vskip 0.3in
\includegraphics[width=0.96\textwidth]{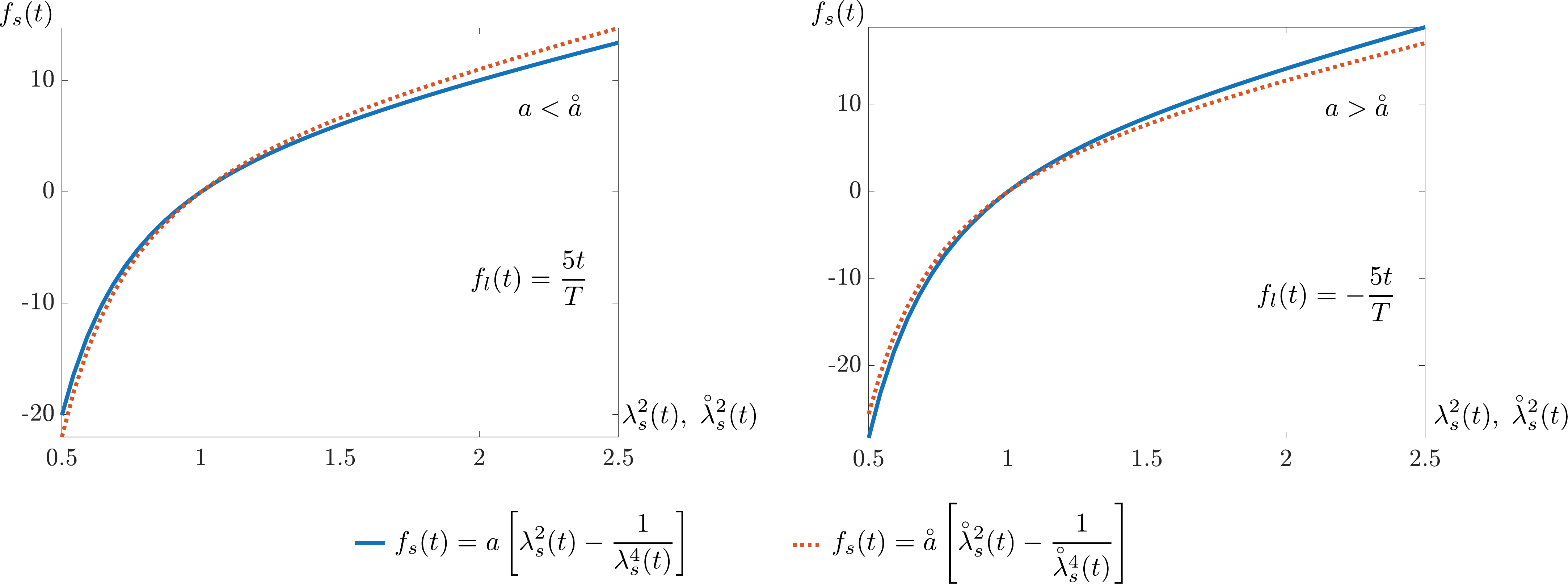}
\vspace*{0.1in}
\caption{Force-stretch relationship for service loading (Example \ref{Ex5}). In the case of $f_l(t)= \frac{5t}{T}$ the $3$D-printed body is less stiff as compared to its corresponding body without any residual stress. When $f_l(t)=-\frac{5t}{T}$, the $3$D-printed body is stiffer than its corresponding stress-free body of the same size.}
\label{fig:serviceresponse}
\end{figure}
\end{example}

\begin{example} \label{Ex6}
Let $R_2=2R_1$, $\velab=\frac{R_1}{2T}$, $\velac=\frac{R_1}{T}$, $T_1= 2T$, and $T_2=3T$ as in Example \ref{Ex5}. When the service load is constant, the Cauchy stress is a function of the radial coordinate alone. We consider the time-dependent loads $f_l(t)= \pm\frac{5t}{T}$ during the accretion-ablation process, and the constant service loads $f_s(t)= \pm 10$. The solid-blue curves in Fig.~\ref{fig:servicestress} represent the following sets 
\begin{equation} 
	\left\{\left(\tilde{r}(R), \frac{\sigma_s^{zz}(R)}{\mu_0}\right):S_1(T) \leq R\leq S_2(T)\right\}\,,
\end{equation}
which show the radial variation of $\sigma_s^{zz}$ resulting from a constant axial force applied to the $3$D-printed body. When a stress-free body of the same size as the $3$D-printed one is subjected to the same service load, it develops a constant $\accentset{\circ}{\sigma}_s^{zz}$ across the cross-section (this is shown by the dashed-red curves in Fig.~\ref{fig:servicestress}). 
\begin{figure}[t!]
\centering
\vskip 0.3in
\includegraphics[width=0.97\textwidth]{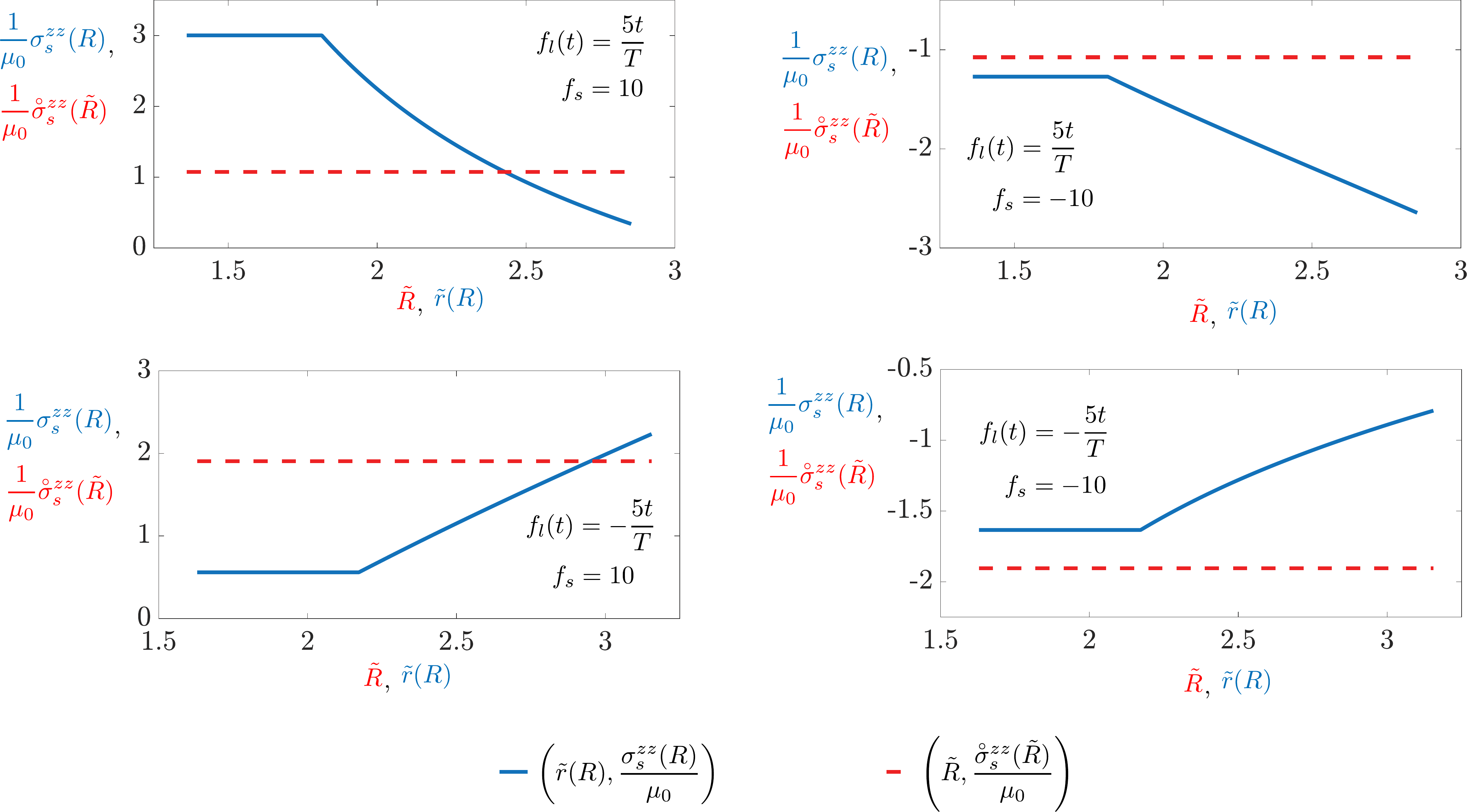}
\vspace*{0.1in}
\caption{The $zz$-component of the Cauchy stress generated by a constant service load (Example \ref{Ex6}). The solid-blue curves show the radial variation of the $zz$-component of the Cauchy stress in the $3$D-printed body. The dashed-red curves represent the same for a stress-free body of the same size and made of the same material as the $3$D-printed one.}
\label{fig:servicestress}
\end{figure}

The $3$D-printed body manufactured under the tensile load $f_l(t)= \frac{5t}{T}$ has tensile residual stress in its inner layers and compressive residual stress in the outer layers. As a result, when a tensile service load is applied, it always develops tensile stress in its inner layers while the stress in the outer layers can be compressive for very small service loads. For larger loads that causes tensile $\sigma_s^{zz}$ throughout the cross-section, $\sigma_s^{zz}$ is greater than $\accentset{\circ}{\sigma}_s^{zz}$ in the inner layers, while it is the opposite for the outer layers. If this body manufactured under the tensile load is subjected to compression, $\sigma_s^{zz}$ can still be tensile in the inner layers for small enough loads. It is possible to have slightly larger compressive service load so that the inner layers have compressive stress, but lesser than that of a stress-free body of same size subjected to the same load, while the outer layers follow the opposite trend. However, if the compressive service load is large enough, the stresses in the $3$D-printed body are more compressive than those of the stress-free body throughout the cross-section.

The residual stress in the $3$D-printed body manufactured under the compressive load $f_l(t)= -\frac{5t}{T}$ is compressive in its inner layers and tensile in the outer layers. If tensile service loads are large enough, $\sigma_s^{zz}$ in the $3$D-printed body is always tensile (with the inner layers being less stressed than the outer ones) but less stressed than the stress-free body of the same size subjected to the same service load. Similarly, if compressive service loads are large enough, $\sigma_s^{zz}$ in the $3$D-printed body is always compressive (with the inner layers being more stressed than the outer ones) but less stressed than the stress-free body of the same size subjected to the same service load.
\end{example}

\section{Conclusions} \label{Sec:Conclusions}

In this paper we presented a geometric formulation of the nonlinear mechanics of accreting-ablating bodies. This theory models the large deformations of bodies that undergo simultaneous accretion and ablation on their boundaries. This is a generalization of the accretion theory that was formulated in \citep{Sozio2019}.
In this formulation the natural (stress-free) configuration of the body is a time-dependent Riemannian manifold.
Formulating the accretion-ablation boundary-value problem requires the construction of the material metric for the accreted portion of the body. For ablation we simply need a map to track the points on the boundary being ablated at any given time. However, we do not need to track the particles after they have left the body. The material metric is an unknown field a priori and is determined after solving the accretion-ablation initial-boundary-value problem.
This theory is not restricted to isotropic materials; the initial body can be made of any anisotropic material. The accreting particles can be anisotropic as well. Also, the material points joining the body can be stressed at their time of attachment.

In the second part of the paper we considered an incompressible thick hollow cylinder undergoing accretion on its outer boundary and ablation on its inner boundary while it is being loaded axially. For the sake of simplicity, we assumed a homogeneous and isotropic material both in the initial body and accreting particles. Also, to simplify kinematics we assumed that the initial body is made of an incompressible solid. It is also assumed that the material that is added to the body during the accretion process is incompressible as well.  
We derived the governing equations for two cases: i) when $t<t_{\text{ablation}}$ a portion of the initial body is still present, and ii) the initial body has been completely ablated. 
Assuming constant accretion and ablation velocities we solved the problem for case i). 
We considered two different loading scenarios: displacement-control and force-control loadings. 
In the case of displacement-control loading we obtained a semi-analytical solution for the problem. Further, we calculated the residual stresses after the completion of the accretion-ablation processes and the removal of external forces. Given a time-dependent axial stretch during loading, we provided analytical expressions for both the residual stretch and axial residual stress.

A future extension of this work would be to study problems where the accretion and ablation velocities are dictated by mass transport and heat transfer. Studying large elastic deformations in accretion-ablation problems coupled with phase changes will be the subject of a future communication.

\section*{Acknowledgement}

This research was partially supported by NSF -- Grant No. CMMI 1939901, and  ARO Grant No. W911NF-18-1-0003.

\bibliographystyle{plainnat}
\bibliography{ref}
\appendix

\section{The relation between $r(R,t)$ and $\lambda(t)$ for an accreting-ablating incompressible thick hollow cylinder under finite extension} \label{AppendixA}

The traction continuity equation \eqref{tractcont} can be rearranged as follows:
\begin{equation} \label{Rearranged}
\int_{S_1(t)}^{R_2}\left[\frac{1}{R}-\frac{R^3 }{r^4(R,t) \lambda^4(t)}\right]\text{d}R +\int_{R_2}^{S_2(t)}\frac{\lambdac^2(R)}{\rac(R)}\left[1-\frac{\rac^4(R)\lambdac^4(R) }{r^4(R,t)\lambda^4(t)}\right]\text{d}R=0\,.
\end{equation}
For the sake of convenience, let us denote
\begin{equation}
\begin{aligned} 
A(t):= &\int_{S_1(t)}^{R_2}\left[\frac{1}{R}-\frac{R^3 }{r^4(R,t) \lambda^4(t)}\right]\text{d}R
 \,,\qquad
B(t):= \int_{R_2}^{S_2(t)}\frac{\lambdac^2(R)}{\rac(R)}\left[1-\frac{\rac^4(R)\lambdac^4(R) }{r^4(R,t)\lambda^4(t)}\right]\text{d}R=0\,.
\end{aligned}
\end{equation}
Observe that \eqref{rPrim} and \eqref{rsecusings1} can be combined to write
\begin{equation}
r^2(R,t) \lambda^2(t) =
\begin{dcases}
	s_1^2(t)\lambda^2(t)-S_1^2(t)+R^2 	\,, &  R \leq R_2\,, \\
	s_1^2(t)\lambda^2(t)-S_1^2(t)+R_2^2 +2\int_{R_2}^{R} \lambdac^2(\zeta)\, \rac(\zeta) \,\text{d}\zeta\,, 
	& R\geq R_2 \,.
\end{dcases}
\end{equation}
Let us define the following two functions
\begin{equation}
	\chi(t):=s_1^2(t)\lambda^2(t)-S_1^2(t)\,,\qquad 
	\psi(R):=\begin{dcases}
	R^2 	\,, &  R \leq R_2\,, \\
	R_2^2 +2\int_{R_2}^{R} \lambdac^2(\zeta)\, \rac(\zeta) \,\text{d}\zeta\,, 
	& R\geq R_2 \,,
\end{dcases}
\end{equation}
so that $r^2(R,t) \lambda^2(t) = \chi(t)+\psi(R)$. Further,
\begin{equation}
\psi'(R)=\begin{dcases}
	2R 	\,, &  R \leq R_2\,, \\
	2\lambdac^2(R)\, \rac(R) \,, 	& R\geq R_2 \,.
\end{dcases}
\end{equation}
Now, one can write
\begin{equation} \label{A-dot}
\begin{aligned}
\dot{A}(t) &=  \int_{S_1(t)}^{R_2} \frac{\partial}{\partial t}\left[\frac{1}{R}-\frac{R^3 }{r^4(R,t) \lambda^4(t)}\right]\text{d}R -  \left[\frac{1}{S_1(t)}-\frac{ S_1^3(t) }{r^4(S_1(t),t) \lambda^4(t)}\right]\dot{S_1}(t) \\
&= -\int_{S_1(t)}^{R_2} \frac{\partial}{\partial t}\left[\frac{R^3 }{[\chi(t)+\psi(R)]^2}\right]\text{d}R -  \velab\left[\frac{1}{S_1(t)}-\frac{ S_1^3(t) }{s_1^4(t) \lambda^4(t)}\right] \\
&= \int_{S_1(t)}^{R_2} \frac{2 \dot{\chi}(t)R^3 }{[\chi(t)+\psi(R)]^3}\text{d}R -  \velab\left[\frac{1}{S_1(t)}-\frac{ S_1^3(t) }{s_1^4(t) \lambda^4(t)}\right] \\
&= \frac{\dot{\chi}(t)}{4}\int_{S_1(t)}^{R_2} \frac{[\psi'(R)]^3 \text{d}R}{[\chi(t)+\psi(R)]^3} -  \velab\left[\frac{1}{S_1(t)}-\frac{ S_1^3(t) }{s_1^4(t) \lambda^4(t)}\right] \,,
\end{aligned}
\end{equation}
and
\begin{equation} \label{B-dot}
\begin{aligned}
	\dot{B}(t) &= - \int_{R_2}^{S_2(t)} \frac{\partial}{\partial t}
	\left[\frac{\rac^3(R)\lambdac^6(R) }{r^4(R,t)\lambda^4(t)}\right]\text{d}R 
	+\frac{\lambdac^2(S_2(t))}{\rac(S_2(t))}\left[1
	-\frac{\rac^4(S_2(t))\lambdac^4(S_2(t))}{r^4(S_2(t),t)\lambda^4(t)}\right] \dot{S_2}(t) \\ 
	&= - \int_{R_2}^{S_2(t)} \frac{\partial}{\partial t}\left[\frac{\rac^3(R)\lambdac^6(R)}
	{[\chi(t)+\psi(R)]^2}\right]\text{d}R\\
	&= \int_{R_2}^{S_2(t)} \frac{2 \dot{\chi}(t) \,\rac^3(R)\lambdac^6(R) }{[\chi(t)+\psi(R)]^3}\text{d}R \\
	&= \frac{\dot{\chi}(t)}{4}\int_{R_2}^{S_2(t)} \frac{[\psi'(R)]^3 \text{d}R}{[\chi(t)+\psi(R)]^3} \,.
\end{aligned}
\end{equation}
Differentiating \eqref{Rearranged} with respect to time and using \eqref{A-dot} and \eqref{B-dot} one obtains
\begin{equation}
	\frac{\dot{\chi}(t)}{4}\int_{S_1(t)}^{S_2(t)} \frac{[\psi'(R)]^3 \text{d}R}{[\chi(t)
	+\psi(R)]^3}  =\velab\left[\frac{1}{S_1(t)}-\frac{ S_1^3(t) }{s_1^4(t) \lambda^4(t)}\right]\,.
\end{equation}
Note that
\begin{equation}
	\frac{1}{S_1(t)}-\frac{ S_1^3(t) }{s_1^4(t) \lambda^4(t)} 
	=  \frac{\chi(t)\,\left[\,\chi(t)+2S_1^2(t)\,\right]\,}{S_1(t)\,\left[\, \chi(t)+S_1^2(t)\,\right]^2}\,,
\end{equation}
and hence
\begin{equation}
	\frac{\dot{\chi}(t)}{4}\int_{S_1(t)}^{S_2(t)} \frac{[\psi'(R)]^3 \text{d}R}{[\chi(t)+\psi(R)]^3}  
	= \frac{\velab \, \chi(t)\,\left[\,\chi(t)+2S_1^2(t)\,\right]\,}{S_1(t)\,\left[\, \chi(t)+S_1^2(t)\,\right]^2}\,.
\end{equation}
Let us define
\begin{equation}
	\gamma(t):= -\int_0^t\frac{4\velab \, \left[\,\chi(\tau)+2S_1^2(\tau)\,\right]\text{d}\tau}{S_1(\tau)\,
	\left[\, \chi(\tau)+S_1^2(\tau)\,\right]^2\, \int_{S_1(\tau)}^{S_2(\tau)} 
	\frac{[\psi'(R)]^3}{[\chi(\tau)+\psi(R)]^3}\text{d}R }\,.
\end{equation}
Thus, we have
\begin{equation}
	\dot{\chi}(t)+\dot{\gamma}(t) \chi(t)=0\,,\quad \text{or}  \qquad
	\frac{\text{d}}{\text{d}t}\left[e^{\gamma(t)}\chi(t)\right]=0\,.
\end{equation}
Integrating this ODE one obtains
\begin{equation}
	e^{\gamma(t)}\chi(t)=e^{\gamma(0)}\chi(0)\,.
\end{equation}
Now since $\gamma(0)=0$, and $\chi(0)=0$ (as $s_1(0)=S_1(0)=R_1$ and $\lambda(0)=1$), we conclude that $\chi(t)=0$, i.e. $s_1^2(t)\lambda^2(t)=S_1^2(t)$ for all $t<\timeablation$. Thus,
\begin{equation}
r^2(R,t) \lambda^2(t) =
\begin{dcases}
	R^2 	\,, &  R \leq R_2\,, \\
	R_2^2 +2\int_{R_2}^{R} \lambdac^2(\zeta)\, \rac(\zeta) \,\text{d}\zeta\,, 
	& R\geq R_2 \,,
\end{dcases}
\end{equation}
is independent of time. Further, substituting $t=\tac(R)$ one obtains
\begin{equation}
\rac^2(R) \lambdac^2(R)= R_2^2 +2\int_{R_2}^{R} \lambdac^2(\zeta)\, \rac(\zeta) \,\text{d}\zeta\,, 
\end{equation}
which when differentiated with respect to $R$ gives
\begin{equation}
	 \frac{\text{d}}{\text{d}R}\left[ \rac(R) \lambdac(R) \right]=\lambdac(R)\,.
\end{equation}
Integrating this along with $\rac(R_2)=R_2$, we finally obtain
\begin{equation}
 \rac(R)=\frac{R_2+ \int_{R_2}^R \lambdac(\xi)\text{d}\xi}{\lambdac(R)}\,,
\end{equation}
and hence
\begin{equation}
r(R,t) =
\begin{cases} 
	\displaystyle \frac{R}{\lambda(t)} \,, & S_1(t) \leq R \leq R_2\,, \\[10pt]
	\displaystyle \frac{\lambdac(R)\, \rac(R) }{\lambda(t)}\,,  & R_2 \leq R \leq S_2(t) \,.
\end{cases}
\end{equation}

\end{document}